\documentclass[useAMS,usenatbib]{mn2e}
\usepackage{epsfig}
\usepackage{amsmath}
\usepackage{natbib}
\usepackage{color}
\usepackage{lscape}
\usepackage[totalwidth=17.8cm, totalheight=24.0cm]{geometry} 
\usepackage{graphicx}

\newcommand{\Msol}{M$_\odot$}

\title[]{Evolutionary Stellar Population Synthesis with MILES. Part
I: The Base Models and a New Line Index System}

\author[]{A. Vazdekis$^{1,2}$\thanks{E-mail: vazdekis@iac.es},
P. S\'anchez-Bl\'azquez$^{1,2}$, 
J. Falc\'on-Barroso$^{1,2}$, 
A. J. Cenarro$^{1,2,3}$,
\newauthor
M. A. Beasley$^{1}$ 
N. Cardiel$^{4}$, 
J. Gorgas$^{4}$, 
R. F. Peletier$^{5}$\\
$^{1}$Instituto de Astrof\'{\i}sica de Canarias, La Laguna, E-38200 Tenerife, Spain\\
$^{2}$Departamento de Astrof\'{\i}sica, Universidad de La Laguna, E-38205 La Laguna, Tenerife, Spain\\
$^{3}$Centro de Estudios de F\'{\i}sica del Cosmos de Arag\'on, C/ General Pizarro 1, E-44001, Teruel, Spain\\
$^{4}$Dept. de Astrof\'{\i}sica, Fac. de Ciencias F\'{\i}sicas, Universidad Complutense de Madrid, E-28040 Madrid, Spain\\
$^{5}$Kapteyn Astronomical Institute, University of Groningen, Postbus 800, 9700 AV, Groningen, Netherlands\\
}
\begin{document}

\date{}
\pagerange{\pageref{firstpage}--\pageref{lastpage}} \pubyear{2009}
\maketitle
\label{firstpage}
\begin{abstract}

We present synthetic spectral energy distributions (SEDs) for single-age,
single-metallicity stellar populations (SSPs) covering the full optical 
spectral range at moderately high resolution (FWHM$=$2.3\AA). These SEDs
constitute our base models, as they combine scaled-solar isochrones with a
empirical stellar spectral library (MILES), which follows the chemical
evolution pattern of the solar neighbourhood. The models rely as much as
possible on empirical ingredients, not just on the stellar spectra, but also
on extensive photometric libraries, which are used to determine the
transformations from the theoretical parameters of the isochrones to
observational quantities. The  unprecedent stellar parameter coverage of the
MILES stellar library allowed us  to safely extend our optical SSP SEDs
predictions from intermediate- to  very-old age regimes, and the metallicity
coverage of the SSPs from super-solar  to \mbox{$\mbox{[M/H]}=-2.3$}. SSPs
with such low metallicities are particularly  useful for globular cluster
studies. We have computed SSP SEDs for a suite of  IMF shapes and slopes. We
provide a quantitative analysis of the dependence of  the synthesized SSP SEDs
on the (in)complete coverage of the stellar parameter  space in the input
library that not only shows that our models are of higher  quality than other
works, but also in which range of SSP parameters our models  are reliable. The
SSP SEDs are a useful tool to perform the stellar populations  analysis in a
very flexible manner. Observed spectra can be studied by means of  full
spectrum fitting or by using line indices. For the latter we propose a new 
Line Index System (named LIS) to avoid the intrinsic uncertainties associated 
with the popular Lick/IDS system and provide more appropriate, uniform,
spectral  resolution. Apart from constant resolution as function of wavelength
the system  is also based on flux-calibrated spectra. Data can be analyzed at
three different  resolutions: 5\,\AA, 8.4\,\AA\ and 14\,\AA\ (FWHM), which are
appropriate for studying globular cluster, low and intermediate-mass galaxies,
and massive galaxies, respectively. Furthermore we provide polynomials to
transform current Lick/IDS line index measurements to the new system. We
provide line-index tables in the new system for various popular samples of
Galactic globular clusters and galaxies. We apply the models to various
stellar clusters and galaxies with high-quality spectra, for which independent
studies are available, obtaining excellent results. Finally we designed a web
page from which not only these models and stellar libraries can be downloaded
but which also provides a suite of on-line tools to facilitate the handling
and transformation of the spectra.

\end{abstract}


\section{Introduction}

To obtain information about unresolved stellar populations in galaxies, one can look at colours,
ranging from the UV to the infrared. Such a method is popular especially when studying
faint objects. Since colours are strongly affected by dust extinction, one prefers to use
spectra when studying brighter objects. These have the additional advantage that
abundances of various elements can be studied, from the strength of absorption lines
(e.g. Rose 1985; S\'anchez-Bl'azquez et al. 2003; Carretero et al. 2004).
Although the UV and the near IR are promising wavelength regions to study, the large
majority of spectral studies is done in the optical, mainly because of our better
understanding of the region. To obtain information, one generally compares model
predictions with measurements of strong absorption line strengths. This method, which is
insensitive to the effects of dust extinction (e.g., MacArthur 2005) very often uses
the Lick/IDS system of indices, which comprises definitions of 25 absorption line
strengths in the optical spectral range, originally defined on a low resolution stellar
library (FWHM$>$8-11\,\AA) (Gorgas et al. 1993, Worthey et al. 1994, Worthey \& Ottaviani
1997). There is an extensive list of
publications in which this method has been applied, mainly for early-type galaxies
(see for example the compilation provided in Trager et al. 1998).

Predictions for the line-strength indices of the integrated light of
stellar clusters and galaxies are obtained with stellar population synthesis
models. These models calculate galaxy indices by adding the contributions of all
possible stars, in proportions prescribed by stellar evolution models. The required 
indices of all such stars are obtained from observed stellar libraries, but since the
spectra in these libraries were often noisy, and did not contain all types of stars
present in galaxies, people used the so-called fitting functions, 
which relate measured line-strength indices to the atmospheric
parameters ($T_{\mbox{\scriptsize eff}}$, $\log g$, \mbox{$\mbox{[Fe/H]}$}) of
library stars. The most widely used fitting functions are those computed on the
basis of the Lick/IDS stellar library (Burstein et al. 1984; Gorgas et al. 1993;
Worthey et al. 1994) There are, however, alternative fitting functions in the
optical range (e.g., Buzzoni 1995; Gorgas et al. 1999; Tantalo, Chiosi \& Piovan 2007;
Schiavon 2007) and in other spectral ranges (e.g., Cenarro et al. 2002;
M\'armol-Queralt\'o et al. 2008). 

The most important parameters that can be obtained using stellar population synthesis are
age (as a proxy for the star formation history) and metallicity (the average
metal content).  There are several methods to perform stellar population analysis based on
these indices. By far the most popular method is to build key diagnostic model grids by
plotting an age-sensitive indicator (e.g. H$\beta$) versus a metallicity-sensitive
indicator (e.g., Mg$b$, $\langle\mbox{Fe}\rangle$), to estimate the age and metallicity
(e.g. Trager et al. 2000; Kuntschner et al. 2006). There are other methods that, e.g.,
employ as many Lick indices as possible  and simultaneous fit them in a
$\chi^2$ sense (e.g., Vazdekis et al. 1997; Proctor et al. 2004). An alternative method
is the  use of Principal Component Analysis (e.g. Covino et al. 1995; Wild et
al. 2009).

In the last decade the appearance of a generation of stellar population models
that predict full spectral energy distributions (SEDs) at moderately high
spectral resolution has provided new means for improving the stellar population
analysis (e.g., Vazdekis 1999, hereafter V99; Bruzual \& Charlot 2003; Le Borgne
et al. 2004). These models employ newly developed extensive empirical stellar
spectral libraries with flux-calibrated spectral response and good atmospheric
parameter coverage. Among the most popular stellar libraries are those of Jones
(1999), CaT (Cenarro et al. 2001a), ELODIE (Prugniel \& Soubiran 2001), STELIB
(Le Borgne et al. 2003), INDO-US (Valdes et al. 2004), and MILES
(S\'anchez-Bl\'azquez et al. 2006d, hereafter Paper~I). Models that employ such
libraries are, e.g., Vazdekis (1999), Vazdekis et al. (2003) (hereafter V03),
Bruzual \& Charlot (2003), Le Borgne et al. (2004). Alternatively, theoretical
stellar libraries at high spectral resolutions have also been developed for this
purpose (e.g., Murphy \& Meiksin 2004; Zwitter et al. 2004;
Rodr\'{\i}guez-Merino et al. 2005; Munari et al. 2005; Coelho et al. 2005;
Martins \& Coelho 2007). Models that use such libraries are, e.g., Schiavon
et al. (2000), Gonz\'alez-Delgado et al. (2005), Coelho et al. (2007).

There are important limitations inherent to the method that prevent us from easily
disentangling relevant stellar population parameters. The most important one is
the well known age/metallicity degeneracy, which not only affects colours but
also absorption line-strength indices (e.g. Worthey 1994). This degeneracy is
partly due to  the isochrones and partly due to the fact that line-strength
indices change with  both age and metallicity. The effects of this degeneracy
are stronger when low resolution indices are used, as the metallicity lines
appear blended. Alternative indices that were thought to work at higher spectral
resolutions, such as those of Rose (1985, 1994), have been proposed to alleviate
the problem.

New indices with greater abilities to lift the age/metallicity degeneracy have
been proposed with the aid of the new full-SED models (e.g., Vazdekis \& Arimoto 1999;
Bruzual \& Charlot 2003; Cervantes \& Vazdekis 2009). These models
allow us to analyze the whole information contained in the observed spectrum at
once. In fact, there is a growing body of full spectrum fitting algorithms
that are being developed for constraining and recovering in part the Star
Formation History (e.g., Panter et al. 2003; Cid Fernandes et al. 2005; Ocvirk
et al. 2006ab; Koleva et al. 2008). Furthermore the use of these SSP SEDs, which
have sufficiently high spectral resolution, has been shown to significantly improve the
analysis of galaxy kinematics both in the optical (e.g., Sarzi et al. 2006) and in the
near-IR (e.g., Falc\'on-Barroso et al. 2003), for both absorption and emission lines. 

Here we present single-age, single-metallicity stellar population (SSP) SEDs
based on the empirical stellar spectral library MILES that we presented in
Paper~I) and Cenarro et al. (2007, hereafter Paper~II). These models represent
an extension of the V99 SEDs to the full optical spectral range. These MILES-models 
are meant to provide better predictions in the optical range for
intermediate- and old-stellar populations. In Paper~I we provided all
relevant details for MILES, which has been obtained at the 2.5~m Issac Newton
Telescope (INT) at the Observatorio del Roque de Los Muchachos, La Palma. The
library is composed of 985 stars covering the spectral range $\lambda\lambda$
3540-7410\,\AA\ at 2.3\,\AA\ (FWHM). MILES stars were specifically selected 
for population synthesis modeling. In Paper~II we present a homogenized
compilation of the stellar parameters ($T_{\mbox{\scriptsize eff}}$, $\log g$,
\mbox{$\mbox{[Fe/H]}$}) for the stars of this library. In fact the parameter
coverage of MILES constitutes a significant improvement over previous stellar
libraries and allows the models to safely extend the predictions to
intermediate-aged stellar populations and to lower and higher metallicities. 

The models that we present here are based on an empirical library, i.e.
observed stellar spectra, and therefore the synthesized SEDs are imprinted
with the chemical composition of the solar neighbourhood, which is the result
of the star formation history experienced by our Galaxy. As the stellar
isochrones  (Girardi et al. 2000) -- i.e. the other main ingredient feeding
the population synthesis code -- are solar-scaled, our models are
self-consistent,  and scaled-solar for solar metallicity. In the low
metallicity regime, however, our models combine scaled-solar isochrones with
stellar spectra that do not show this abundance ratio pattern (e.g.,
Edvardsson et al. 1993; Schiavon 2007). In a second paper we will present
self-consistent models, both scaled-solar and $\alpha$-enhanced,  for a range
in metallicities. For this purpose we have used MILES, together with
theoretical model atmospheres, which are coupled to the appropriate stellar
isochrones. 

We note, however, that the use of our (base) models, which are not truly
scaled-solar for low metallicities, does not affect in a significant manner the
metallicity and age estimates obtained from galaxy spectra in this metallicity
regime (e.g., Michielsen et al. 2008). For the high metallicity regime, where
massive galaxies reside and show enhanced \mbox{$\mbox{[Mg/Fe]}$} ratios, the
observed line-strengths for various popular indices (e.g., Mg$b$, Fe5270) are
significantly different from the scaled-solar predictions. It has been shown
that by changing the relative element-abundances, and making them
$\alpha$-enhanced, both the stellar models (e.g., Salaris, Groenewegen \& Weiss
2000) and  the stellar atmospheres (e.g. Tripicco \& Bell 1995; Korn, Maraston
\& Thomas 2005;  Cohelo et al. 2007) are affected. However it has also been
shown that base models, such as the ones we are presenting here, can be used to
obtain a good proxy for the \mbox{$\mbox{[Mg/Fe]}$} abundance ratio, if 
appropriate indices are employed for the analysis (e.g. Yamada et al. 2006). In
fact we confirm the results of, e.g., S\'anchez-Bl\'azquez et al. (2006b), de la
Rosa et al. (2007) and Michielsen et al. (2008), who obtain a linear relation
between the proxy for \mbox{$\mbox{[Mg/Fe]}$}, obtained with scaled-solar
models, and the abundance ratio estimated with the aid of models that
specifically take into account non-solar element mixtures (e.g., Tantalo et al.
1998; Thomas et al. 2003; Lee \& Worthey 2005; Graves \& Schiavon 2008).

In Section \ref{sec:models} we describe the main model ingredients, the details
of the stellar spectral library MILES and the steps that have been followed for 
its implementation
in the stellar population models. In Section \ref{sec:MILES_SSP_SEDs} we present
the new single-age, single-metallicity spectral energy distributions, SSP SEDs,
and provide a quantitative analysis for assessing their quality. In Section
\ref{sec:system} we show line-strengths indices as measured on the newly
synthesized SSP spectra and propose a new reference system, as an alternative to the
Lick/IDS system, to work with these indices. In Section \ref{sec:colours}
we provide a discussion about the colours measured on the new SSP SEDs. In Section
\ref{sec:applications} we apply these models to a set of representative stellar
clusters and galaxies to illustrate the capability of the new models. In Section
\ref{sec:web} we present a webpage to facilitate the use and handling of our
models. Finally in Section \ref{sec:conclusions} we provide a summary.

\section{Models}
\label{sec:models}

The SSP SEDs presented here represent an extension to the full optical spectral
range of the V99 models, as updated in V03. We briefly summarize here the main
ingredients and the relevant aspects of this code. 

\subsection{Main ingredients}
\label{sec:ingredients}

 We use  the solar-scaled theoretical isochrones of Girardi et
al. (2000). A wide range of ages and metallicities are covered, including the
latest stages of stellar evolution. A synthetic prescription is used
to include the thermally pulsing AGB regime 
to the point of complete envelope ejection.
The isochrones are computed for six
metallicities $Z=$0.0004, 0.001, 0.004, 0.008, 0.019 and 0.03, where 0.019
represents the solar value. In addition we include an updated version of the
models published in (Girardi et al. 1996) for $Z=$0.0001. The latter calculations 
are now compatible with Girardi et al. (2000). The range of initial stellar masses 
extends from  0.15 to
7${\rm M}_{\odot}$. The input physics of these models has been updated with
respect to Bertelli et al. (1994) with an improved version of the equation of
state, the opacities of Alexander \& Ferguson (1994) (which results in a RGB that is
slightly hotter than in Bertelli et al.) and a milder convective overshoot
scheme. A helium fraction was adopted according to the relation:
$Y\approx0.23+2.25Z$. For the TP-AGB phase Girardi et al. (2000) adopt a simple
synthetic prescription that, for example, does not take into account the third
dredge-up. An improved treatment for this difficult stellar evolutionary phase
has been recently included  by Marigo et al. (2008).
The effects of such improvements become relevant for the
near-IR spectral range (see e.g., Marigo et al. 2008; Maraston 2005).

The theoretical parameters of the isochrones $T_{\mbox{\scriptsize eff}}$, $\log g$,
\mbox{$\mbox{[Fe/H]}$} are transformed to the
observational plane by means of empirical relations between colours and 
stellar parameters (temperature, metallicity and gravity), instead of using  theoretical
stellar atmospheres. We mostly use the metallicity-dependent empirical
relations of Alonso, Arribas \& Mart\'{\i}nez-Roger (1996, 1999; respectively,
for dwarfs and giants). Each of these libraries are composed of $\sim$~500 stars
and the obtained temperature scales are based on the IR-Flux method, i.e. only
marginally dependent on model atmospheres. We use the empirical compilation of
Lejeune, Cuisinier \& Buser (1997, 1998) (and references therein) for the
coolest dwarfs ( $T_{\mbox{\scriptsize eff}}\le4000\,K$ and giants
($T_{\mbox{\scriptsize eff}}\le3500\,K$) for solar metallicity. For these low
temperatures we use a semi-empirical approach to other metallicities on the
basis of these relations and the  model atmospheres of Bessell et al. (1989,
1991) and the library of Fluks et al. (1994). The empirical compilation of
Lejeune et al. (1997,1998) was also used for stars with temperatures above
$\sim$8000\,K. We apply the metal-dependent bolometric corrections given by
Alonso, Arribas \& Mart\'{\i}nez-Roger (1995, 1999; respectively, for dwarfs and
giants). For the Sun we adopt $BC_{\odot}=-0.12$ and a bolometric magnitude of
4.70.

We have computed our predictions for several IMFs: the two power-law
IMFs described in Vazdekis et al. (1996) (i.e unimodal and bimodal), which are
characterized by its slope $\mu$ as a free parameter, and the multi-part
power-law IMFs of Kroupa (2001) (i.e. universal and revised). In our notation
the Salpeter (1955) IMF corresponds to a unimodal IMF with $\mu=1.3$. Further
details of the IMF definitions are given in Appendix~A of V03.  We set the lower
and upper mass-cutoff of the IMF to 0.1 and 100\,\Msol, respectively.  

\subsection{MILES stellar library}
\label{sec:MILES}

\begin{figure*}
\includegraphics[angle=270,width=7.in]{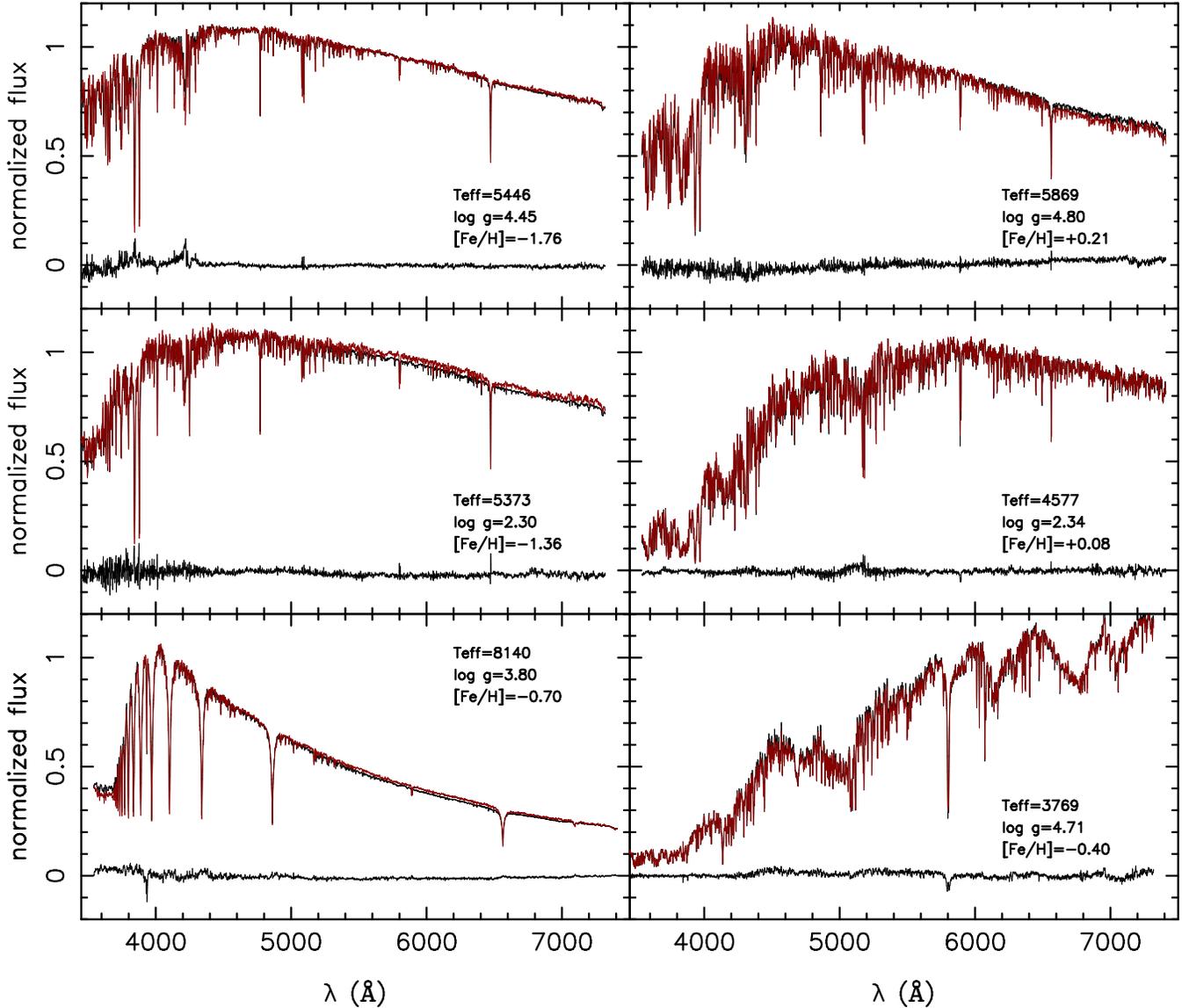}
\caption{MILES stars for various representative spectral types are plotted in
the different panels in black. In each panel we overplot the spectrum corresponding
to a star with identical atmospheric parameters that we have computed on the
basis of the MILES database (except for the star itself) following the algorithm 
described in Section~\ref{sec:synthesis}. The residuals are given in the bottom of each
panel. (see the text for details).}
\label{fig:interpolator}
\end{figure*}

\begin{figure*}
\includegraphics[angle=270,width=7.in]{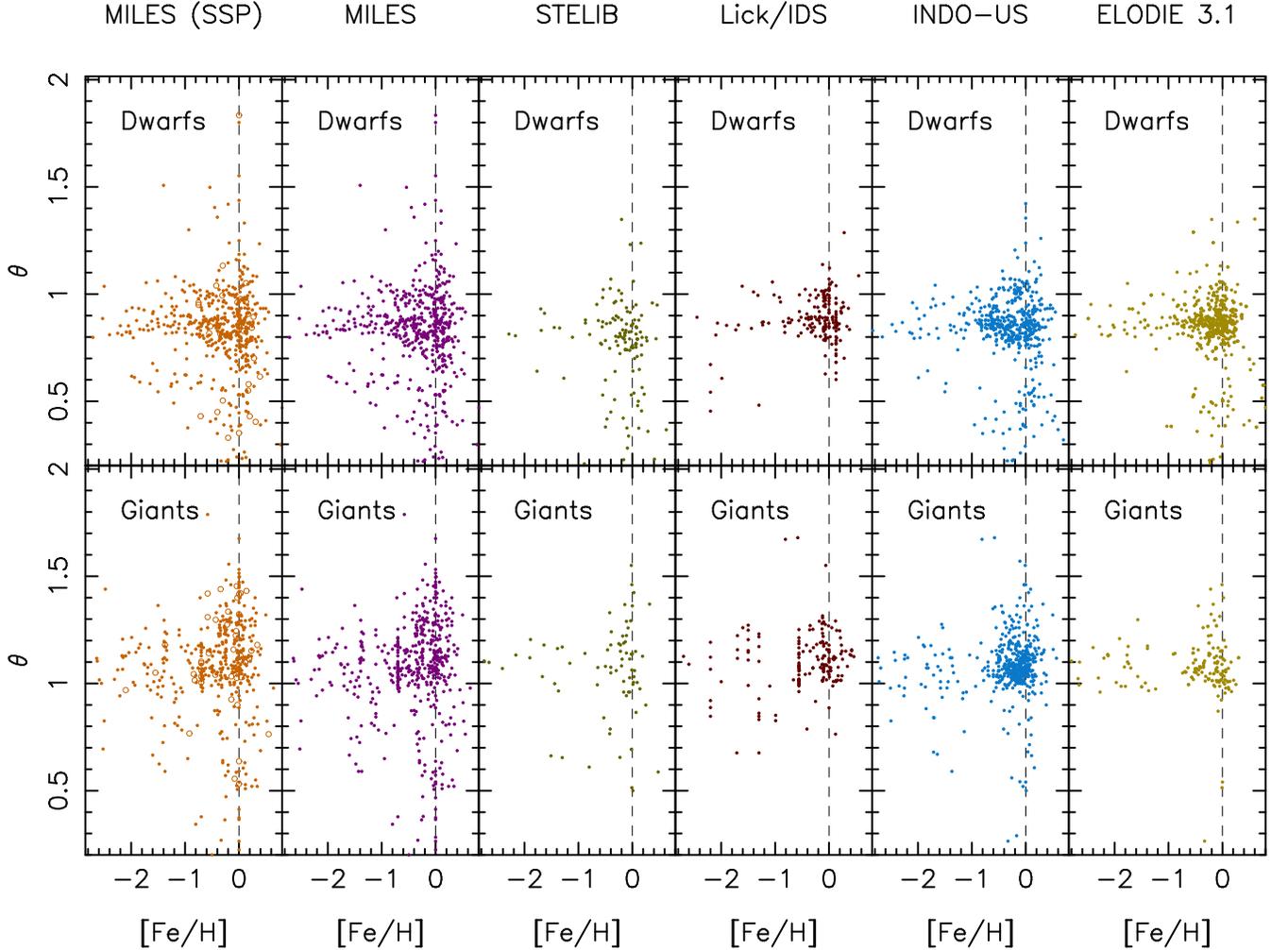}
\caption{The fundamental parameter coverage of the subsample of selected MILES 
stars is shown in the left panels for dwarfs (top) and giants (bottom). Stars
with decreased weights are plotted with open circles. The parameters of the
original MILES sample are plotted in the second column of the panels. We show, for
comparison, the parameter coverage of various popular stellar libraries (STELIB,
Lick, INDO-US and ELODIE). For
reference we indicate in each panel the solar metallicity (thin dashed vertical
line).}
\label{fig:MILES-parameters}
\end{figure*}

To compute the SSP SEDs in the full optical spectral range we use the MILES
library, for which the main characteristics are given in Paper~I. The stars of
the library were selected to optimize the stellar parameter coverage required
for population synthesis modelling. This is, in fact, one of the major advantages
of this library. Another important feature of the library is that the stellar
spectra were carefully flux-calibrated. To reach this goal, all the stars were 
observed through a wide (6$''$) slit to avoid selective flux losses due to the 
differential refraction effect, in addition to the higher resolution setups used 
to achieve the blue and red parts of the stellar spectra. Telluric absorptions 
that were present in the redder part of the spectra were properly removed. We 
refer to Paper~I for the technical details on how these aspects were tackled. 

To prepare this library for its implementation in the models we identified those
stars whose spectra might not be properly representative of a given set of
atmospheric parameters. We used the SIMBAD database for identifying the
spectroscopic binaries in the MILES sample. A number of these stars were removed
from the original sample as they were found non essential as MILES contains
sufficient stars with similar atmospheric parameters. We also checked for those
stars with high signal of variability ($\Delta V > 0.10$~mag), according to the
Combined General Catalogue of Variable Stars of Kholopov et al. (1998) (the
electronically-readable version provided at CDS). Some stars with signs of
variability were removed, when no such high variability was expected according to
their spectral types and atmospheric parameters and, at the same time, we were
able to identify alternative stars in the library with similar parameters and no
such sign of variability. Various stars were discarded because emission lines
were detected in their spectra. For various technical reasons we also discarded
from the original MILES list an additional number of stars: those with very low
signal-to-noise in the blue part of the spectrum or for which the atmospheric
parameters were lacking or had problems in the continuum. 

Each stellar spectrum of this selected subsample of stars was compared to a
synthetic spectrum of similar atmospheric parameters computed with the algorithm
described in V03 (Appendix B) (see also \ref{sec:synthesis}) employing the MILES
stellar library. To perform this test we first excluded from the list the target
star for which we wished to synthesize a similar spectrum. We then compared the
synthesized stellar spectrum with the observed one. This comparison was found to
be very useful for identifying stellar spectra with possible problems and,
eventually, discard them. In case of doubt, particularly for those 
regions of parameter space with poorer coverage, 
we compared the observed spectrum of our star with
the spectra of those stars, with the closest atmospheric parameters, which were
selected by the algorithm to synthesize our target star. As an example of this
method, Fig.~\ref{fig:interpolator} shows the results for stars of
varying temperature, gravity and metallicity. 

According to these careful inspections and tests we marked 135 stars of the
MILES database. While 60 of these stars were removed from the original sample,
we kept 75 stars, which were found to be useful for improving the coverage of
certain regions of parameter space. For the latter, however, we decreased 
the weight
with which they contribute when we synthesize a stellar spectrum for a given set of
atmospheric parameters. In practise this is performed by artificially decreasing
their signal-to-noise, according to the prescription adopted in
\ref{sec:synthesis}. For the interested reader we provide explanatory notes for
these stars in an updated table of the MILES sample that can be found in
{\bf http://miles.iac.es}.

\subsubsection{Stellar atmospheric parameter coverage}
\label{sec:coverage}
Figure\,\ref{fig:MILES-parameters} shows the parameter coverage of the selected
subsample of MILES stars for dwarfs and giants (separated 
at $\log g=3.0$) compared with the original MILES sample. The 75 stars with decreased weights are
also indicated. The parameters of MILES plotted here are the result of an
extensive compilation from the literature, homogenized by taking as a reference
the stars in common with Soubiran, Katz \& Cayrel (1998), which have very well
determined atmospheric parameters. We refer the reader to Paper~II for an
extensive description of the method and the adopted parameters. For
comparison we show the parameter coverage of some popular stellar libraries
(STELIB: LeBorgne et al. 2003; Lick/IDS: Gorgas et al. 1993, Worthey et al. 1994; INDO-US:
Valdes et al 2004; and ELODIE 3.1: Prugniel et al. 2007). 
For ELODIE 3.1 we use the parameters determined with the TGMET software (see Prugniel \& Soubiran 2004 for details)
to maximize the number of plotted stars. We adopt the mean value for those stars with repeated observations.

Fig.~\ref{fig:MILES-parameters} shows that, at solar metallicity,  all types of stars are well
represented  in all libraries. At lower metallicities, however, MILES shows a
much better  coverage of the parameter space for both, dwarfs and giants, than other libraries. Particularly
relevant is the presence of dwarfs with temperatures above
$T_{\mbox{\scriptsize eff}}>6000\,K$ and metallicities 
\mbox{$\mbox{[M/H]}<-0.5$}. With these stars we are in position to overcome a
major limitation of current population synthesis models based on empirical
stellar libraries providing with predictions for metal-poor stellar populations 
in the age range  0.1 -- 5\,Gyr. Furthermore, there are
sufficient stars to compute models for metallicities as low as
\mbox{$\mbox{[M/H]}=-2.3$}, particularly for very old stellar populations where
the temperature of the turnoff is lower. 

Another advantage with respect to the other libraries is the coverage of
metal-rich dwarf and giant stars that allow us to safely compute SSP SEDs for
\mbox{$\mbox{[M/H]}=+0.2$}.  Also for even higher metallicities MILES shows 
a good stellar parameter coverage and, therefore, it is possible to compute SEDs
for \mbox{$\mbox{[M/H]}\sim+0.4$}. These predictions will be presented
--and we will make them available in our website -- in a future   paper, where we feed the models with
the Teramo isochrones (Pietrinferni et al. 2004), that reach these high metallicity values.

These supersolar metallicity predictions are particularly
relevant for studying massive galaxies and it is the first time they 
can be safely computed, without extrapolating the behaviour of the 
spectral characteristics of the stars at higher metallicities as it 
is done, e.g., in models based on the
Lick/IDS empirical fitting functions (e.g. Worthey 1994; Thomas et al.\ 2003). Finally, it is worth noting that these
new model predictions also extend the range of SSP metallicities and ages of
our previous model SEDs based on the Jones (1999) (V99) and Cenarro et al.
(2001a) (V03) stellar libraries. A quantitative analysis of the quality of the
models based on the atmospheric parameter coverage of the employed libraries  is
provided in Section~\ref{sec:quality}.  

\begin{figure*}
\includegraphics[angle=0,width=7.0in]{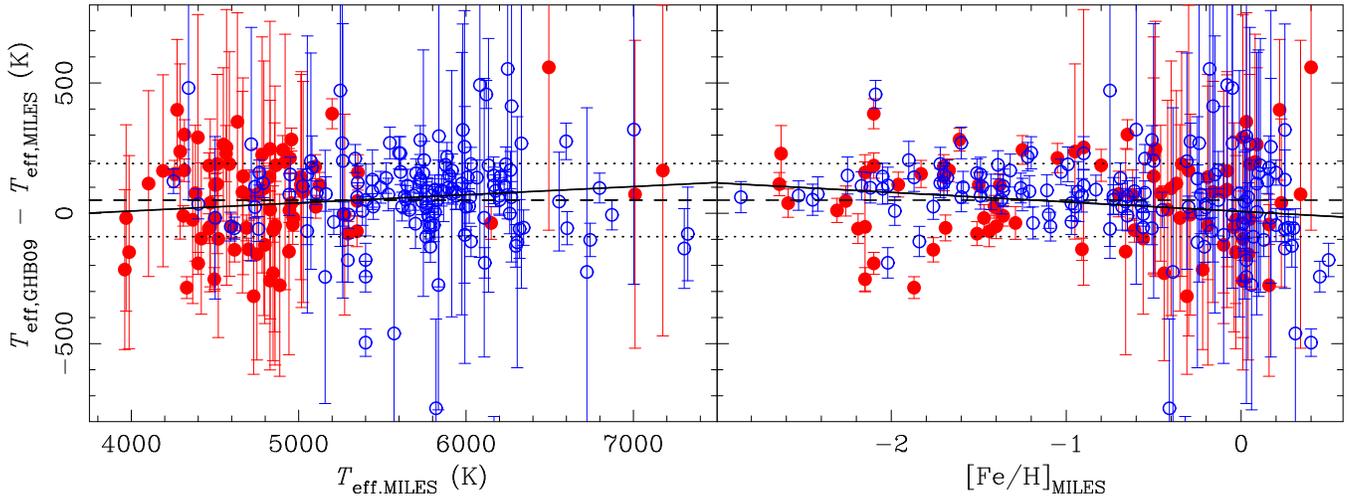}
\caption{Comparison of MILES temperatures with those of Gonz\'alez-Hern\'andez
\& Bonifacio (2009) as a function of temperature (left panel) and metallicity
(right panel). Giants are represented by red solid circles (81 stars) whereas
dwarfs are represented by open blue circles (134 stars). The error weighted
linear fit to all the stars is plotted in the two panels with a solid line. The
dashed line in each panel shows the mean error weighted offset obtained from all
the stars $\Delta T_{\mbox{\scriptsize eff}} = 51(\pm10)$\,K, whereas the dotted 
lines represent the standard deviation (140\,K) from this offset.}
\label{fig:GonzalezBonifacio}
\end{figure*}

\subsection{SSP spectral synthesis}
\label{sec:synthesis}

We use the method described in V99 and V03 for computing the SSP SEDs on the
basis of the selected subsample of MILES stars. In short, we integrate the
spectra of the stars along the isochrone taking into account their number per
mass bin according to the adopted IMF. For this purpose, each requested stellar
spectrum is normalized to the corresponding flux in the V band, following
the prescriptions adopted in our code, which are based on the
photometric libraries described in \S~\ref{sec:ingredients}.
We refer the interested reader to the papers previously mentioned for a full
description of the method. 

The SSP SED, $S_{\lambda}(t,{\rm [M/H]})$, is calculated as follows:

\begin{eqnarray}
S_{\lambda}(t,{\rm [M/H]})&=&\int_{m_{\rm l}}^{m_{\rm t}}
S_{\lambda}(m,t,{\rm [M/H]})N(m,t)\times \nonumber \\
&& {F_V}(m,t,{\rm [M/H]})dm,
\label{eq:SSP}
\end{eqnarray}


\noindent  where $S_{\lambda}(m,t,{\rm [M/H]})$ is the empirical spectrum,
normalized in the V band, corresponding to a star of mass $m$ and metallicity
[M/H], which is alive at the age assumed for the stellar population $t$.
$N(m,t)$ is the number of this type of star, which depends on the adopted IMF.
$m_{\rm l}$ and $m_{\rm t}$ are the stars with the smallest and largest stellar
masses, respectively, which are alive in the SSP. The upper mass limit depends
on the age of the stellar population. Finally, ${F_V}(m,t,{\rm [M/H]})$ is its
flux in the V band, which comes from transforming the theoretical parameters of
the isochrones. It is worth noting that the latter is performed on the basis of
empirical photometric stellar libraries, as described in
Section~\ref{sec:ingredients}, rather than relying on theoretical stellar
atmospheres, as it is usually done in other stellar population synthesis codes. 
This avoid  errors coming from theoretical uncertainties as the incorrect 
treatment of convection, turbulence, non-LTE effects, incorrect or imcomplete
line list, etc (see Worthey \& Lee 2006).

We obtain the spectrum for each requested star with a given
atmospheric parameters interporlating the spectra of adjacent 
stars  using the algorithm described in V03. The code
identifies the MILES stars whose parameters are enclosed within a given box
around the point in parameter space ($\theta_{0}$, $\log g_{0}$, [M/H]$_{0}$). When
needed, the box is enlarged in the appropriate directions until suitable
stars are found (for example, less and more metal-rich). 
This is done by dividing the
original box in 8 cubes, all with one corner at that point. This reduces the errors
in case of  gaps and asymmetries in the distribution of
stars around the point. The larger the density of stars around the requested
point, the smaller the box is. The sizes of the smallest boxes are determined by
the typical uncertainties in the determination of the parameters (Cenarro et al. 2001b). 
In each of the boxes, the stars are combined taking into account their parameters and the
signal-to-noise of their spectra. Finally,  the combined spectra in the different boxes 
are used to obtain a spectra with the required atmospheric parameters by weighting 
for the appropiate quantity.  For a full description of the algorithm we refer the reader to
V03 (Appendix B). 

It is worth noting that, within this scheme, a stellar spectrum is computed
according to the requested atmospheric parameters, irrespective of the
evolutionary stage. We are aware that this approach is not fully appropriate for
the TP-AGB phase, where O-rich, C-rich and stars in the superwind phase are
present. In fact, in M\'armol-Queralt\'o et al. (2008) we have found non
negligible differences in the CO bandhead at 2.3\,$\mu m$ between 19 AGB stars, all from
the MILES library, and RGB stars of similar parameters.
This, in principle, could introduce an error in our predictions  
as  very few  C-rich, O-rich or superwind phase TP-AGB stars  can be found in MILES. However, the
contribution of such stars to the total flux budget in the V-band is just 
a few percent, mainly for stellar populations of intermediate-ages (0.1 --
1.5\,Gyr; see e.g., Bruzual 2007).

\subsubsection{Effects of systematic variations in the adopted stellar parameters} 
\label{sec:GonzalezBonifacio}

Percival \& Salaris (2009) have recently shown that systematic
uncertainties associated with the three fundamental stellar atmospheric
parameters might have a non negligible impact on the resulting SSP
line-strength indices. The interested reader is refered to that paper for
details and for a suite of tests showing the effects of varying these
parameters. 

In particular, a relatively small offset in the effective temperature 
of 50-100~K, which is of the order of the systematic errors 
in the conversion from temperature to colours used here (Alonso et al. 1996)
may change the age of a 14\,Gyr stellar population by 2-3 Gyr and
alliviate the so-called zero-point problem for which the ages of the 
globular clusters (GCs) are older than the most recent estimations of the age 
of the Universe.
Furthermore, they show that, in many cases, there is a mistmatch  in scales
between the underlying models (the isochrones) and the adopted stellar 
library in the stellar population models.
We showed already in Paper~II that the Alonso et al. (1996, 1999) photometric
library did not show any offset with the MILES atmospheric parameters. Therefore, 
there is not a missmatch in scales between the models and the stellar library in the 
models presented here.

However, inspired by this work we decided to compare our stellar parameters
with those from several recently published works, e.g., 
Gonz\'alez-Hern\'andez \& Bonifacio
(2009), Casagrande et al. (2006), Ram\'{\i}rez \& Mel\'endez (2005),
to see if our parameters should be corrected. 
We show here, for ilustrative purposes,  the comparison of our parameters with those using
the new temperature scale of Gonz\'alez-Hern\'andez \& Bonifacio
(2009), whom derived Teff using the
infrared flux method (Blackwell et al.\ 1990), 
as it is the work with a larger number of stars in common with MILES (215) and 
because it is the one showing larger differences.

In Fig.\,\ref{fig:GonzalezBonifacio} we plot the difference in temperature
for the dwarf and giant stars in common between Gonz\'alez-Hern\'andez \& Bonifacio (2009) 
and MILES as a function of  temperature and metallicity.
We performed linear fits weighted with the errors obtaining the following relations:

\begin{equation} 
\Delta T_{\mbox{\scriptsize eff}} = -116(\pm80) +
0.0312(\pm0.0148){T_{\mbox{\scriptsize eff}}}_{\rm MILES}
\label{eq:HBTeff}
\end{equation}  

\noindent and metallicity

\begin{equation} 
\Delta T_{\mbox{\scriptsize eff}} = 7(\pm17) - 36.56(\pm11.79)\mbox{\mbox{[Fe/H]}}
\label{eq:HBFeH}
\end{equation}  

\noindent obtaining an rms of 138\,K in the two cases. The  mean
temperature offset of 59\,K and 54\,K for dwarfs and giants, respectively. This
is in good agreement with the offsets obtained by Gonz\'alez-Hern\'andez \&
Bonifacio (2009) when comparing their temperature scale with that of Alonso et
al. (1996) and Alonso et al. (1999).  Finally we did not find any significant
offset in metallicity or gravity among these samples. 

We computed an alternative SSP model SED library by transforming the MILES
temperatures to match this scale. These models can be used to better assess the
uncertainties involved in the method. We comment on these models in Sections
\ref{sec:colours} and \ref{sec:GonzalezBonifacioSSPs}. However, we do
not find any strong reason for adopting the Gonz\'alez-Hern\'andez \& Bonifacio
(2009) since the accuracy at high metallicities is worse than in the Alonso et
al. (1996,1999) (the temperature scale of Gonz\'alez-Hern\'andez \& Bonifacio
(2009) has been optimized for low metallicities while the errors in the
temperature determination for metal rich giant stars are very large).
Furthermore, as we already mentioned above, the temperature scale of the
isochrones and  stellar libraries in our models do not show any offset.


\section{MILES SSP SEDs}
\label{sec:MILES_SSP_SEDs}

\begin{figure*} 
\includegraphics[width=7.0in,angle=-90]{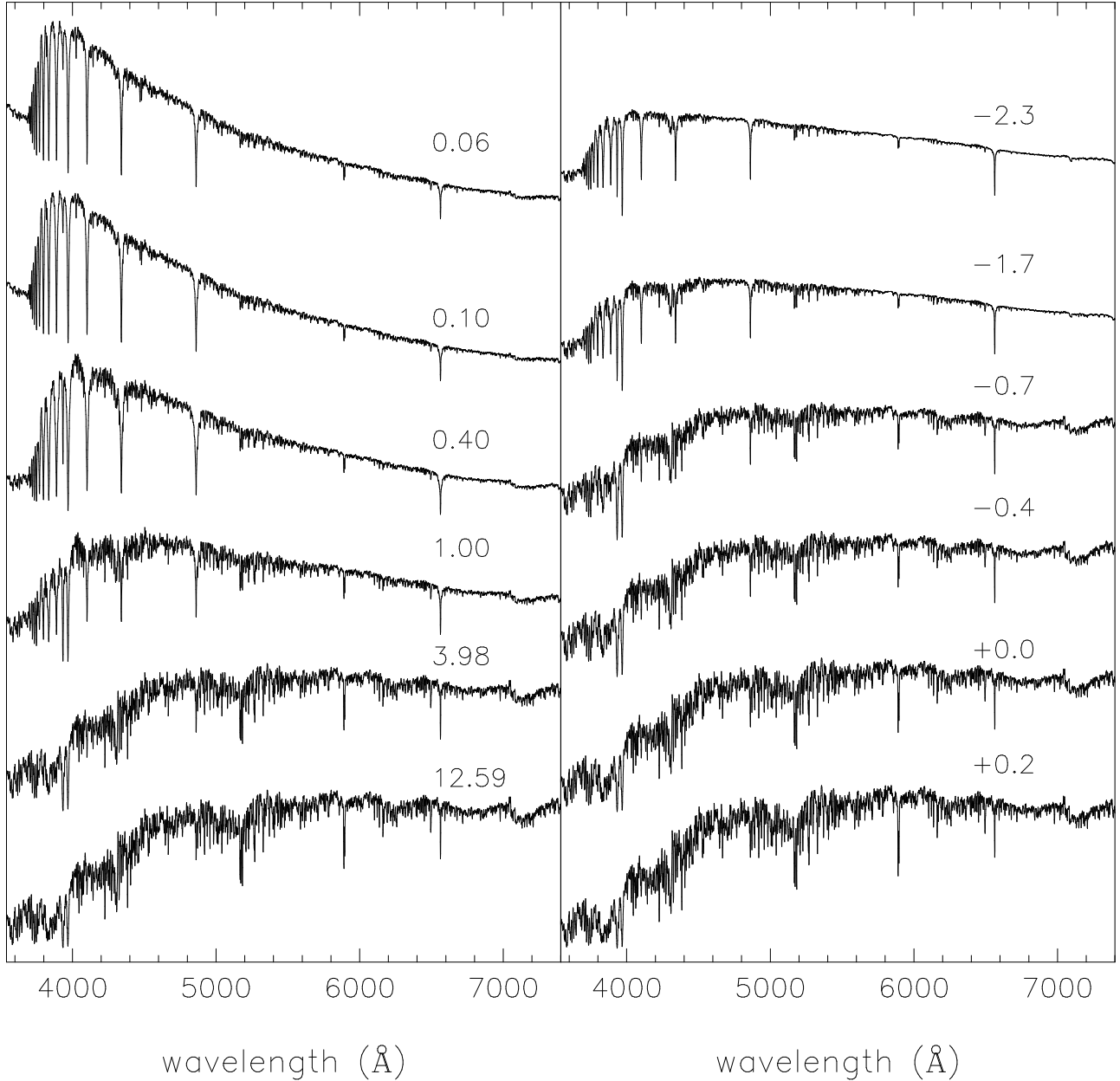}
\caption{Left panel: Solar metallicity SSP spectra of different ages (in Gyr),
increasing from top to bottom. Right panel: SSP spectra of 10\,Gyr and
different metallicities, increasing from top to bottom. All the SSP spectra
are plotted at the nominal resolution of the models (i.e. FWHM=2.3\,\AA). All
these models are computed adopting a Kroupa Universal IMF. For clarity,
the SSP spectra have been shifted by arbitrary amounts. } 
\label{fig:SSP_spectra} 
\end{figure*}

\begin{figure*}
\includegraphics[width=5.0in,angle=-90]{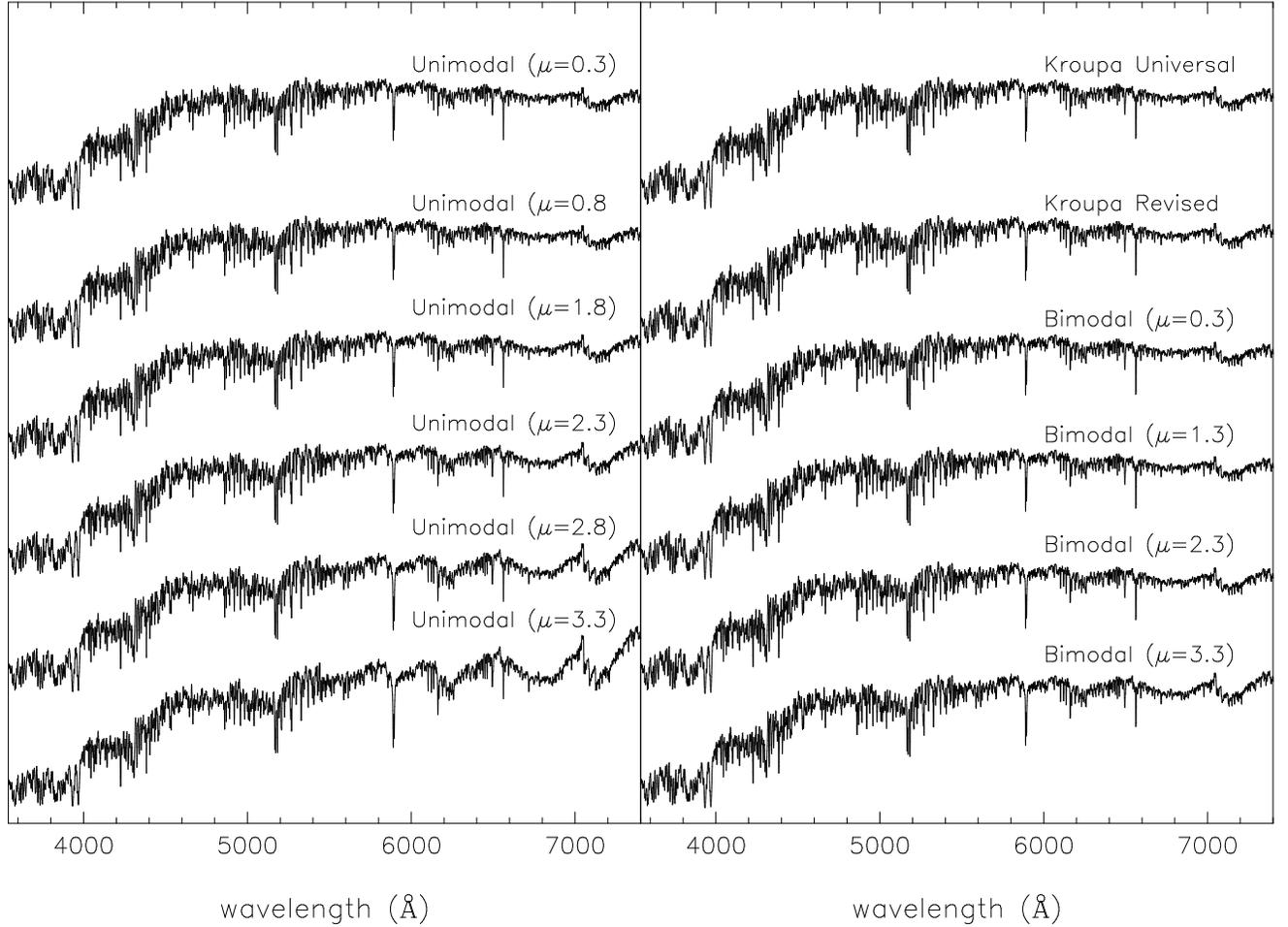}
\caption{Left panel: solar metallicity SSP spectra of 10\,Gyr, 
computed using a Unimodal IMF of increasing slope from top to bottom. The
Salpeter case, i.e. $\mu=$1.3, has been plotted in the right panel of
Fig.~\ref{fig:SSP_spectra}. The SSP spectra are plotted at the nominal
resolution of the models (i.e. FWHM=2.3\,\AA). Right panel: solar metallicity
SSP spectra of 10\,Gyr for different IMF shapes. From top to bottom: Kroupa
Universal, Kroupa Revised and Bimodal with increasing slopes, as indicated in
the panel. All the spectra have been shifted by arbitrary amounts for clarity.}
\label{fig:SSP_spectra_IMF}
\end{figure*}

\begin{table*}
\centering
\caption{\label{tab:SEDproperties}Spectral properties and parameter coverage of the synthesized MILES SSP SEDs}
\begin{tabular}{lc}
\hline 
\multicolumn{2}{c}{Spectral properties}\\
\hline                   
Spectral range     & $\lambda\lambda$ 3540.5-7409.6\,\AA \\	
Spectral resolution& FWHM $=2.3$\,\AA, $\sigma=54$\,km~s$^{-1}$\\	
Linear dispersion  & 0.9\,\AA/pix (51.5\,km~s$^{-1}$)\\
Continuum shape    & Flux-scaled                    \\
Telluric absorption residuals & Fully cleaned \\
Units &F$_{\lambda}$/L$_{\odot}$\AA$^{-1}$M$_{\odot}^{-1}$, L$_{\odot}=3.826\times10^{33}{\rm erg.s}^{-1}$ \\
\hline
\multicolumn{2}{c}{SSPs parameter coverage}\\
\hline
IMF type           & Unimodal, Bimodal, Kroupa universal, Kroupa revised\\
IMF slope (for unimodal and bimodal) & 0.3 -- 3.3 \\
Stellar mass range& 0.1 -- 100\,M$_{\odot}$ \\
Metallicity        & $-2.32$, $-1.71$, $-1.31$, $-0.71$, $-0.41$, $0.0$, $+0.22$\\
Age (\mbox{$\mbox{[M/H]}=-2.32$})& $10.0 < t < 18$\,Gyr	(only for IMF slopes $\le$ 1.8)   \\
Age (\mbox{$\mbox{[M/H]}=-1.71$})& $0.07 < t <18$\,Gyr	   \\
Age (\mbox{$-1.31\le\mbox{[M/H]}\le+0.22$})& $0.06 < t < 18$\,Gyr	   \\
\hline
\end{tabular}
\end{table*}

Table~\ref{tab:SEDproperties} summarizes the spectral properties of the newly
synthesized SSP SEDs. The nominal resolution of the models is FWHM$=$2.3\,\AA,
which is almost constant along the spectral range. For applications requiring to
know the resolution with much greater accuracy, we refer the reader to Fig.~4 of
Paper~I. This resolution is poorer than that of the models we presented in V99
(FWHM=1.8\,\AA), but this is compensated by the large spectral range
$\lambda\lambda$ 3540.5-7409.6\,\AA, with flux-calibrated response and no
telluric residuals. For computing the SEDs we have adopted a total initial mass of
1\,M$_{\odot}$. Table~\ref{tab:SEDproperties} also summarizes the SSP parameters
for which our predictions can be safely used. We discuss this issue in
Section~\ref{sec:quality}.  

\subsection{Behaviour of the SSP SEDs} 
\label{sec:SEDsbehaviour}

In Fig.~\ref{fig:SSP_spectra} we  show the new 
spectral library of SSP models for different ages and metallicities. 
All the spectra are plotted at the nominal 
resolution of the models (FWHM=2.3\AA).  The SSP age-sequence shows
weaker 4000\,\AA\ break and stronger Balmer line strengths as the age decreases.
The largest Balmer line strengths, however, are reached for 
ages of $\sim$0.4 Gyr. On the other hand the
metallic absorption lines become more prominent with increasing age. It can be 
seen that the continuum of the models with larger ages is heavily lowered by the
strengthening of the metallic features (line blanketing). 
This is the reason why the Lick/IDS indices were defined 
using psedocontinua instead of real continua  (Worthey et al. 1994). 
In the  SSP SED metallicity-sequence it can be seen that 
Balmer line-strengths decrease, and metallicity-sensitive features
get stronger for   increasing metallicity. 
These two SSP SED sequences clearly illustrate
the effects of the age/metallicity degeneracy. 

In Fig.~\ref{fig:SSP_spectra_IMF} we illustrate the effects of varying the IMF
slope ($\mu$) and the IMF shape. The left panel shows a
sequence of SSP spectra of solar metallicity and 10\,Gyr, where the slope is
varied for the Unimodal IMF. For old stellar populations this IMF type shows the
largest variations as the fraction of low mass stars vary dramatically when
varying the IMF slope. 
We see that the reddest part of the spectrum shows the
largest sensitivity to this parameter. These variations do no evolve
linearly as a function of $\mu$. In fact the SSP SEDs corresponding to the
flatter IMFs look rather similar. On the other hand, 
the effects caused by varying $\mu$
quickly become stronger for the steepest IMFs, notably redward of $\sim$5500\,\AA,
including the Na line at $\sim$5800\,\AA\ and the TiO molecular band around
$\sim$6200\,\AA\ and $\sim$6800\,\AA. The right panel shows that the effects of
varying the IMF slope are much less significant for the Bimodal IMF. This is
because for this IMF type the contribution from stars with masses lower than
0.6\,M$_{\odot}$ is decreased. Since this IMF more closely resembles
the Kroupa IMFs, it is probably more realistic. 
Finally, we note that the SSP SEDs for the two Kroupa
(2001) IMFs and the Bimodal IMF with slope 1.3 (as well as the Salpeter IMF) do
not show significant differences. 

\subsection{Reliability of the SSP SEDs}
\label{sec:quality}

\begin{figure*}
\includegraphics[width=7.in]{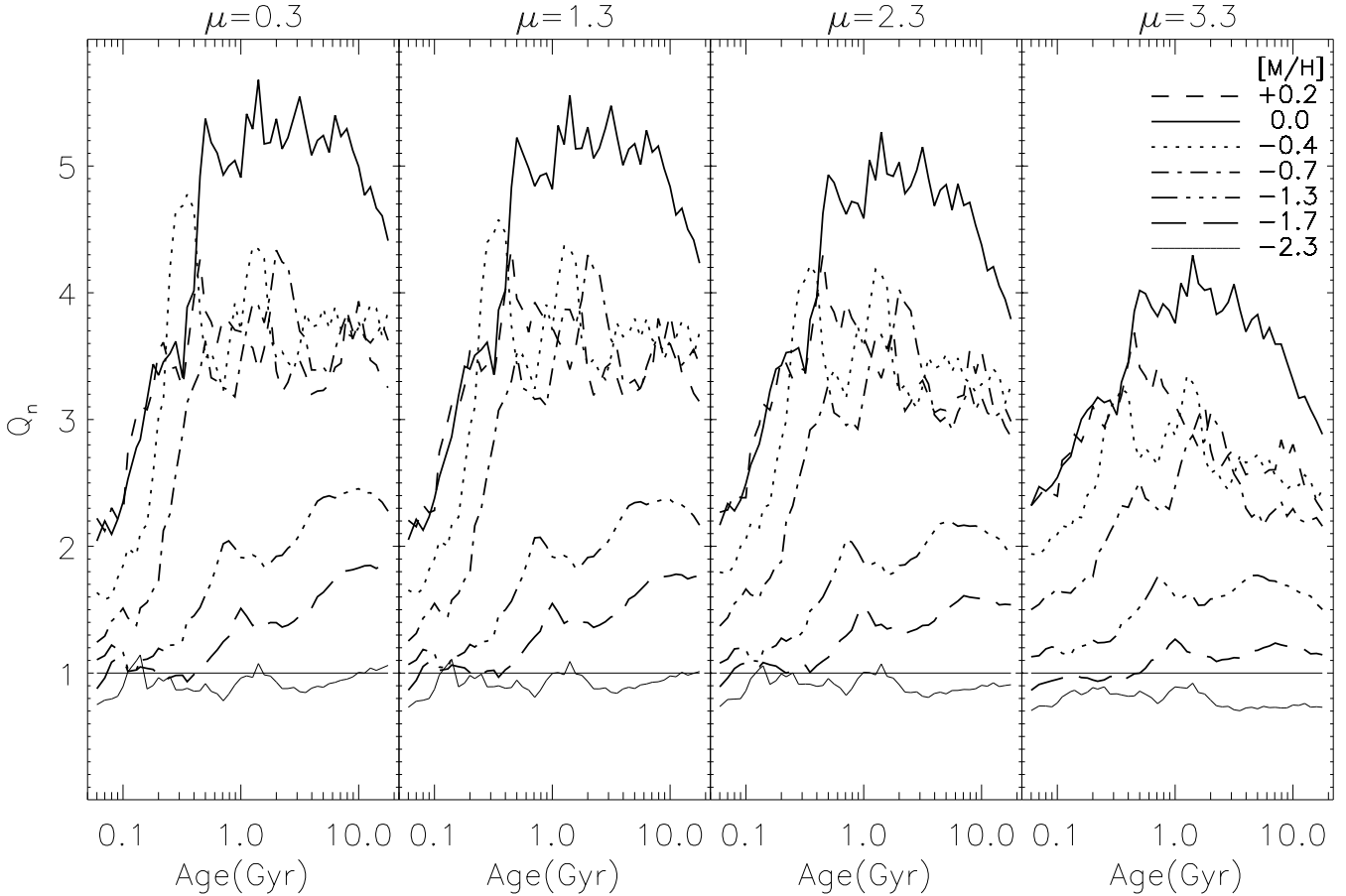}
\caption{The quality parameter, Q$_n$, as a function of the SSP age
(in Gyr) for different metallicities, indicated by the line types as quoted in
the inset. For computing the SEDs we use a Unimodal IMF with the slope
indicated on the top of each panel. SSPs with Q$_n$ values larger than 1 can be
safely used (see the text for details).}
\label{fig:quality_MILES}
\end{figure*}

In this section we try to provide some quantitative way of estimating the
reliability and the quality of the synthesized SSP SEDs, and to estimate the size of
the errors due to an incomplete coverage of the stellar parameter space of the input
library.  We perform this evaluation as a
function of SSP age, metallicity and IMF. 
To do this, we have taken advantage of our algorithm for
synthesizing a representative stellar spectrum for a given set of atmospheric
parameters (see Section \ref{sec:synthesis}) to compute for each SSP a
parameter, $Q$, as follows 

\begin{equation}
Q=\frac{
\sum_{i=1}^{n_{m_{\rm t}}}\left [ {
\frac{x_{s}\sum_{j=1}^{8} N_i^j}{\sum_{j=1}^{8}{\sqrt{
\left(\frac{{\phi_{\theta}}_i^j}{\sigma_{{\theta}_{\rm m}}}\right)^2+
\left(\frac{{\phi_{\log g}}_i^j}{\sigma_{{\log g}_{\rm m}}}\right)^2+
\left(\frac{{\phi_{\rm [M/H]}}_i^j}{\sigma_{\rm [M/H]_{\rm m}}}\right)^2}}}}\right ]{N_i {F_V}_i}}
{\sum_{i=1}^{n_{m_{\rm t}}}{N_i {F_V}_i}}
\label{eq:Q}
\end{equation}

\noindent We integrate the term enclosed within the square-brackets along the
isochrone, from the smallest stellar mass, $m_{\rm l}$, to the most massive
star, $m_{\rm t}$, which is alive at the age of the stellar population $t$,
weigthing by the flux in the V band, ${F_V}_i$, and the number of stars, $N_i$,
according to the adopted IMF. $N_i^j$ is the number of stars that have been
found within cube $j$ in parameter space 
(where $j=1,\ldots,8$, see Section~\ref{sec:synthesis}).
For a star with given $\theta$, $\log g$ and [M/H], the size of these cubes is given
by ${\phi_{\theta}}_i^j$,${\phi_{\log g}}_i^j$ and
${\phi_{\rm [M/H]}}_i^j$. $\sigma_{{\theta}_{\rm m}}$, $\sigma_{{\log g}_{\rm
m}}$ and $\sigma_{{\rm [M/H]}_{\rm m}}$ is the minimum uncertainty in the
determination of $\theta$, $\log g$ and \mbox{$\mbox{[M/H]}$}, respectively. Following V03 we
adopt ${\sigma_{\theta}}_{\rm m} = 0.009$, ${\sigma_{\log g}}_{\rm m} = 0.18$
and ${\sigma_{\rm [M/H]}}_{\rm m} = 0.09$. The size of the smallest cube is
obtained by multiplying the minimum uncertainty by the factor $x_{s}$.
Therefore, according to the square-bracketed term, for each star along the
isochrone, the larger the total number of stars found in these 8 cubes and the
smaller the total volume, the larger the quality of the synthesized stellar
spectrum is. 

We also compute, for each SSP, the minimum acceptable value for this parameter,
$Q_m$. Our algorithm ensures that at least we find stars in 3 (out of 8) cubes
to synthesize a stellar spectrum (see Appendix~B of V03 for details).
Furthermore the maximum enlargement that is allowed for each cube is assumed to
be 1/10 of the total range covered by the stellar parameters in our library.
Therefore the sizes for the largest cubes are ${\sigma_\theta}_{\rm M} = 0.17$,
${\sigma_{\log g}}_{\rm M} = 0.51$ and ${\sigma_{\rm [M/H]}}_{\rm M} = 0.41$.
Note that for ${\sigma_{\theta}}$ we also assume the condition that  $60\le
T_{\mbox{\scriptsize eff}} \le 3350\,K$, which corresponds to the adopted
${\sigma_{\theta}}$ limiting values when transforming to the
$T_{\mbox{\scriptsize eff}}$ scale. Therefore to compute $Q_m$ we will assume
for each stellar spectrum along the isochrone the following three conditions
that must be satisfied: {\it i)} we find stars in three cubes, {\it ii)} there
is a single star for each of these three cubes, which implies that $N_m =
\sum_{j=1}^{8} N_i^j$=3, and {\it iii)} it is necessary to open the size of these
cubes, generically, $f_{p} \times {\sigma_p}_{\rm M}$, $f_{p}=$0.34, i.e.
the star is found to be located within 1\,$\sigma$ of ${\sigma_p}_{\rm M}$,
considering the two directions at either side of the requested parametric point.
By substituting these values in Eq.\ref{eq:Q} we obtain  

\begin{equation}
Q_{m}=\frac{
\frac{N_{m}}{8{f_p}}\sum_{i=1}^{n_{m_{\rm t}}}
\frac{N_i {F_V}_i}{{\sqrt{
\left(\frac{{\sigma_{\theta}}_M}{\sigma_{{\theta}_{\rm m}}}\right)^2+
\left(\frac{{\sigma_{\log g}}_M}{\sigma_{{\log g}_{\rm m}}}\right)^2+
\left(\frac{{\sigma_{\rm [M/H]}}_M}{\sigma_{\rm [M/H]_{\rm m}}}\right)^2}}}}
{\sum_{i=1}^{n_{m_{\rm t}}}{N_i {F_V}_i}}
\label{eq:Q_m}
\end{equation}

\noindent Then a quality measure of the SSP SED, due to the atmospheric
parameter coverage of the stellar library feeding the models, is given by the
normalized parameter 

\begin{equation} 
Q_{n}=\frac{Q}{Q_m} 
\label{eq:Q_n} 
\end{equation}

\noindent Although the absolute value of $Q_n$ depends on the assumptions
adopted for computing $Q_m$, the $Q_n$ parameter allows us to compare the
quality of a given SSP SED compared to the SEDs computed for other SSP
parameters. In addition, $Q_n$ allows us to compare the quality of SSP SEDs
synthesized on the basis of different stellar spectral libraries, given the fact
that the stellar spectra of these libraries are of sufficiently high quality.
Furthermore these prescriptions are the same  that we use for computing a
stellar spectrum of given atmospheric parameters, which have been extensively
tested in V03.

Figure~{\ref{fig:quality_MILES}} shows the value of $Q_n$ as a function of the
SSP age (in Gyr) for different metallicities (different line types) and adopting
a Unimodal IMF, with its slope increasing from the left to the right panel. 
SSP SEDs
with $Q_n$ values above 1 can be considered of sufficient quality, and therefore
they can be safely used. The panels show that, as expected, the higher quality
is achieved for solar metallicity. For the Salpeter IMF, i.e. the second panel,
$Q_n$ reaches a value of $\sim$5 for stellar populations in the range 1 --
10\,Gyr.  Outside this age range the quality starts to decrease as low-mass and
hotter MS stars are less numerous in the MILES database. 

A similar behaviour is found in Fig.~ {\ref{fig:quality_MILES}} for the SSPs of
metallicities \mbox{$\mbox{[M/H]}=+0.22$}, $-0.41$ and $-0.71$, all with $Q_n$
values in the range 3 -- 4. Despite the fact that for all these metallicities the
value of $Q_n$ sharply decreases toward younger stellar populations,
the computed SSP SEDs can be considered safe. For lower metallicities we obtain
significantly lower $Q_n$ values, which gradually decrease toward younger
ages. For old stellar populations we obtain $Q_n$ values around 2 and around
1.5 for \mbox{$\mbox{[M/H]}=-1.31$} and \mbox{$\mbox{[M/H]}=-1.71$},
respectively. Interestingly, according to our quantitative analysis, the SSP
SEDs synthesized for \mbox{$\mbox{[M/H]}=-2.32$} and ages above 10\,Gyr are
barely acceptable for IMFs with slopes $\mu=0.3$ and $\mu=1.3$. 
These SEDs are particularly useful for globular cluster
studies. Note the sharp increase of $Q_n$ observed for ages around 1\,Gyr. This
should be mostly attributed to the fact that for stars of temperatures larger
than 9000\,K we do not take into account the metallicity of the available stars
for computing a stellar spectrum of given atmospheric parameters (see V03 for
details). Therefore, the quality of the SEDs synthesized for the younger stellar
populations is lower than stated by $Q_n$. 

Fig.~ {\ref{fig:quality_MILES}} also shows that $Q_n$ decreases slightly with
increasing IMF slope. In fact, the largest values are obtained for the flatter
IMF ($\mu=$0.3), as a result of the smaller contribution of low mass stars, 
which are
less abundant in MILES. This explains why $Q_n$ drops significantly
for the steepest IMF. It is worth noting that the SSP SEDs that are synthesized 
for the
lowest metallicity cannot be considered safe for IMFs steeper than Salpeter.

\begin{figure}
\includegraphics[width=3.2in]{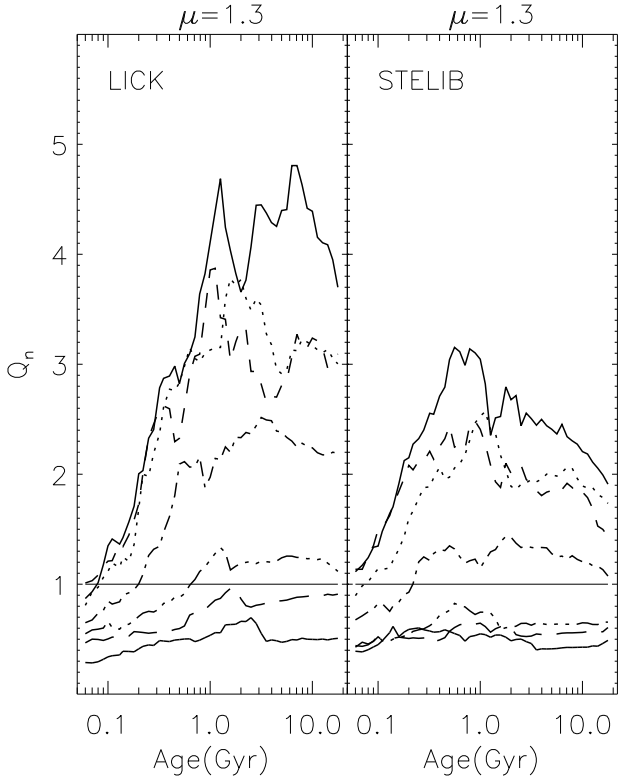}
\caption{Same test as in Fig.~\ref{fig:quality_MILES} but for SSP SEDs computed
on the basis of the atmospheric parameters corresponding to the Lick (left
panel) and STELIB (right panel) stellar spectral libraries (see the text for
details). To perform this test we selected the MILES spectra for the stars in
common with MILES, whereas for the rest of the stars we synthesized their
spectra with our interpolating algorithm, using the MILES spectral database
(see the text for details).}
\label{fig:quality_LICK_STELIB}
\end{figure}

For comparison we show in Fig.~ {\ref{fig:quality_LICK_STELIB}} the $Q_n$ values
obtained for two of the most popular libraries that are used in stellar
population synthesis, i.e. Lick and STELIB. The first library is mainly used to
predict Lick/IDS line indices via the empirical fitting functions of Worthey et
al. (1994) (e.g., Worthey 1994; Vazdekis et al. 1996; Thomas et al. 2003).
STELIB is the library feeding the models of Bruzual \& Charlot (2003). For this
purpose we used MILES to assemble a stellar spectral library resembling Lick,
and another one resembling STELIB. For each of these libraries we selected the
MILES spectra for the stars in common with MILES with listed stellar
parameters (i.e. 213 for Lick and 117 for STELIB). For the remaining stars, which
are not available in MILES, we synthesized their spectra with our interpolating
algorithm of Section \ref{sec:synthesis}, using MILES. 

The two panels of Fig.~ {\ref{fig:quality_LICK_STELIB}} show that the obtained
$Q_n$ values are significantly lower than the ones obtained for MILES for all
metallicities and ages. The left panel shows that the SSP predictions based on
the Lick library are not reliable for metallicities lower than $\sim-$1.3. The
right panel shows that this limit occurs at higher metallicity ($\sim-$0.7) for
any prediction based on STELIB. Overall the STELIB SSP models are of much lower
quality than the predictions based on the Lick library, judging from the
parameter coverage of these libraries.  Note that there is a large dispersion in
the $Q_n$ values obtained for old stellar populations. This feature is less
prominent in Fig.~\ref{fig:quality_MILES}, due to the fact that the distribution of
stellar parameters is far more homogeneous for MILES.

Table~\ref{tab:SEDproperties} summarizes the ranges in age, metallicity and IMF
slopes for which our MILES SEDs can be considered safe. We do, however, strongly
recommend to inspect Fig.~\ref{fig:quality_MILES} for a more accurate
description of the quality of the SSP SEDs. Although not shown, we find that the
$Q_n$ values obtained for the two Kroupa IMF shapes and the Bimodal IMF with
slope 1.3 are virtually identical to those shown in the second panel of Fig.~
\ref{fig:quality_MILES}. For the Bimodal IMF the $Q_n$ values also decrease with
increasing IMF slope, but the effect is much lower than is the case for the
Unimodal IMF.     

\subsection{Model SED comparisons}
\label{sec:modelcomparisons}

For an extensive comparison of our newly synthesized SSP SEDs with the models
of other authors we refer the reader to the recently published paper of Koleva
et al. (2008) and Cid-Fernandes \& Gonz\'alez (2010)\footnote{We provided 
these authors with a preliminary 
version of our MILES models to perform this comparison}. Koleva et al. compared three SSP spectral
libraries: GALAXEV (Bruzual \& Charlot 2003), Pegase-HR (Le Borgne et al. 2004)
and ours. Whereas the two former codes use the Padova 1994 stellar evolution
models (Bertelli et al. 1994), ours employ the Padova 2000 models (Girardi et
al. 2000), whose red giant branch are about 50--200\,K hotter. Each of these
codes use different prescriptions for transforming the theoretical stellar
parameters to the observational plane. The main difference is that our
models mostly rely on empirical relations, rather than on model atmospheres.
However, the key difference between these models is the stellar libraries that
they use, i.e., STELIB, ELODIE (version 3.1) and MILES for
GALAXEV, PEGASE-HR and ours, respectively. Koleva et al. (2008) do this
comparison by means of the full spectrum fitting approach, using two different
algorithms: NBURSTS (Chilingarian et al. 2007), which performs a parametric
non-linear fit, and STECKMAP (Ocvirk et al. 2006ab), which employs a
non-parametric formalism. 

Koleva et al. (2008) find that the quality of the models is very dependent on the
atmospheric parameter coverage of the stellar spectral libraries. The two
fitting methods show consistent results when our SEDs are compared to the
PEGASE-HR SEDs, obtaining residuals of the order of 1 per cent. As expected, the
largest discrepancies are found for low metallicities (\mbox{$\mbox{[M/H]}<-0.7$}), as shown
in their Fig.~ 2. However in Fig.~\ref{fig:MILES-parameters} we show that MILES
provides a significantly better coverage than ELODIE 3.1 for this
low-metallicity regime both, for dwarfs and giants. Furthermore the last two
panels of Fig.~ 2 of Koleva et al. show that the discrepancy between these two
model sets are larger for SSPs with ages smaller than $\sim$5\,Gyr, which can be
explained by the lack of hot metal-poor dwarfs in the ELODIE library. The fits
performed by these authors reveals that the worst results are obtained when
either PEGASE-HR or our models are compared with the SEDs of Bruzual \& Charlot
(2003). Koleva et al. conclude that this is due to the poor stellar atmospheric
coverage of STELIB, particularly for the 
non solar metallicity regime, as is also
shown in our Fig.~\ref{fig:MILES-parameters}. This has been recently confirmed
by Cid-Fernandes \& Gonz\'alez (2009) by comparing the model spectra to Milky
Way globular clusters.  

\begin{figure}
\includegraphics[width=3.3in]{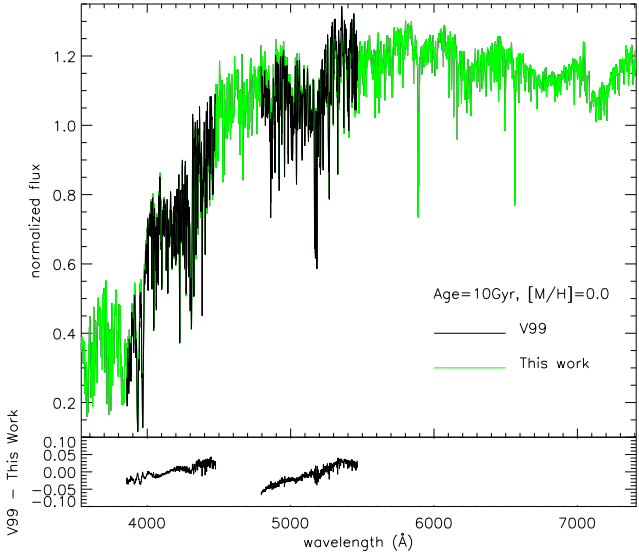}
\caption{SSP SEDs of solar metallicity and 10\,Gyr computed with the MILES (thin
solid line) and Jones (1999) stellar libraries, i.e. the V99 model (thick dotted
line). We adopted a Unimodal IMF of slope 1.3. The V99 model, which has a higher
resolution (FWHM$=$1.8\,\AA) than MILES, was smoothed to match the
resolution of the MILES model (FWHM$=$2.3\,\AA). The residuals are plotted in
the lower panel with the same scale.}
\label{fig:V99vsMILES_0.0}
\end{figure}

\begin{figure}
\includegraphics[width=3.3in]{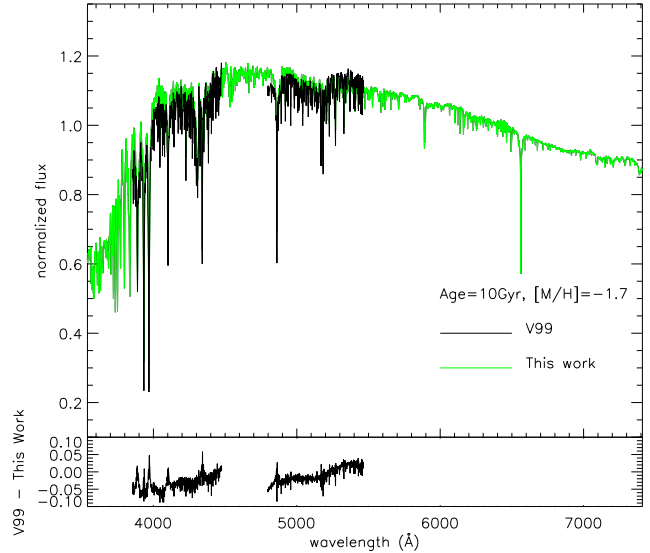}
\caption{Similar comparison as in Fig.~\ref{fig:V99vsMILES_0.0} but for metallicity
\mbox{$\mbox{[M/H]}=-1.7$}.}
\label{fig:V99vsMILES_1.7}
\end{figure}

When comparing our MILES models with the models that we presented in V99 we
show in Fig.~\ref{fig:V99vsMILES_0.0} a representative SSP spectrum of solar
metallicity and 10\,Gyr, with the corresponding V99 model over plotted. The
latter is an updated version of the original V99 models, which has been
computed with the same code used to synthesize the MILES SSP SEDs. The V99
SSP spectrum, with a resolution of 1.8\,\AA\,(FWHM), was smoothed to match the
resolution of the MILES model, i.e., 2.3\,\AA\,(FWHM). The residuals are
plotted on a similar scale. The residuals show a systematic trend in the two
narrow spectral ranges covered by the V99 models. This must be attributed to
errors in the   flux-calibration quality of the latter, as was reported in V99.
The same pattern is seen in the
residuals of Fig.~\ref{fig:V99vsMILES_1.7}, which shows a similar comparison
at metallicity \mbox{$\mbox{[M/H]}=-1.7$}. Contrary to Fig.~
\ref{fig:V99vsMILES_0.0}, the residuals of Fig.~\ref{fig:V99vsMILES_1.7} show
much larger variations at shorter wavelength scales. This translates into
larger index strength variations in this  low metallicity regime, showing that 
the
Jones (1999) library feeding the V99 models lacks important stars. This
comparison illustrates how relevant it is to synthesize SSP SEDs with stellar
libraries with good atmospheric parameters coverage, such as MILES.

\section{Defining a new standard system for index measurements}
\label{sec:system}

A major application of the SSP SEDs is to produce key line-strength indices that
can be compared to the observed values. In practise this has been the most
popular approach for studying in detail the stellar content of galaxies. So far
the most widely used indices are those of the Lick/IDS system (e.g., Gonz\'alez
1993; Trager et al. 1998, 2000; Vazdekis et al. 1997; J\o rgensen 1999;
S\'anchez-Bl\'azquez et al. 2003, 2006a,b,c, 2009). The standard Lick/IDS system of
indices was defined on the basis of a stellar spectral library (Gorgas et al.
1993; Worthey et al. 1994), which contains about 430 stars in the spectral range
$\lambda\lambda$4000-6200\,\AA. The Lick/IDS system has been very popular
because the stellar library contains stars with a fair range of stellar
parameters and, therefore, is appropriate to build stellar population models. The
most popular application of this library has been the use of empirical fitting
functions, which relate the index strengths, measured at the Lick/IDS
resolution, to the stellar atmosphere parameters. Because the value of the
Lick/IDS indices depend on the broadening of the lines, authors with spectral
data who wish to use the population models based on the Lick/IDS fitting
functions (e.g., Worthey 1994; Vazdekis et al. 1996; Thomas et al. 2003), or
compare them with previously published data, need to transform their
measurements into the Lick/IDS system. As the resolution of the Lick/IDS library
(FWHM$\sim$8-11\AA) is much lower than what is available with modern
spectrographs, the science spectra are usually broadened to the lower Lick/IDS
resolution. This is performed by convolving with a Gaussian function, whose
width 
varies as a function of wavelength, as the resolution of the Lick/IDS system,
apart from being low, also suffers from an ill-defined wavelength dependence (see
Worthey \& Ottaviani 1997). This effect is particularly significant for systems
with small intrinsic velocity dispersions, such as globular clusters and 
dwarf galaxies, for which part of
the valuable information contained in the higher resolution galaxy spectra is
lost. 

After that, it is still necessary to correct the measurements for the line
broadening due to the velocity dispersion of the stars in the integrated spectra
of galaxies. The correction should be calculated by finding a good stellar 
template of the galaxy by
combining an appropriate set of stellar or synthetic spectra which matches the
observed spectrum. This is because this correction is highly sensitive to the
strength of the index. Despite of this, many studies still use single
polynomial obtained with a single star as a template, that may have very 
different index-values
than those on the observed spectra (see Kelson et al. 2006 for a discussion of
the systematic effects associated to this procedure). 

Due to the fact that the shape of the continuum in the Lick/IDS stars is not
properly calibrated, the use of the Lick system requires a
conversion of the observational data to the instrumental response curve of Lick/IDS
dataset (see the analysis by Worthey \& Ottaviani 1997) that will
affect, predominantly, the broad indices. This is usually done by observing a
number of Lick stars with the same instrumental configuration as for the
science objects. Then, by comparing with the tabulated Lick index measurements,
the authors find empirical correction factors for each index.  This experiment
shows that even for the narrow indices, differences exist between the Lick/IDS
stars and other libraries (see Worthey \& Ottavianni 1997; Paper~I). The origin of these differences is not clear, but they need to be
corrected.

Furthermore, the spectra of this library have a low effective signal-to-noise
ratio due to the significant flat-field noise (Dalle  Ore et al. 1991; Worthey
et al. 1994; Trager et al. 1998). This translates into larger random errors
in the indices, much larger than in present-day galaxy data. It is worth noting
that the accuracy of the measurements based on the Lick system is often limited
by these transformations, rather than by the quality of the galaxy data.

\subsection{A new Line Index System: LIS}
\label{sec:LIS}

With the advent of new stellar libraries, such as MILES, with better resolution,
better signal-to-noise, larger number of stars  and a very complete coverage of
the atmospheric parameter space, it is about time to revisit the standard
spectrophotometric system at which the indices are measured. It is not our
objective to tell the authors at which resolution the indices need to be measured.
In fact, we recommend to use the flexibility provided by the model SSP SEDs and
compare the measurements  performed on  each object with models with the same
total broadening. However, this might not be a valid option in many cases, e.g.,
when a group of measurements is to be represented together in the same
index-index diagram.  It is also convenient to agree on a standard resolution at
which to measure line-strengths, so that measurements can be compared easily
with previous
studies, with colleagues, and to avoid systematic errors that might affect the
conclusions.  One
of the reasons why the Lick/IDS system is still popular nowadays is because most
previous studies have calculated their indices in this system. Although it is a
good standard procedure to compare new measurements with previous works, the
fact that those are transformed into the Lick/IDS systems perpetuates the 
problem. For this reason, we propose in this section a new Line Index System,
hereafter LIS, with three new spectral resolutions at which to measure the Lick
indices.  Note that this new system should not be restricted to the Lick set of
indices in a flux calibrated system. In fact, LIS can be used for any index in the 
literature (e.g., for the Rose
(1994) indices), including newly defined indices (e.g., Cervantes \& Vazdekis 2009).  

We provide conversions to transform the data from the Lick/IDS system to
LIS as well as tables with index measurements in the new system for popular
samples of  Milky Way globular clusters, nearby elliptical galaxies and bulges.
For this purpose we use index tables already published by other authors, and
transform them into the new system. To calculate the conversion, we have broadened
the MILES stars in common with the Lick/IDS library (218 stars in total) to the
new standard resolutions defined for both, globular clusters and galaxies, and
compare the indices measured in both datasets. Third-order polynomials were
fitted in all cases. These transformations are given in
Appendix~\ref{ap:Transformations}. In a forthcoming paper, we will present
empirical fitting functions for the Lick indices at these new standard
resolutions proposed here. A preliminary version of these fitting functions can
be found in Mart\'{\i}n-Hern\'andez et al. (2007), which were computed at the
resolution 2.3\,\AA\ (FWHM). We concentrate here on the Lick/IDS suite of indices
because it has been the most widely used in the literature. 

The definition of this new system using the MILES stellar library has two main
practical advantages:

\begin{itemize}

\item The resolution of the MILES stars is constant as a function of wavelength. 
As a result,
the user will not have to degrade the spectra with a wavelength dependent
function. 

\item Additional offsets will not have to be applied to the data, as long as the
measurements are flux calibrated (in a relative sense). 
Therefore the system is universal,
and the index measurements straightforward to perform.

\end{itemize}

It is very difficult to choose an appropriate resolution at which to define a new
system, as it depends on the intrinsic broadening of the objects and the
instrumental  resolution.  The Lick/IDS indices have been measured in a variety
of objects, from globular clusters to giant elliptical galaxies, covering a very
wide range in velocity dispersion, and in data with a variety of instrumental
resolutions. The appropriate resolution also depends on which
indices are going to be measured; while some Lick/IDS indices do not change their
sensitivity to age and metallicity with broadening, others (e.g. Ca4227,
Fe4531) do. For these reasons,  we consider it appropriate to define three 
different standard resolutions. The first one has been specifically chosen for
stellar clusters and dwarf galaxy data, while the other two are more appropriate
for working with massive galaxy spectra.

\begin{figure}
\includegraphics[width=2.8in,angle=-90]{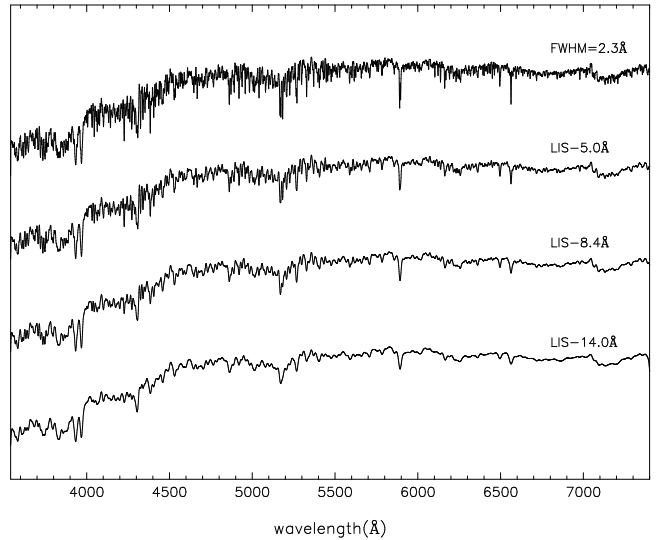}
\caption{SSP SED of solar metallicity, 10\,Gyr and Kroupa Universal IMF, plotted
for different resolutions. From top to bottom, FWHM$=$2.3, 5.0, 8.4 and
14.0\,\AA. The latter three resolutions correspond to the new system of indices
proposed here (LIS).}
\label{fig:LIS_resolutions}
\end{figure}

\subsubsection{LIS-5.0\AA.}

When choosing a standard resolution at which to compare indices of GCs, we were
guided by the resolution of existing high-quality GC data. Schiavon et al.
(2005) presented a spectral library of 40 Galactic GCs with a resolution of
$\sim3.1$\,\AA (FWHM). This represents an important comparison database for
extragalactic GC systems, and a logical upper limit for our choice of
resolution. However, the majority of extragalactic spectral data for GCs have
lower resolution. Data from the Gemini/GMOS collaboration  (e.g., Norris et al.
2008; Pierce et al. 2006ab) exhibits a typical FWHM of $\sim$5\,\AA.  Very
similar to that are the VLT/FORS spectra presented by Puzia et al. (2004). 
Keck/LRIS data from the
SAGES\footnote{http://www.ucolick.org/$\sim$brodie/sages/SAGES$\_$Welcome.html} group
has been generally taken at slightly higher resolutions  $\sim$2.8--4\,\AA\ 
(e.g. Strader et al. 2005; Beasley et al. 2006; Cenarro et al. 2007;  Chomiuk et
al. 2008; Beasley et al. 2009). Since it is considerably more straightforward to
degrade higher resolution data to lower resolution than it is the attempt to
correct lower resolution indices to higher resolution, we suggest using a
resolution (FWHM) of 5\,\AA\ in GC index comparisons, a choice which encompasses
all of the above data. This choice reflects a factor of 2 improvement in
resolution over the Lick/IDS system.

Authors willing to compare their data with the models presented here will only
have to broad their spectra to a resolution of 5\,\AA~(FWHM) (this corresponds
to $\sigma=127\,{\rm km\,s}^{-1}$ at 5000\,\AA) and measure the indices in
there. If they wish to compare with previous published data in the Lick/IDS
system,  they can use the conversion provided in Table~\ref{ap:Transformations}.

\subsubsection{LIS-8.4\AA.}

For galaxies we have decided to define two different standard resolutions. The
first one, discussed in this subsection, is fixed at resolution FWHM=8.4\,\AA\
(this corresponds to $\sigma=214\,{\rm km\,s}^{-1}$ at 5000\,\AA). The reasons
for this choice are twofold: 

$\bullet$ It matches, roughly, the resolution of the Lick/IDS system in
the H$\beta$ -- Mg triplet region. The indices in this regions (e.g., H$\beta$,
Mg$b$, Fe5270) are the most widely used in the literature and, therefore, the
measurements in the new system will not differ considerably from previous
measurements.

$\bullet$  200$\,{\rm km\,s}^{-1}$ is roughly the mean velocity dispersion for the
early-type galaxies in  the Sloan Digital Sky Survey (Bernardi et al.
2003a,b,c).

This system is particularly appropriate for studies of dwarf and
intermediate-mass galaxies, where the total broadening of the spectra
($\sigma_{\rm inst}^2+\sigma_{\rm gal}^2$) is not higher than  
$\sigma= 214\,{\rm km\,s}^{-1}$ at 5000\,\AA. 
In this case, the author will only have to broad their
spectra to the proposed total $\sigma$ using a Gaussian broadening function and
measure the indices.

If, on the contrary, the total broadening of the object is higher than
this value, the author has the option of using the system LIS-14.0\AA, or 
to correct his/her measurements to a total broadening of 8.4\,\AA (FWHM)
using polynomials (see, e.g., Kuntschner 2000; S\'anchez-Bl\'azquez et al. 2006a).

In a similar way as for LIS-5\,\AA, previous data in the Lick system can be
converted into this new system using the polynomials given in 
Table~\ref{conversion.gal}. 

\subsubsection{LIS-14.0\AA.} 

The third system is defined at a resolution of FWHM=14\,\AA\  (this corresponds
to $\sigma=$357\,kms$^{-1}$ at 5000\,\AA). LIS-14.0\AA\ has been designed especially
for the study of massive galaxies. In the spectra obtained for these galaxies
the broadening is dominated by the velocity dispersion of the stars. The total
broadening is higher than the standard Lick/IDS resolution for which, in order
to compare with the models, usually one had to perform a correction due to
velocity dispersion of the galaxy. As mentioned above, this correction is one of
the largest sources of systematic errors in the measurement of the indices (e.g.
Kelson et al. 2006).

For studies that only analyse the stronger Lick indices in a sample of galaxies
with velocity  dispersions $\sigma>$ 200 kms$^{-1}$, we suggest to broaden all the
spectra by an amount such that the total resolution 
($\sigma_{\rm total}=\sqrt{\sigma_{\rm inst}^2+\sigma_{\rm gal}^2+\sigma_{\rm
broad}^2}$) matches  FWHM=14\,\AA, before measuring the indices, where 
$\sigma_{\rm inst}$ represents the
instrumental resolution and $\sigma_{\rm gal}$ the velocity dispersion of the
galaxy. This will help us to avoid to have to perform any further correction to the indices
due to broadening.

Fig~\ref{fig:LIS_resolutions} shows the synthetic spectra of an SSP with
10\,Gyr and solar metallicity broadened to the resolution of the different LIS-
systems.
The polynomials required to transform the indices on the Lick/IDS system to this
resolution are provided in Table~\ref{conversion.gal3}.

\subsection{Brief notes about system transformation}

Most spectrographs keep the resolution constant in units of wavelength, rather than
velocity. Therefore, the broadening of the spectra,  if done in linear
wavelength scale, should be always  performed using gaussians of a fixed number of
pixels. If the broadening of the data has to be made with a Gaussian with  a
constant width in kms$^{-1}$. then the broadening in the  spectra should be
performed on a spectrum with a logarithmic wavelength scale.

It is common practise in studies of stellar populations to observe stars in
common with the Lick/IDS library. This is done to derive offsets to
transform the data onto the spectrophotometric system of the Lick
library\footnote{For most indices, the corrections are better described by
linear relations instead of offsets, see  appendix in S\'anchez-Bl\'azquez et
al. (2009)}. We stress here that this is necessary, {\it only} because the 
Lick/IDS spectra suffer from a number of calibration problems. However, these
offsets will not correct properly any uncertainty in the  calibration of the
science spectra, as the position of the line-index on the detector changes 
for extragalactic objects due to their redshift.

\begin{landscape}
\begin{table*}
\tiny
\begin{minipage}{200mm}
\caption{Lick indices measured in the Line Index System for globular clusters and galaxies from selected references.}
\label{table.index}
\begin{tabular}{lrrrrrrrrrrrrr}
\hline\hline
\multicolumn{13}{c}{Puzia et al. (2002) [LIS-5.0\AA]}\\
\hline\hline
Name  & H$\delta_A$ & H$\delta_F$ & CN$_1$ & CN$_2$  & Ca4227 & G4300 & H$\gamma_A$ & H$\gamma_F$ & Fe4383 & Ca4455 & Fe4531 & C4668 & H$\beta$\\ 
\hline
NGC5927 &$ -2.2576$ & 0.1134 &$  0.0975$  &$  0.1362$ & 0.8685 & 4.9847 &$ -4.1324$ & $-1.2751$ & 2.9922 & 0.6743 & 2.6763 & 3.0431 & 1.7502\\ 
   ~    &$  0.0388$ & 0.0269 &$  0.0009$  &$  0.0012$ & 0.0176 & 0.0282 &$  0.0408$ & $ 0.0277$ & 0.0493 & 0.0247 & 0.0424 & 0.0629 & 0.0335\\ 
NGC6218 &$  3.1849$ & 2.8266 &$ -0.0543$  &$ -0.0298$ & 0.2969 & 3.3555 &$  1.4944$ & $ 1.9157$ & 0.1026 & 0.0319 & 1.3028 & 0.1446 & 2.8334\\ 
   ~    &$  0.0186$ & 0.0123 &$  0.0007$  &$  0.0007$ & 0.0098 & 0.0251 &$  0.0248$ & $ 0.0161$ & 0.0358 & 0.0127 & 0.0344 & 0.0535 & 0.0278\\ 
NGC6284 &$  2.0326$ & 2.6239 &$ -0.0210$  &$  0.0056$ & 0.3717 & 3.8327 &$  0.2104$ & $ 1.3103$ & 0.8237 & 0.1649 & 1.8133 & 0.5586 & 2.5439\\ 
   ~    &$  0.0250$ & 0.0187 &$  0.0007$  &$  0.0008$ & 0.0128 & 0.0248 &$  0.0288$ & $ 0.0207$ & 0.0442 & 0.0161 & 0.0363 & 0.0575 & 0.0272\\ 
NGC6356 &$ -1.0805$ & 0.6216 &$  0.0606$  &$  0.0891$ & 0.6710 & 5.4774 &$ -3.8769$ & $-0.9782$ & 2.3657 & 0.4230 & 2.6000 & 1.7702 & 1.7484\\ 
   ~    &$  0.0186$ & 0.0117 &$  0.0006$  &$  0.0007$ & 0.0100 & 0.0142 &$  0.0214$ & $ 0.0139$ & 0.0272 & 0.0122 & 0.0253 & 0.0440 & 0.0189\\ 
NGC6388 &$ -0.2366$ & 1.0513 &$  0.0602$  &$  0.0917$ & 0.5320 & 4.8704 &$ -2.6754$ & $-0.1589$ & 2.4337 & 0.3673 & 2.5714 & 1.7768 & 2.1259\\ 
   ~    &$  0.0100$ & 0.0062 &$  0.0003$  &$  0.0004$ & 0.0049 & 0.0092 &$  0.0128$ & $ 0.0078$ & 0.0158 & 0.0066 & 0.0120 & 0.0196 & 0.0098\\ 
NGC6441 &$ -0.2459$ & 1.1013 &$  0.0682$  &$  0.0997$ & 0.6307 & 4.8525 &$ -2.8379$ & $-0.1968$ & 2.7413 & 0.3912 & 2.6228 & 1.7832 & 2.0552\\ 
   ~    &$  0.0260$ & 0.0176 &$  0.0007$  &$  0.0009$ & 0.0150 & 0.0237 &$  0.0271$ & $ 0.0200$ & 0.0333 & 0.0157 & 0.0320 & 0.0501 & 0.0240\\ 
NGC6528 &$ -2.0478$ & 0.3512 &$  0.1078$  &$  0.1389$ & 1.0559 & 5.6030 &$ -5.6367$ & $-1.6013$ & 4.6496 & 0.7368 & 2.9746 & 4.4146 & 1.8889\\ 
   ~    &$  0.0439$ & 0.0294 &$  0.0011$  &$  0.0013$ & 0.0194 & 0.0303 &$  0.0452$ & $ 0.0302$ & 0.0509 & 0.0276 & 0.0447 & 0.0796 & 0.0351\\ 
NGC6553 &$ -2.2176$ & 0.8167 &$  0.1465$  &$  0.1810$ & 1.2451 & 5.7732 &$ -5.8327$ & $-1.6070$ & 3.9143 & 0.6909 & 3.3030 & 3.7248 & 2.0026\\ 
   ~    &$  0.0705$ & 0.0432 &$  0.0017$  &$  0.0022$ & 0.0328 & 0.0476 &$  0.0639$ & $ 0.0425$ & 0.0777 & 0.0460 & 0.0672 & 0.1140 & 0.0521\\ 
NGC6624 &$ -0.9553$ & 0.6427 &$  0.0650$  &$  0.0977$ & 0.6539 & 5.3411 &$ -3.7187$ & $-0.8093$ & 2.4164 & 0.4004 & 2.5801 & 1.8939 & 1.7645\\ 
   ~    &$  0.0184$ & 0.0136 &$  0.0005$  &$  0.0006$ & 0.0092 & 0.0144 &$  0.0209$ & $ 0.0161$ & 0.0252 & 0.0128 & 0.0218 & 0.0366 & 0.0199\\ 
NGC6626 &$  2.3537$ & 2.4215 &$ -0.0247$  &$  0.0026$ & 0.3655 & 3.7828 &$  0.1746$ & $ 1.2743$ & 0.7371 & 0.0931 & 1.6467 & 0.6247 & 2.3904\\ 
   ~    &$  0.0222$ & 0.0178 &$  0.0006$  &$  0.0009$ & 0.0113 & 0.0260 &$  0.0278$ & $ 0.0171$ & 0.0381 & 0.0124 & 0.0351 & 0.0571 & 0.0278\\ 
NGC6637 &$ -1.1022$ & 0.3993 &$  0.0417$  &$  0.0691$ & 0.5761 & 5.5794 &$ -3.9200$ & $-1.0808$ & 2.1029 & 0.3539 & 2.4863 & 1.7632 & 1.7367\\ 
   ~    &$  0.0155$ & 0.0107 &$  0.0004$  &$  0.0005$ & 0.0071 & 0.0105 &$  0.0175$ & $ 0.0127$ & 0.0214 & 0.0096 & 0.0200 & 0.0303 & 0.0141\\ 
NGC6626 &$  2.3537$ & 2.4215 &$ -0.0247$  &$  0.0026$ & 0.3655 & 3.7828 &$  0.1746$ & $ 1.2743$ & 0.7371 & 0.0931 & 1.6467 & 0.6247 & 2.3904\\ 
   ~    &$  0.0222$ & 0.0178 &$  0.0006$  &$  0.0009$ & 0.0113 & 0.0260 &$  0.0278$ & $ 0.0171$ & 0.0381 & 0.0124 & 0.0351 & 0.0571 & 0.0278\\ 
NGC6637 &$ -1.1022$ & 0.3993 &$  0.0417$  &$  0.0691$ & 0.5761 & 5.5794 &$ -3.9200$ & $-1.0808$ & 2.1029 & 0.3539 & 2.4863 & 1.7632 & 1.7367\\ 
   ~    &$  0.0155$ & 0.0107 &$  0.0004$  &$  0.0005$ & 0.0071 & 0.0105 &$  0.0175$ & $ 0.0127$ & 0.0214 & 0.0096 & 0.0200 & 0.0303 & 0.0141\\ 
NGC6981 &$  1.5589$ & 1.7715 &$ -0.0262$  &$ -0.0091$ & 0.3862 & 3.5881 &$  0.4667$ & $ 1.2872$ & 0.1925 & 0.0895 & 1.3977 & 0.3620 & 2.5091\\ 
   ~    &$  0.0179$ & 0.0127 &$  0.0006$  &$  0.0007$ & 0.0091 & 0.0208 &$  0.0220$ & $ 0.0164$ & 0.0360 & 0.0126 & 0.0333 & 0.0475 & 0.0255\\ 
\hline\hline
 Name       & Fe5015 & Mg$_1$  & Mg$_2$ & Mgb   & Fe5270 & Fe5335 & Fe5406 & Fe5709  &  Fe5782 & Na5895& TiO$_1$ & TiO$_2$\\
\hline
NGC5927 & 4.9167 & 0.0756 & 0.1999 & 3.5827 & 2.2866 & 2.0628 & 1.3108 &$  0.5205$ & 0.8942  & 4.4516 & 0.0377&$  0.0895$ \\
   ~    & 0.0715 & 0.0007 & 0.0009 & 0.0358 & 0.0442 & 0.0524 & 0.0381 &$  0.0340$ & 0.0317 & 0.0396 & 0.0008 &$  0.0008$ \\
NGC6218 & 2.8312 & 0.0232 & 0.0582 & 1.1352 & 0.8644 & 1.0777 & 0.4432 &$ -0.1952$ & 0.2400 & 1.1520 & 0.0101 &$  0.0045$ \\
   ~    & 0.0691 & 0.0006 & 0.0008 & 0.0343 & 0.0311 & 0.0490 & 0.0289 &$  0.0345$ & 0.0275 & 0.0387 & 0.0009 &$  0.0010$ \\
NGC6284 & 3.2827 & 0.0375 & 0.0854 & 1.5008 & 0.9413 & 1.2086 & 0.6797 &$  0.1218$ & 0.3272 & 2.2529 & 0.0083 &$  0.0044$ \\
   ~    & 0.0752 & 0.0007 & 0.0009 & 0.0323 & 0.0364 & 0.0454 & 0.0328 &$  0.0350$ & 0.0227 & 0.0420 & 0.0008 &$  0.0008$ \\
NGC6356 & 4.1852 & 0.0647 & 0.1602 & 2.8145 & 1.7640 & 1.8961 & 1.0925 &$  0.4241$ & 0.5489 & 3.1437 & 0.0224 &$  0.0502$ \\
   ~    & 0.0446 & 0.0005 & 0.0006 & 0.0214 & 0.0231 & 0.0326 & 0.0210 &$  0.0217$ & 0.0191 & 0.0262 & 0.0005 &$  0.0006$ \\
NGC6388 & 4.2186 & 0.0502 & 0.1310 & 2.1997 & 1.9315 & 1.9042 & 1.1532 &$  0.5123$ & 0.6698 & 3.6860 & 0.0219 &$  0.0447$ \\
   ~    & 0.0244 & 0.0002 & 0.0003 & 0.0107 & 0.0158 & 0.0146 & 0.0125 &$  0.0106$ & 0.0096 & 0.0143 & 0.0003 &$  0.0003$ \\
NGC6441 & 4.3328 & 0.0640 & 0.1586 & 2.7556 & 2.0015 & 1.9567 & 1.1619 &$  0.5665$ & 0.8016 & 3.9727 & 0.0128 &$  0.0513$ \\
   ~    & 0.0591 & 0.0006 & 0.0007 & 0.0269 & 0.0379 & 0.0432 & 0.0253 &$  0.0258$ & 0.0239 & 0.0265 & 0.0006 &$  0.0006$ \\
NGC6528 & 5.2490 & 0.1033 & 0.2382 & 3.7516 & 2.4408 & 2.5976 & 1.7286 &$  0.8399$ & 0.8249 & 5.1220 & 0.0575 &$  0.1196$ \\
   ~    & 0.0807 & 0.0009 & 0.0009 & 0.0395 & 0.0457 & 0.0479 & 0.0476 &$  0.0311$ & 0.0364 & 0.0402 & 0.0010 &$  0.0009$ \\
NGC6553 & 5.7870 & 0.0897 & 0.2324 & 3.9036 & 2.7011 & 2.5835 & 1.3858 &$  0.7923$ & 1.1831 & 3.7845 & 0.0523 &$  0.1340$ \\
   ~    & 0.0971 & 0.0011 & 0.0013 & 0.0514 & 0.0702 & 0.0753 & 0.0521 &$  0.0424$ & 0.0454 & 0.0563 & 0.0010 &$  0.0009$ \\
NGC6624 & 4.3141 & 0.0628 & 0.1554 & 2.7574 & 1.8628 & 1.8745 & 1.1159 &$  0.5191$ & 0.6466 & 2.5670 & 0.0337 &$  0.0593$ \\
   ~    & 0.0422 & 0.0004 & 0.0005 & 0.0166 & 0.0243 & 0.0342 & 0.0256 &$  0.0173$ & 0.0214 & 0.0219 & 0.0005 &$  0.0005$ \\
NGC6626 & 3.3106 & 0.0364 & 0.0811 & 1.4305 & 1.1532 & 1.1603 & 0.7001 &$  0.1977$ & 0.4725 & 1.9530 & 0.0187 &$  0.0363$ \\
   ~    & 0.0669 & 0.0005 & 0.0006 & 0.0299 & 0.0322 & 0.0365 & 0.0272 &$  0.0304$ & 0.0269 & 0.0322 & 0.0007 &$  0.0006$ \\
NGC6637 & 4.0857 & 0.0501 & 0.1388 & 2.5752 & 1.6781 & 1.6068 & 0.9602 &$  0.3732$ & 0.4894 & 2.4638 & 0.0263 &$  0.0418$ \\
   ~    & 0.0365 & 0.0004 & 0.0005 & 0.0164 & 0.0212 & 0.0237 & 0.0170 &$  0.0143$ & 0.0144 & 0.0225 & 0.0004 &$  0.0005$ \\
NGC6626 & 3.3106 & 0.0364 & 0.0811 & 1.4305 & 1.1532 & 1.1603 & 0.7001 &$  0.1977$ & 0.4725 & 1.9530 & 0.0187 &$  0.0363$ \\
   ~    & 0.0669 & 0.0005 & 0.0006 & 0.0299 & 0.0322 & 0.0365 & 0.0272 &$  0.0304$ & 0.0269 & 0.0322 & 0.0007 &$  0.0006$ \\
NGC6637 & 4.0857 & 0.0501 & 0.1388 & 2.5752 & 1.6781 & 1.6068 & 0.9602 &$  0.3732$ & 0.4894 & 2.4638 & 0.0263 &$  0.0418$ \\
   ~    & 0.0365 & 0.0004 & 0.0005 & 0.0164 & 0.0212 & 0.0237 & 0.0170 &$  0.0143$ & 0.0144 & 0.0225 & 0.0004 &$  0.0005$ \\
NGC6981 & 2.7705 & 0.0265 & 0.0532 & 1.1555 & 0.9544 & 0.7882 & 0.3914 &$ -0.1326$ & 0.1239 & 1.2737 & 0.0070 &$ -0.0034$ \\
   ~    & 0.0619 & 0.0006 & 0.0007 & 0.0304 & 0.0323 & 0.0394 & 0.0218 &$  0.0333$ & 0.0151 & 0.0367 & 0.0006 &$  0.0008$ \\
\hline
\multicolumn{13}{l}{The complete version of this table can be found in the electronic version of the journal.}  
\end{tabular}                                                                                                   
\end{minipage}                                                                                                  
\end{table*}                                                                                                    
\end{landscape}

\subsection{Reference tables} 

\begin{figure}
\includegraphics[angle=-90,width=3.2in.]{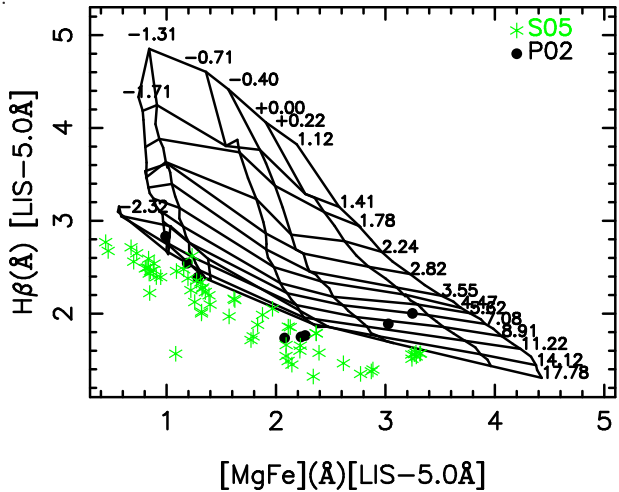}
\caption{ H$\beta$ vs. [MgFe] diagnostic diagram in the LIS-5.0\AA\, system.
Asterisks represent the globular cluster sample of Schiavon et al. (2005),
whereas the solid circles represent the clusters of Puzia et al. (2002). We used
the indices listed in Table~\ref{table.index} (the electronic version). For the
models we used a Kroupa Universal IMF.}
\label{fig:lis5}
\end{figure}

One of the reasons many authors keep using the Lick/IDS system is to compare
with previous studies, which are usually presented in this system. We provide
here with a set of indices, obtained from several references already transformed
to the new systems. The data are available at the CDS and in the electronic version
of the paper.

We have selected several samples of early-type galaxies and globular clusters
from the literature and transformed the line-strength indices to the new
systems defined here. This is intended to provide a benchmark of indices
properly transformed into LIS and measured on high quality data that
other authors can use to compare with. The selected samples are the following:

\begin{figure*}
\includegraphics[angle=-90,width=7.0in.]{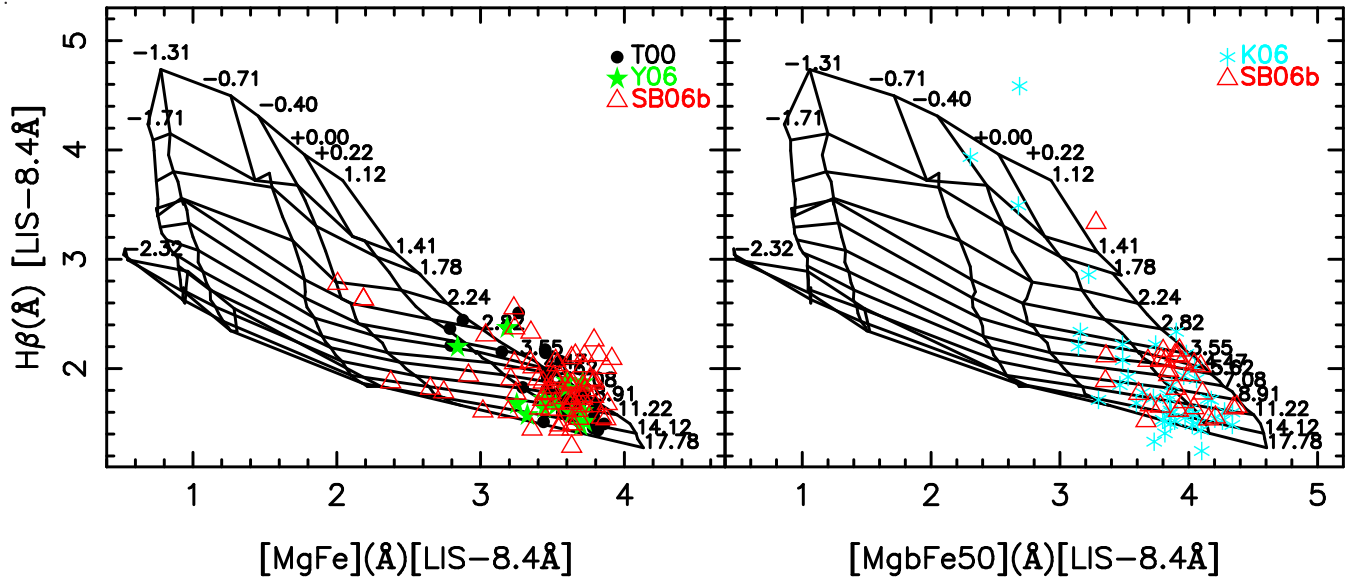}
\caption{H$\beta$ vs. [MgFe] and H$\beta$ vs. [MgFe50] diagnostics diagrams in
LIS-8.4\AA. In the left panel we plot the galaxy samples of Trager et al. (2000)
(solid circles), Yamada et al. (2006) (stars) and S\'anchez-Bl\'azquez et al.
(2006a) (open triangles). In the right panel we plot the galaxy sample of
Kuntschner et al. (2006) (asterisks) and 36 galaxies from the sample of
S\'anchez-Bl\'azquez et al. (2006a) (open triangles), as for all these galaxies
the two iron lines required to calculate the [MgFe] index fall outside the
covered spectral range (i.e. Fe5270 and Fe5335).}
\label{fig:lis8.4}
\end{figure*}

\begin{figure*}
\includegraphics[angle=-90,width=7.0in.]{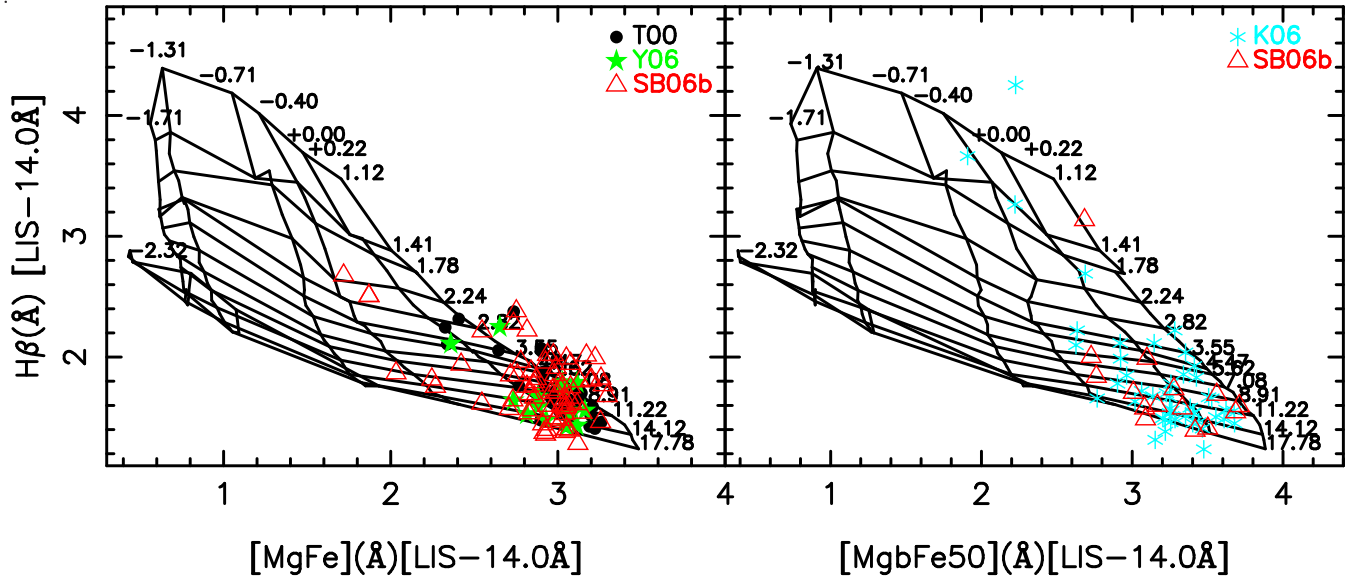}
\caption{Same as in Fig.~\ref{fig:lis8.4} but for LIS-14.0\AA.}
\label{fig:lis14}
\end{figure*}

\begin{itemize} 

\item{Puzia et al. (2002).} 
This sample comprises 12 Galactic globular clusters. Lick indices were measured
from long-slit spectra obtained with the Boller \& Chivens Spectrograph mounted
on the 1.5m telescope in La Silla. The wavelength coverage of the spectra ranges
from 3400 to 7300\,\AA\ with a final resolution of $\sim$6.7\,\AA. All the
spectra were observed with a slit with of 3$\farcs$0. We transformed the 
Lick indices into the new system using the transformations provided in this paper 
(Table \ref{conversion.}).

\item{Schiavon et al. (2005).}
The sample consists of 40 galactic globular clusters, obtained with the Blanco
4-m telescope at the Cerro Tololo observatory. The spectra cover the range
$\sim$3350-6430\AA\ with 3.1\,\AA\ (FHWM) resolution. The signal-to-noise ratio
of the flux calibrated spectra ranges from 50 to 240\,\AA$^{-1}$ at 4000\,\AA\ 
and from 125 to 500\,\AA$^{-1}$ at 5000\,\AA.  The sample has been carefully
selected to contain globular clusters with a range of ages and metallicities.
The indices provided in Table~\ref{table.index} have been 
 measured  directly in the spectra degraded  to the LIS-5.0\AA\ resolution. 

\item{Trager et al. (2000).}  
This sample consist of 40 galaxies drawn from that of Gonz\'alez (1993)
intended to cover, relatively uniformly, the full range of velocity dispersion,
line-strength and color displayed by local elliptical galaxies. The
line-strength indices (i.e.  H$\beta$, Mgb, Fe5270, Fe5335) were measured from
long-slit data obtained with the CCD Cassegrain Spectrograph on the 3m Shane
Telescope of Lick Observatory. The spectral features were measured within
r$_e$/8 aperture. The indices were transformed into the new system applying the 
transformations from Tables~\ref{conversion.gal} and \ref{conversion.gal3}.

\item{S\'anchez-Bl\'azquez et al. (2006a).} This sample consists of 98 early-type galaxies including ellipticals and
lenticulars, spanning a wide range in central velocity dispersion ($\sigma$).
The sample contains galaxies in different environments: isolated galaxies,
galaxies in groups, and in several clusters. The central spectra were extracted
within an aperture 0.6~kpc (assuming H$_0$=70 kms$^{-1}$). The spectra were  corrected
from the presence of nebular emission using {\tt GANDALF} (Sarzi et al. 2006), instead of using the [OIII]$\lambda$-based correcion
published in the original paper. We use the models provided in this paper as templates.
The spectral range covers from  $\sim$3500/3700
(depending on the run) to $\sim$5200/5700\,\AA, with spectral resolution ranging
from 3.6 to 6.56\,\AA\ (FWHM) (see original paper for more details). The indices
provided in Table~\ref{table.index} have been measured in the emission-free spectra previously 
degraded at the different LIS resolutions. For those galaxies with a $\sigma$ larger
than the LIS resolution an additive  correction was made using the kinematic templates (the combination of SSP-models that better
reproduced the galaxy spectra obtained in the measurement of $\sigma$ using {\tt ppxf} (Cappellari \& Emsellem 2004)), 
following the prescriptions in Kelson et al.\ (2006).
The correction was obtained as the difference in the indices measured in these templates degraded at  the LIS-resolution
 and at the $\sqrt{\sigma_{gal}^2+\sigma_{ins}^2}$,
where $\sigma_{\rm gal}$ is the velocity dispersion of the galaxy and $\sigma_{\rm inst}$ the instrumental resolution.

\item{Yamada et al. (2006).}
The sample consists of 14 galaxies, morphologically classified as E in the Virgo
cluster. The galaxies where selected with three criteria: (1) they had to be
classified as elliptical galaxy in the RC3 catalogue (de Vaucouleurs et al.
1991); (2) they had to be free from strong emission lines; (3) they have to
cover a wide range of luminosity evenly distributed along the color-magnitude of
the Virgo cluster. These extremely high quality spectra, which have been
extracted within an aperture of r$_e$/10, have signal-to-noise ratio well above
$>$100\,\AA$^{-1}$ for all the galaxies, and about $\sim$500\,\AA$^{-1}$ for some of
them. The spectra cover a wavelength range from 3800 to 5800\,\AA, and the
resolution ranges from 2.0\,\AA\ to 3.1\,\AA, depending on the observational
runs, which have been performed at the William Herschel Telescope (WHT; 4.2m) and Subaru (8.2m)
telescopes. For full details of the sample we refer the reader to the original
paper. Three  galaxies have a velocity dispersion larger than the resolution 
of the LIS-8.4\AA\ and the indices in this system were corrected in the same way as in the SB06b sample

\item{Kuntschner et al. (2006).} 
This sample consists of 48 E/S0
galaxies in the SAURON survey of early-type galaxies (de Zeeuw et al. 2002) at
the William Herschel 4.2m telescope in La Palma. We use here the  line strength measurements
(i.e. H$\beta$, Fe5015, Mg$b$) averaged within
r$_e$/8 aperture. The spectral coverage of the observed SAURON
integral-field observations ranges from 4800 to 5380\,\AA\ at resolution
$\sim$4.2\,\AA\ (FWHM). The indices of Table~\ref{table.index} were obtained applying the 
transformations of Table~\ref{conversion.gal} and \ref{conversion.gal3} to the indices transformed into 
the Lick systems provided in the original paper.

Table~\ref{table.index} shows a portion of the tables to illustrate its format
and content. The table contains the Lick/IDS indices measured at the different
LIS resolutions with the index definitions provided in Trager et al. (1998). The
full tables are available in the electronic version of the paper.

Fig.~\ref{fig:lis5} shows the commonly used index-index diagnostic diagram
H$\beta$ vs. [MgFe] in the new LIS-5.0\AA\, system for the Puzia et al. (2002)
and Schiavon et al. (2005) globular cluster samples. We use the indices in
Table~\ref{table.index}. Note that we do not plot the model indices for SSPs
younger than 10\,Gyr for \mbox{$\mbox{[M/H]}=-2.3$}, according to our Q$_n$
parameter analysis. In Fig.~\ref{fig:lis8.4} and Fig.~\ref{fig:lis14} we show
the H$\beta$ vs. [MgFe] and H$\beta$ vs. [MgFe50] diagrams for the four galaxy
samples both in LIS-8.4\AA\, and LIS-14.0\AA. Note that for all the galaxies of
Kuntschner et al. (2006) and for 36 galaxies in the S\'anchez-Bl\'azquez et al.
(2006a) sample it was not possible to plot the [MgFe] index, as either one or
the two iron indices (Fe5270, Fe5335) required to calculate this index were
outside the covered spectral range. Note that we used {\tt GANDALF} (Sarzi et
al. 2006) to remove nebular emission from the galaxy spectra of
S\'anchez-Bl\'azquez et al. (2006a) and Yamada et al. (2006). We refer the
reader to all these original works for a detailed discussion of the obtained
results.
\end{itemize}

\begin{figure*}
\resizebox{0.23\textwidth}{!}{\includegraphics[angle=-90]{resolution.age.panel1.ps}}
\resizebox{0.23\textwidth}{!}{\includegraphics[angle=-90]{resolution.age.panel2.ps}}
\resizebox{0.23\textwidth}{!}{\includegraphics[angle=-90]{resolution.age.panel3.ps}}
\resizebox{0.23\textwidth}{!}{\includegraphics[angle=-90]{resolution.age.panel4.ps}}
\caption{Time evolution of several Lick line-strength indices measured on our
new SSP SEDs for solar metallicity and Kroupa Universal IMF, smoothed to different
resolutions. Solid line: resolution of LIS-5.0\,\AA\, (FWHM = 5\,\AA); 
dashed line: resolution of LIS-8.4\,\AA\, (FHWM = 8.4\,\AA); 
dashed-dotted line: resolution of LIS-14.0\AA\, (FWHM = 14\,\AA).}
\label{models-age}
\end{figure*} 

\begin{figure}
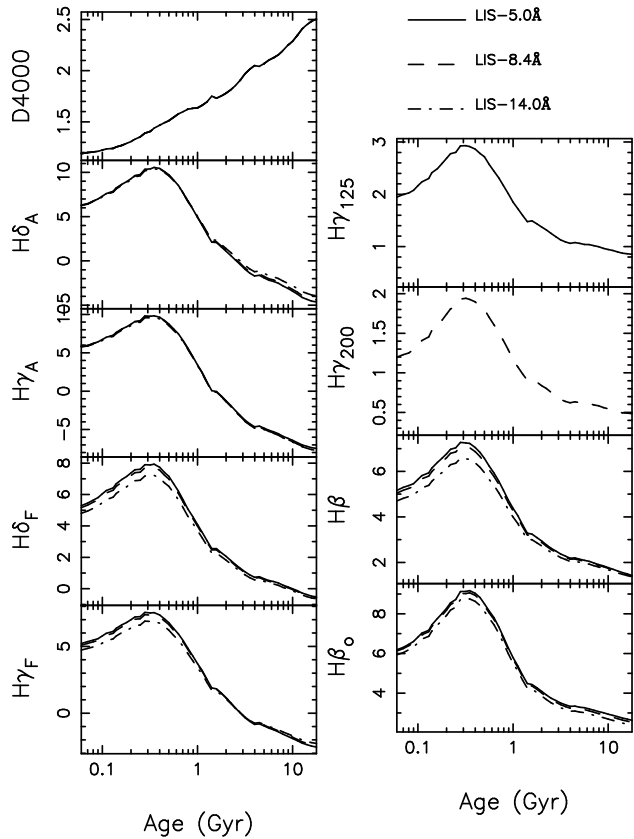

\resizebox{0.23\textwidth}{!}{\includegraphics[angle=-90]{resolution.age.panel5.ps}}
\resizebox{0.23\textwidth}{!}{\includegraphics[angle=-90]{resolution.age.panel6.ps}}
\caption{Time evolution of various Balmer line-strength indices measured on the
SSP SEDs of solar metallicity and Kroupa Universal IMF smoothed at different resolutions 
as in Fig.~\ref{models-age}.}
\label{models-age-Balmer}
\end{figure}

\begin{figure*}
\resizebox{0.23\textwidth}{!}{\includegraphics[angle=-90]{resolution.z.panel1.ps}}
\resizebox{0.23\textwidth}{!}{\includegraphics[angle=-90]{resolution.z.panel2.ps}}
\resizebox{0.23\textwidth}{!}{\includegraphics[angle=-90]{resolution.z.panel3.ps}}
\resizebox{0.23\textwidth}{!}{\includegraphics[angle=-90]{resolution.z.panel4.ps}}
\caption{Variation of line-strength indices with metallicity predicted by our
single stellar population  models of 10\,Gyr and Kroupa Universal IMF. 
Different line-styles show
predictions for indices measured at different  spectral resolution, as in
Fig.~\ref{models-age}.} 
\label{models-z}
\end{figure*}
\begin{figure}
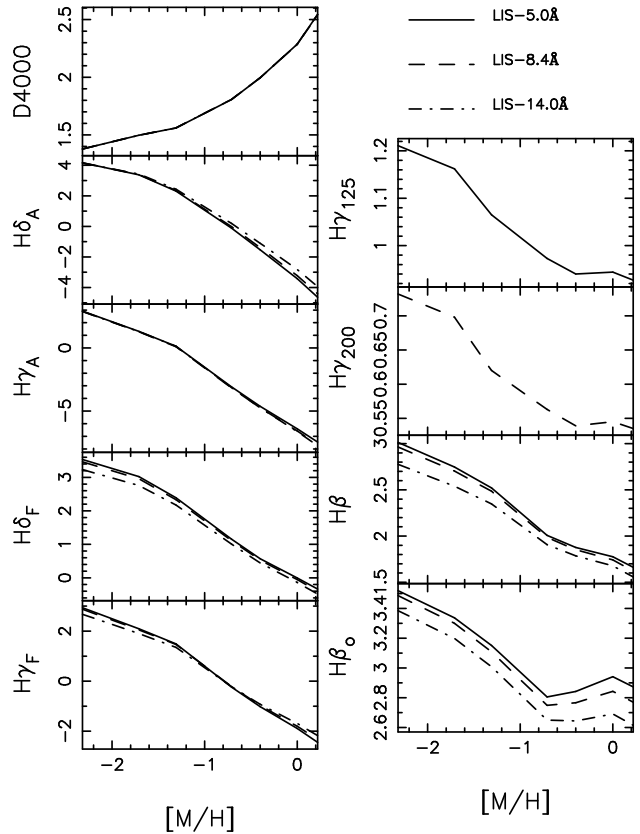

\resizebox{0.23\textwidth}{!}{\includegraphics[angle=-90]{resolution.z.panel5.ps}}
\resizebox{0.23\textwidth}{!}{\includegraphics[angle=-90]{resolution.z.panel6.ps}}
\caption{Variation of Balmer line-strength indices with metallicity predicted by
our single stellar population  models of 10\,Gyr and Kroupa Universal IMF. 
Different line-styles show
predictions for indices measured  at different  spectral resolution, as in
Fig.~\ref{models-age}.}
\label{models-z-Balmer}
\end{figure}

\subsection{MILES SSP SED index measurements}

Figure \ref{models-age} shows the predicted time evolution 
of our solar metallicity and Salpeter IMF models for a selection of
line-strength indices measured at four different spectral resolutions: the
nominal resolution of the models, i.e. FWHM = 2.3\,\AA, and the resolutions of
LIS-5.0\AA, LIS-8.4\AA, and LIS-14.0\AA\  (i.e., FWHM$=$5, 8.4 and 14\,\AA,
respectively). It can be seen that the sensitivity of some indices to variations
in resolution is barely noticeable. This is the case for the wide molecular
index definitions, i.e., CN, Mg and TiO. For some lines, however, broadening the
spectra means loosing part of the information contained in them (e.g. Ca4227).
The most critical situation when analyzing data by means of diagnostic
diagrams based on these indices happens when the relative variation of the index
with age depends on resolution. Authors should keep this in mind in order to
choose the right system and diagnostics to work with the data. Our
recommendation is to use the most straightforward approach possible, which is to
work with the model SEDs smoothed to match the total resolution of the data.
This approach allows the user to take full advantage of the whole information
contained in the data (see V99 for an extensive discussion). Interestingly we
note that the time evolution for most Lick indices shows a minimum for
intermediate-age stellar populations in the range 0.1--1.0\,Gyr, as 
already mentioned in Sec.~\ref{sec:SEDsbehaviour}. This is
particularly evident for the CN indices.  

Fig.~\ref{models-age-Balmer} is similar to Fig.~\ref{models-age} but here several
Balmer index definitions are shown. 
We plot the standard Lick H$\beta$ index and the four
higher-order Balmer indices of Worthey \& Ottaviani (1997). We also include two
of the age indicators of Vazdekis \& Arimoto (1999), the recently defined
H$\beta_{o}$ index of Cervantes \& Vazdekis (2009) and the D4000 break (e.g.,
Gorgas et al. 1999). We see that for all the Balmer index definitions the
dependence on resolution is marginal or virtually nil (e.g. the 4000\,\AA\
break). The two H$\gamma_{\sigma}$ of Vazdekis \& Arimoto (1999) are
specifically defined to be insensitive to resolution. For this reason we only
plot, for H$\gamma_{125}$, the values corresponding to the resolution of
LIS-5.0\AA\, and for H$\gamma_{200}$ the ones for LIS-8.4\AA\, as these two
resolutions fall within the stability range of these indices, respectively.  We
did not include the third index defined in Vazdekis \& Arimoto (1999),
H$\gamma_{275}$, as the resolution of LIS-14.0\AA\ is larger than is allowed
for this index. Interestingly, being almost negligible, we see that, whereas the
sensitivity to resolution for the new H$\beta_o$ index is slightly larger than
that for the standard H$\beta$ Lick index for old stellar populations, the
situation is reverse for ages smaller than $\sim$1\,Gyr. The higher-order Balmer line index
definitions of Worthey \& Ottaviani (1997) also show larger resolution
sensitivity for younger stellar populations. Note that all Balmer indices,
except the D4000 break, peak at $\sim$0.3\,Gyr and, therefore, the fits obtained
on the basis of these indices can have two solutions. Kauffmann et al. (2003)
take advantage of the behaviour of these indices to define diagnostics for
constraining the Star Formation History.

Figure~\ref{models-z} and Fig.~\ref{models-z-Balmer} shows the variation of
the line-strength indices as a function of metallicity predicted by our new
models  for stellar populations of 10\,Gyr and Salpeter IMF. As in the
previous figures the predictions are shown for three different resolutions;
FWHM= 5.0, 8.4 and 14\,\AA. We see a resolution-dependent differential
behaviour as function of metallicity for several indices, such as for example
Ca4227, Fe5335, Fe5782. This shows that a lot of care has to be taken to match
the resolution of data and models when comparing both as, otherwise, 
artificial trends can be derived (see Kelson et al. 2006).
Fig.~\ref{models-z-Balmer} shows the somewhat opposite dependence 
on the resolution of two H$\beta$ index definitions: H$\beta_o$ shows larger
resolution dependence than H$\beta$ for higher metallicities. Note however the
insensitivity to metallicity achieved by this H$\beta_o$ definition for
\mbox{$\mbox{[M/H]}>-1$}, which is not seen for the commonly used H$\beta$
Lick index. Such insensitivity is also shown by the two H$\gamma$ indices of
Vazdekis \& Arimoto (1999) plotted here. We are not  describing here in
detail the index bahaviour as function of metallicity and age, but  the
metallicity insensitivity of the two TiO molecular indices for
\mbox{$\mbox{[M/H]}>-0.5$} is particularly noteworthy.

\subsection{H$\beta$: comparison with the fitting function predictions}

\begin{figure}
\includegraphics[angle=-90,width=3.2in.]{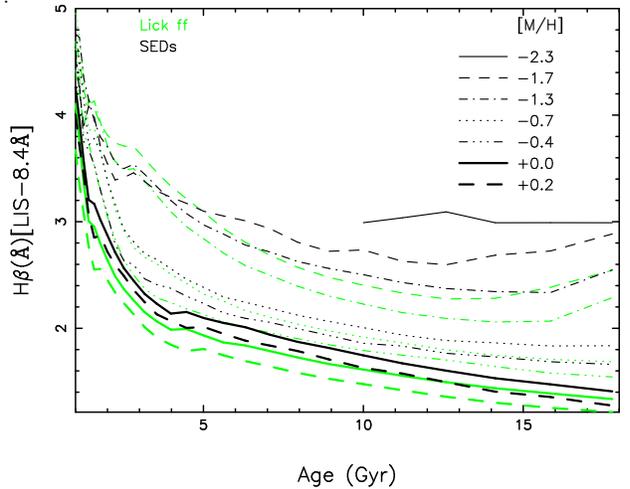}

\caption{Comparison of the H$\beta$ index computed on the basis of the empirical
fitting functions of Worthey et al. (1994),  using the models of Vazdekis et al.
(1996), as updated in this work and transformed to the LIS-8.4\AA\ system
following Table~\ref{conversion.gal}, with the index measured on the SSP SEDs,
once smoothed to the LIS-8.4\AA\, resolution.  For the lowest metallicity we
only plot the results for SSPs older than 10\,Gyr according to our qualitative
analysis shown in the second panel of Fig.~\ref{fig:quality_MILES}. However, for
the sake of the discussion we also show the predictions based on the Lick/IDS
fitting functions for \mbox{$\mbox{[M/H]}=-1.7$}  despite we consider they do
not have enough quality due to the low number of stars at those metallicities
(see the text for details).}

\label{fig:hbeta_transf}
\end{figure}

In Fig.~\ref{fig:hbeta_transf} we focus specifically on the behaviour of the
most popular age indicator, the Lick H$\beta$ index, as a function of age
(plotted in linear scale, for ages $>$1\,Gyr) and metallicity. In this figure we
compare the index computed on the basis of the empirical fitting functions of
Worthey et al. (1994), using the Vazdekis et al. (1996) code, as updated in
this work and transformed to the LIS-8.4\AA\, system following
Table~\ref{conversion.gal}, with the H$\beta$ index measured on the SSP SEDs,
once smoothed to the LIS-8.4\AA\, resolution. We see that the models based on
MILES provide larger values for solar and supersolar metallicities and for
\mbox{$\mbox{[M/H]}<-1$} than  those based on the Lick/IDS fitting functions.
This implies that when using MILES models the obtained ages are older by about
$\sim$2\,Gyr for SSPs around $\sim$10\,Gyr. 

Another interesting effect at metallicities \mbox{$\mbox{[M/H]}<-1$}  is that,
for ages above $\sim$12\,Gyr, H$\beta$ starts increasing instead of decreasing
as expected. This effect, not seen at higher metallicities, is attributed to the
contributions of the Horizontal Branch (HB) stars (e.g. Maraston \& Thomas
2000). Therefore the H$\beta$ index, rather than indicating the age of the
stellar population due to their sensitivity to the $T_{\mbox{\scriptsize eff}}$
of the turnoff stars, become sensitive to the HB morphology. This result is in
agreement with Maraston \& Thomas (2000). We see, however, that this sudden
increase in the strength of the H$\beta$ index is less visible for the new
predictions for the three lowest metallicities plotted in this figure. In fact,
the effect virtually disappears for \mbox{$\mbox{[M/H]}<-2$}.  This result would
have important implications for the interpretation of integrated spectra of
composite stellar systems, for which the Balmer lines, such as the H$\beta$
index, are the primary age diagnostics. In fact, Maraston \& Thomas (2000) have
suggested that a an spread in metallicity of an old stellar populations can
account for most of the dispersion of H$\beta$ values observed in the
ellipticals family -- due to the higher H$\beta$ index of the metal poor
stars--  without the need of a young population as had been previously 
suggested in other studies (e.g. Gonz\'alez 1993). As we obtain with MILES
higher H$\beta$ strengths, the required fraction for the metal-poor component
should be smaller than is suggested by Maraston \& Thomas (2000).

It is very important to
note that in Fig.~\ref{fig:hbeta_transf} we are comparing the same synthesis
code, with identical HB prescriptions. The different behaviour at low metallicities
between the two sets of predictions should be fully attributed, therefore, to the limitations of
the Lick stellar library, as it becomes evident when comparing the $Q_n$ values
obtained for these two sets of predictions in this low metallicity regime (see
the second panel of Fig.~\ref{fig:quality_MILES} and first panel of
Fig.~\ref{fig:quality_LICK_STELIB})\footnote{In fact, guided by the obtained $Q_n$
parameter, we only plotted the H$\beta$ values for ages larger than 10\,Gyr for
the MILES models of \mbox{$\mbox{[M/H]}=-2.3$}. 
However we plotted the H$\beta$ values obtained
for the Lick/IDS fitting functions for \mbox{$\mbox{[M/H]}=-1.7$}, 
despite the fact that their
corresponding $Q_n$ values are lower than 1 (see the first panel of
Fig.~\ref{fig:quality_LICK_STELIB})}. 

\section{MILES SSP Colours}
\label{sec:colours}

One of the main advantages of MILES is that the stellar spectra were carefully
flux-calibrated. In fact we obtained for each star of the library a wide-slit
spectrum to avoid selective flux losses due to the differential refraction
effect (see Paper~I for details). This allows us to measure accurate colours
from the synthesized SSP SEDs. This is performed by convolving the SED with
the corresponding filter responses. To obtain the Johnson $B-V$ broad band
colour we use the filters of Buser \& Kurucz (1978), with the zero points set
with Vega. To obtain the  zero point for the $B$ and $V$ filters, we used two
different Vega SEDs from Hayes (1985) and from Bohlin \& Gilliland
(2004), the latter based on high signal to noise Space Telescope Imaging
Spectroph (STIS) observations and
extrapolated to higher wavelengths using Kurucz model atmospheres. After
normalising the synthetic spectra to the SED of Vega we measured the 
$B-V$ colours by applying the selected filter transmission curves.

\begin{figure}
\includegraphics[width=3.6in.]{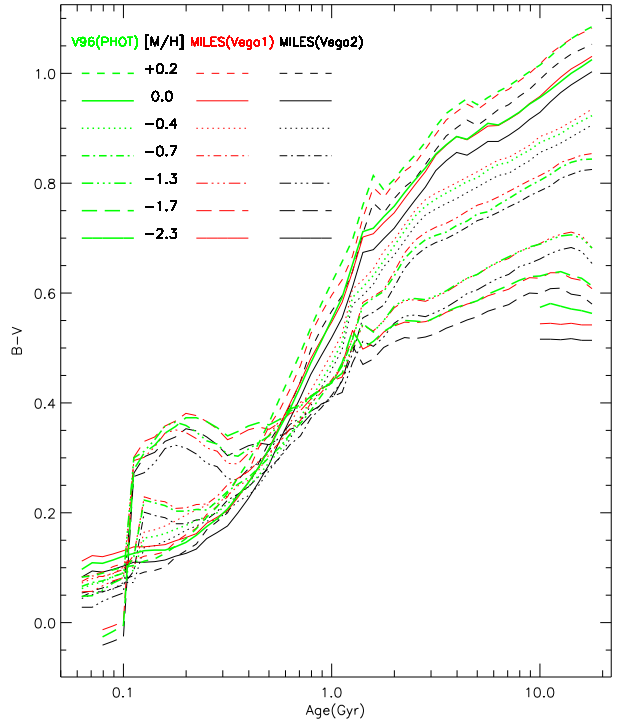}

\caption{We plot the broad band $B-V$ colour derived from the SSP SEDs
for different ages and metallicities (as indicated within the panel) and the
Kroupa Universal IMF with the zero point set with two different Vega spectra
in black (very thin) and red (thicker) (see the text for details). The thickest green line
represents the
photometric colour computed by Vazdekis et al. (1996), as updated in this
work, which is based on extensive empirical photometric stellar libraries.}

\label{fig:colours}
\end{figure}

The results are shown in Fig.~\ref{fig:colours}. We also plot, for comparison,
the photometric version of this colour following Vazdekis et al. (1996)  which
is computed with the Alonso et al. (1996, 1999) calibrations. The figure
shows a good agreement for nearly all ages and metallicities, within the typical
photometric uncertainties derived from the zero point of the $B$ and $V$
filters, as shown by the colours differences using  two different normalizations
of the Vega spectrum. We obtain rather similar time evolution colour curves for
all the metallicities, with an offset of $\sim$0.02\,mag. For the SSPs with
\mbox{$\mbox{[M/H]}=-2.32$} the offset can be as large as $\sim$0.05\,mag (MILES
SEDs colours slightly bluer), which is well below the photometric errors of the
colours. Note, however, that for this metallicity the MILES models are barely
acceptable, as shown by the Q parameter (see Fig.~\ref{fig:quality_MILES}). This
is because of the stellar parameter coverage of the MILES library and the
fact that colour transformation are not accurate at such low metallicity.

A rather
similar offset ($\sim$0.015\,mag.) has been quoted in Paper~I (see their
Fig.~8) by comparing the $B-V$ colours derived from MILES stellar spectra to
the values tabulated in the Lausanne photometric database (Mermilliod et al.
1997) for the same stars. As discussed in Paper~I this offset sets an upper
limit to the systematic uncertainties of our photometry. For the sake of
clarity we do not show in Fig.~\ref{fig:colours} the colours obtained from the
model SEDs computed with the temperature scale of Gonz\'alez-Hern\'andez \&
Bonifacio (2009). These models provide redder $B-V$ colour by $\sim$0.02\,mag, i.e.
similar to the zero point uncertainties. This is because for a given
temperature the stellar population model selects cooler spectra as the
Gonz\'alez-Hern\'andez \& Bonifacio (2009) temperatures are hotter than MILES
by 51\,K.   

The results presented here show that the colours derived from our models,
either those predicted via photometric libraries or the ones measured from our
MILES SSP SEDs, are consistent. Such agreement shows the reliability of the
temperatures adopted in Paper~II, which are consistent with the temperature
scales of Alonso et al. (1996,1999), i.e. the photometric libraries that feed
our models. We also provide, in electronic form and on the web (see Section~\ref{sec:web}),
colours measured directly in the MILES spectra using combinations of various filter systems that are
commonly used in the literature (e.g., SDSS, HST). 

\section{Applications to stellar cluster and galaxy data}
\label{sec:applications}

We compare, in this section, our new model SEDs with a set of representative stellar
clusters of varying age and metallicity. We also illustrate the use and
capabilities of these models for the study of  galaxies with varying
age, metallicity, abundance ratio and mass, by giving a few examples for some
well-studied galaxies.

\subsection{Stellar Clusters}
\label{sec:clusters}

\begin{figure} 
\includegraphics[angle=0,width=3.3in.]{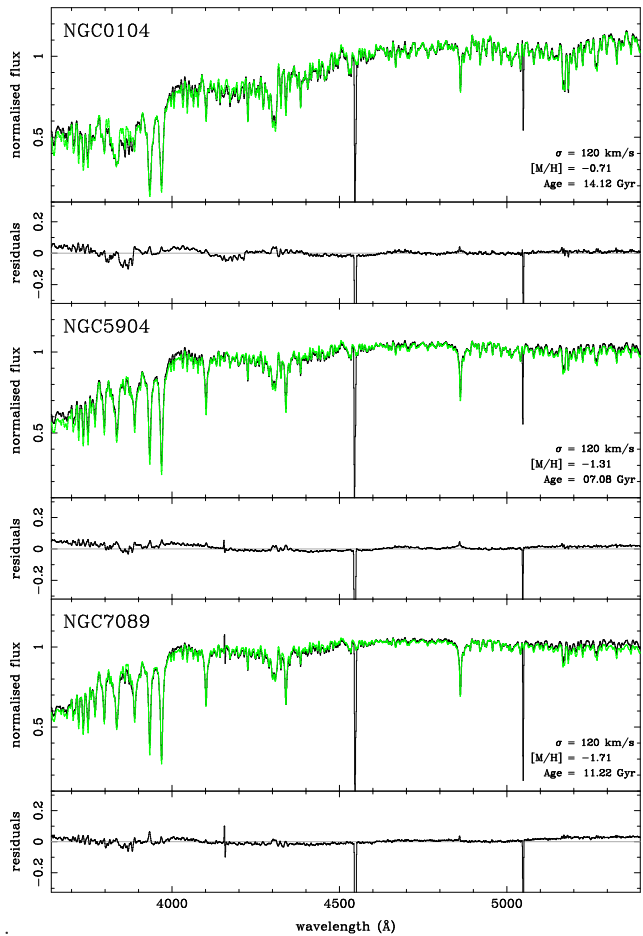}
\caption{Spectra of various representative Galactic globular clusters with
varying metallicity, from Schiavon et al. (2005), plotted in black. For each
cluster we overplot in green the best fitting SSP SED. We adopt for the models a
Kroupa Universal IMF. The age and metallicity of the SSP are indicated within each
panel. The SSP SEDs were smoothed to match the instrumental resolution of the
observations (i.e. $\sigma = 120\,{\rm km\,s}^{-1}$). The obtained residuals are
plotted at the same scale. Note that no continuum removal was applied when
performing the fit.} 
\label{fig:GCs} 
\end{figure}

Galactic globular clusters allow us to uniquely test our evolutionary synthesis
models, since independent age and metallicity estimates are available through
deep CMD analyses. These studies have concluded that, in most cases, globular
clusters can be considered as formed in a single burst, which is characterized
by a single age and a single metallicity, i.e. an SSP. Note, however, that there
are clusters for which two stellar components have been identified (e.g., Meylan
2003) and that stochastic effects (e.g., Cervi\~no et al.
2002) may affect significantly their integrated spectra. 

We have selected three representative Galactic globular clusters, with varying
metallicity, from the sample of Schiavon et al. (2005), which is composed of 40
clusters. These authors obtained integrated spectra of high signal-to-noise
(S/N$\sim$250) covering the range $\lambda\lambda$3350--6430\,\AA\ with 3.1\,\AA\
(FHWM) resolution. Fig.~\ref{fig:GCs} shows the spectra and the best SSP SED
fits overplotted. We used, for this purpose, our SSP SED library with Kroupa
Universal IMF. The SSP SEDs were conveniently smoothed to match the instrumental
resolution of the observations (i.e. $\sigma \sim 120\,{\rm km\,s}^{-1}$). We
normalized the stellar cluster spectra and the model SEDs at 5000\,\AA, but we
did not modify the spectral shape of the data (and the models) to perform the
fits. The best fit is understood as the one that minimizes the residuals, which
are also plotted in the figure at the same scale. The obtained ages and
metallicities fully agree with the CMD-derived estimates of the Harris (1996)
catalog. For 47\,Tuc we obtained the same fit as in V99. For NGC\,5904 we
derived a
younger age, in agreement with the very deep CMD study of Marin-Franch et al.
(2009) who have estimated for this cluster a relative age of 0.83, i.e. the
cluster is located within their young GC branch. We also found a younger age for
this cluster, by means of a line-strength analysis that included the newly
defined H$\beta_o$ age indicator of Cervantes \& Vazdekis (2009). The higher
H$\beta$ value obtained for this cluster is attributed in part to the presence
of an increased fraction of Blue Stragglers (see Cenarro et al. 2008).    

The residuals of Fig.~\ref{fig:GCs} do not show any significant colour
difference in the covered spectral range, but some very low frequency
variations that can be fully attributed to differences in the flux calibration.
In fact, note that these low frequency residuals are qualitatively identical
among the  three clusters. However the residuals show differences in various
molecular bands, mainly in the blue part of the spectra for all the clusters.
In particular, CN features at $\sim$3860\,\AA\ and $\sim$4150\,\AA, are much
stronger in the clusters than in the models. We also see a residual in the G
band at $\sim$4300\,\AA, being this absorption weaker in the data. The
CN-strong absorption, and its anticorrelation with CH, is a well known feature
in the Galactic globular cluster family, which has been shown in both their
individual stars (e.g., Norris \& Freeman 1979) and its integrated spectra
(e.g. Rose \& Tripicco 1986). A similar result is obtained in V99 for 47\,Tuc,
but using a different input stellar spectral library and also a different
spectrum for the cluster. Such CN-strong absorption feature has also been
reported for extragalactic globular clusters, like NGC\,1407 (Cenarro et al.
2007b). Their Lick/IDS C$_2$4668 vs. CN$_2$ diagram shows the same
anticorrelation with respect to the scaled-solar abundance ratios. Overall, we
find that the residuals blueward of $\sim$4300\AA\ are significantly larger
than for the redder part of the spectra for all the clusters. We do not discard
possible effects due to the fact that the Milky Way globular clusters show
oxygen-enhanced abundance ratios for low metallicities. In fact Cassisi et al.
(2004) have shown that such effects become relevant for the spectral ranges
covered by the $B$, and mostly $U$, broadband filters. Therefore an
appropriate modelling for these clusters requires working with such
$\alpha$-enhanced stellar evolutionary isochrones as well as the use of stellar
spectra with similar abundance pattern. In addition we also require to use
specific stellar spectra with CN-strong absorption features and CN-enhanced
isochrones. 

\begin{figure}
\includegraphics[angle=270,width=3.3in.]{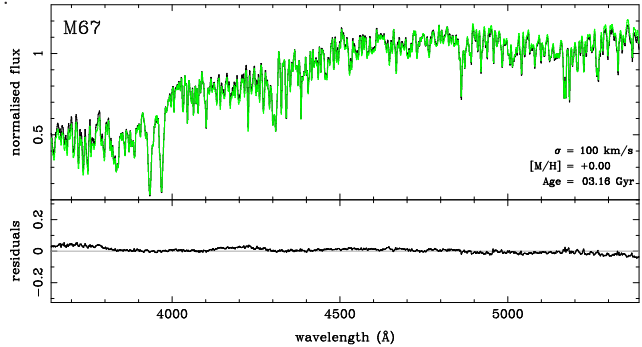}
\caption{The spectrum of the Milky Way open cluster M\,67 (Schiavon et
al. 2004) plotted in black. We overplot the best fitting SSP SED in green,
with its age and metallicity indicated within the panel. We adopt a Kroupa Universal
IMF. The model has been smoothed to match the instrumental resolution of the
observed spectrum (also quoted). The residuals are plotted in the bottom panel
with the same scale. Note that no continuum removal was applied for performing
the fit. }
\label{fig:M67}
\end{figure}

In Fig.~\ref{fig:M67} we show our SED fit to the integrated spectrum of the
standard open cluster M\,67 (Schiavon, Caldwell \& Rose 2004). Unlike for the
globular clusters of Fig.~\ref{fig:GCs} these authors obtained the integrated
spectrum for M\,67 by co-adding individual spectra of cluster members, weighted
according to their luminosities and relative numbers. The obtained age and
metallicity is in good agreement with these authors (see also Schiavon 2007) and
with our isochrone fitting results (Vazdekis et al. 1996) within the
uncertainties. The residuals only show a low frequency pattern, which is related
to differences in the flux calibration. However the higher frequency residuals
are negligible. It is particularly interesting that, unlike in
Fig.~\ref{fig:GCs}, we do not see the residuals blueward of 4300\,\AA. This
shows that all relevant spectral types are present in the MILES library.
Therefore the residuals obtained for Galactic globular clusters can be
attributed to the non-solar abundance ratios present in these clusters. However
we propose to use our base models in the same way that we describe in Section
\ref{sec:alpha} to deal with such abundance patterns.

\begin{figure}
\includegraphics[angle=270,width=3.3in.]{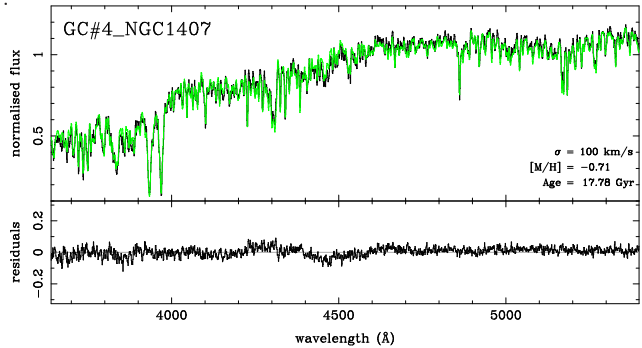}
\caption{The spectrum a bright globular cluster of the galaxy NGC\,1407, from
Cenarro et al. (2007b), plotted in black.  We overplot the best fitting SSP SED in green,
with its age and metallicity indicated within the panel. We adopt a Kroupa Universal
IMF. The model has been smoothed to match the instrumental resolution of the
observed spectrum (also quoted). The residuals are plotted in the bottom panel
with the same scale. Note that no continuum removal was applied for performing
the fit. }
\label{fig:GCCenarro}
\end{figure}

We illustrate in Fig.~\ref{fig:GCCenarro} the potential use of these models for
the study of extragalactic globular cluster systems. We plot a Keck/LRIS
spectrum of a globular cluster of the massive elliptical galaxy NGC\,1407
(Cenarro et al. 2007b), with mean S/N$\sim$40. The model SEDs were properly
smoothed to match the resolution of the observed spectrum ($\sigma \sim 100\,{\rm
km\,s}^{-1}$). We obtain an age of 17.78 Gyr and a metallicity of 
\mbox{$\mbox{[M/H]}$}$=-0.71$ in agreement, within the errors, with
the values derived from the detailed line-strength analysis performed by these authors.

\subsubsection{Using line-strength indices to estimate GC ages and
metallicities} 
\label{sec:GonzalezBonifacioSSPs}

\begin{figure*}
\includegraphics[angle=0,width=7.0in.]{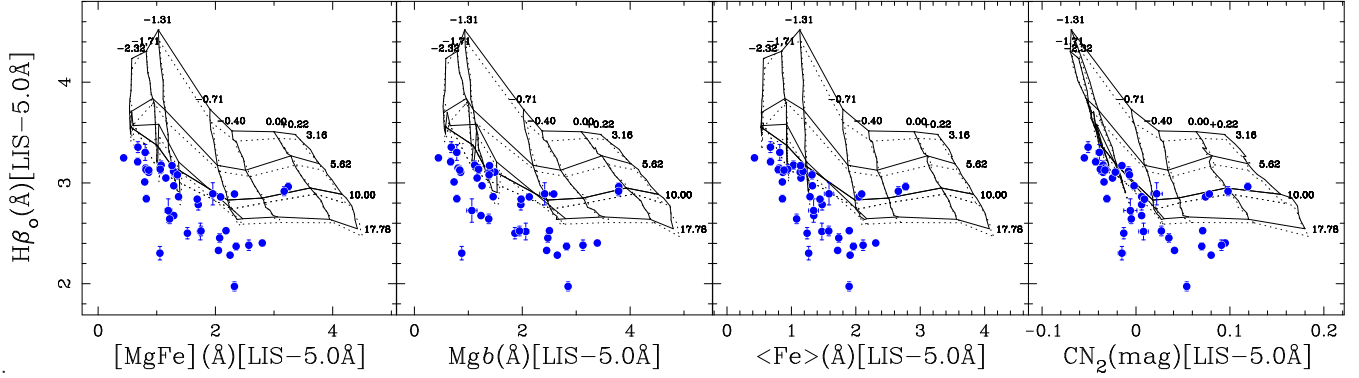}
\caption{The H$\beta_o$ age indicator of Cervantes \& Vazdekis (2009) is plotted
versus various metallicity indices, all measured at LIS-5.0\AA\, resolution, for
the Galactic globular cluster sample of Schiavon et al. (2005). Overplotted are
our reference models (solid lines), which adopt the MILES temperatures, and
another set of models that adopt the recently published temperature scale of
Gonz\'alez-Hern\'andez \& Bonifacio (2009) (dotted lines). SSP metallicities are
quoted on the top of the model grid, just above the line corresponding to the
youngest stellar populations (i.e. 3.16\,Gyr). We indicate the age of the SSPs
(in Gyr) for the most metal-rich models (i.e.\mbox{$\mbox{[M/H]}$}$=+0.22$).}
\label{fig:GonzalezBonifacioSSPs}
\end{figure*}

The line-strengths derived from our base models have
already been compared to the line-strengths of the integrated spectra of
Galactic globular clusters (e.g., Mendel et al. 2007; Cenarro et al. 2008;
Cervantes \& Vazdekis 2009). These comparisons confirm the well known model
zero-point problem affecting the age estimates obtained from the Balmer lines,
i.e. the derived ages are larger than the CMD-derived ages. Note that the same
problem appears when using other models. For an extensive description and
discussion of this problem we refer the reader to the above papers and to, e.g.,
Gibson et al. (1999), Vazdekis et al. (2001a), and Schiavon et al (2002).

In Fig.~\ref{fig:GonzalezBonifacioSSPs} we plot several line-strength diagnostic
diagrams to obtain the ages and metallicities of the Galactic globular cluster
sample of Schiavon et al. (2005). We use, for this purpose, our reference models
and the alternative set of models described in Section
\ref{sec:GonzalezBonifacio}, which adopt the recently published temperature
scale of Gonz\'alez-Hern\'andez \& Bonifacio (2009). We see that the  data fall
outside the model grid towards ages larger than the oldest predicted SSP (see
also Fig.~\ref{fig:lis8.4}). This is older than the ages derived from CMDs. If
we use the model grid that adopt the Gonz\'alez-Hern\'andez \& Bonifacio (2009)
temperature scale we obtain, however,  lower H$\beta_o$ values, i.e.,  we
alleviate in part the model zero-point problem obtaining slighly younger ages
for the globular clusters than with the reference models, using the  MILEs
temperatures.  This is because when synthesizing a SSP SED the code assigns a
cooler spectrum (i.e. with lower Balmer index values) to a requested star of
given parameters, as the Gonz\'alez-Hern\'andez \& Bonifacio (2009) temperatures
are $\sim$51\,K hotter than in MILES. Note also that these models provide
slightly larger values for the metallicity indices. However, as pointed out in
Section \ref{sec:GonzalezBonifacio}, we do not find a strong physical reason to
use the Gonz\'alez-Hern\'andez \& Bonifacio (2009) temperature scale,
particularly for metal-rich objects. Note that
Fig.~\ref{fig:GonzalezBonifacioSSPs} also shows that a significantly larger
temperature offset ($\sim$150\,K), i.e., much larger than the current
uncertainties in the temperature scale, is required to be able to overcome the
model zero-point. Therefore other aspects, such as those discussed in Vazdekis
et al. (2001a) and Schiavon et al. (2002) should be considered.

It is remarkable the large age sensitivity of the H$\beta_o$ index, which allows
us to obtain nearly orthogonal model grids, particulalry for
\mbox{$\mbox{[M/H]}$}$>-0.71$. We see that our age estimates for the most
metal-rich globular clusters do not depend on the metallicity indicator in use.
However the obtained metallicities might differ among the plotted diagnostic
diagrams, indicating a departure from the scaled-solar abundance pattern. In
fact, based on the H$\beta_o$ vs. CN2 diagram we obtain for many globular
clusters metallicities that are significantly larger than the ones inferred from
the H$\beta_o$ vs. [MgFe] plot. Therefore these clusters show CN2 values that
are larger than expected according to their total metallicities, as traced by
the [MgFe] index (e.g. Thomas et al. 2003). Note that characteristic CN-strong
pattern emerged in the residuals obtained from the full-spectrum fits shown in
Fig.~\ref{fig:GCs}.

Finally, note that these index-index diagrams show that there are a group of
clusters with apparently younger ages. Cenarro et al. (2008) found a clear
correspondence between the H$\beta$ values and the specific frequency of Blue
Stragglers (BSS), such that the higher the BSS ratio the younger the apparent
spectroscopic age. We refer the reader to that paper for further details. 

\subsection{Galaxies}
\label{sec:galaxies}

\begin{figure*}
\includegraphics[angle=0,width=7.0in.]{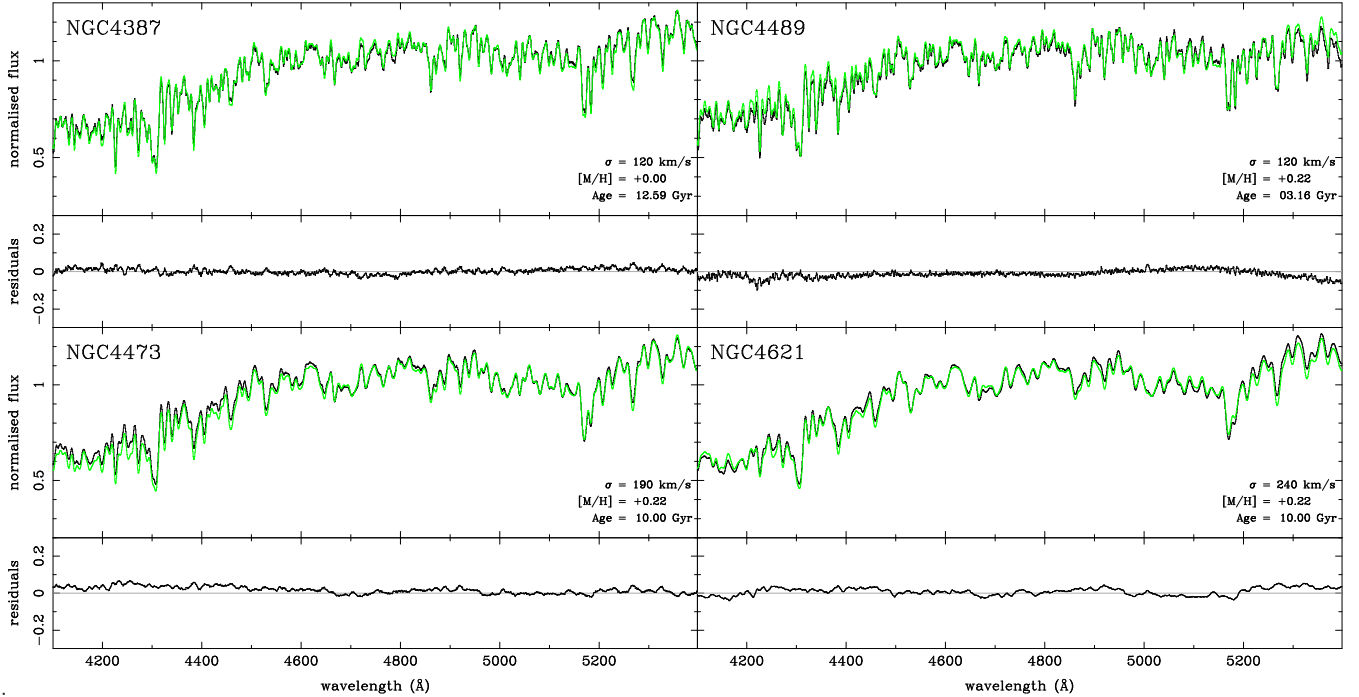}
\caption{Various Virgo elliptical galaxies from the sample of Yamada et al.
(2006), with S/N $>$100\,\AA$^{-1}$ at the H$\gamma$ feature, are plotted in
black. For each galaxy we overplot in green the SSP SED, with Kroupa Universal
IMF, corresponding to the age and metallicity (indicated within each panel),
which have bee  derived by these authors via the H$\gamma_{\sigma}$ vs. [MgFe]
diagnostic diagram. The model spectra were smoothed to match the total
resolution of the spectra (also indicated within the panels). Note that no
continuum correction was applied. The residuals are plotted within the panels
with the same scale.}
\label{fig:galaxies}
\end{figure*}

In this section we show the use of our models for studying galaxy spectra. 
With this aim we have chosen four galaxies (out of 14) of the Virgo early-type
galaxy sample of Yamada et al. (2006), with spectra of extraordinarily high S/N
(S/N $>$100\,\AA$^{-1}$ at the H$\gamma$ feature), extracted within an aperture
of r$_e$/10. Our subsample of galaxies  (NGC\,4387, NGC\,4473, NGC\,4489 and
NGC\,4621) cover a range in age, metallicity, abundance ratio and mass, as these
galaxies were selected alongside the Colour-Magnitue Relation of the Virgo cluster. The resolution of
the Subaru spectrum for NGC\,4489 is FWHM$=$3.1\,\AA, whereas for the other
three galaxies, observed at the William Herschel Telescope (Vazdekis et al.
2001b), the resolution is FWHM$=$2.4\,\AA. Such high quality spectra were
obtained to be able to measure the H$\gamma_{\sigma}$ age indicator of Vazdekis
\& Arimoto (1999), which is virtually insensitive to metallicity. Mean
luminosity weighted ages and metallicities were obtained by plotting
H$\gamma_{\sigma}$ versus the total metallicity indicator [MgFe] (Gonzalez 1993;
Vazdekis et al. 2001b; Thomas et al. 2003). The derived values are tabulated in
Table~5 of Yamada et al. (2006). We did not attempt here to fit the spectra of
these galaxies, but in Fig.~\ref{fig:galaxies} we overplot for each galaxy the
SSP SED with the closest age and metallicity, as listed in their Table~5. The
model SEDs were smoothed to match the total resolution of the galaxy spectra
(i.e. taking into account the instrumental resolution and galaxy velocity
dispersion), which are indicated in the panels. 

The residuals of the comparison for the low-velocity dispersion ($\sigma$), old
galaxy, NGC\,4387, are very small. For the
other galaxy with low $\sigma$ but young age, NGC~4489, the residuals 
in the blue part are as small as those obtained for NGC\,4387. However,  we see
a pattern in the residuals at longer wavelengths. If the stars of this galaxy
would have formed in two bursts then a characteristic residual pattern would be
seen, namely a surplus of light in the  blue and a deficit in the red, which is
not the case. The good flux-calibration of the models presented here allows this
type of analysis. Unlike the
other galaxies, the spectrum of NGC\,4489 was taken in one of the Subaru
observing runs, for which the correction of the continuum shape to a flux scale
is less accurate than is achieved for the observing runs at the WHT. The same
applies to the feature seen in the residuals at $\sim$4225\,\AA.

\subsubsection{Dealing with abundance ratios using the base models}
\label{sec:alpha}

For the most massive galaxies, NGC\,4473 and NGC\,4621 we see in Fig.~\ref{fig:galaxies} positive
residuals for the Fe features, e.g., Fe4383, Fe5270, Fe5335, and negative
residuals for the Mg feature at $\sim$5170\,\AA. This shows the well known
\mbox{$\mbox{[Mg/Fe]}$} enhancement, which is characteristic of massive galaxies (e.g.,
Peletier 1989; Worthey, Faber \& Gonz\'alez 1992; Gonz\'alez 1993).

\begin{figure}
\includegraphics[angle=-90,width=3.3in.]{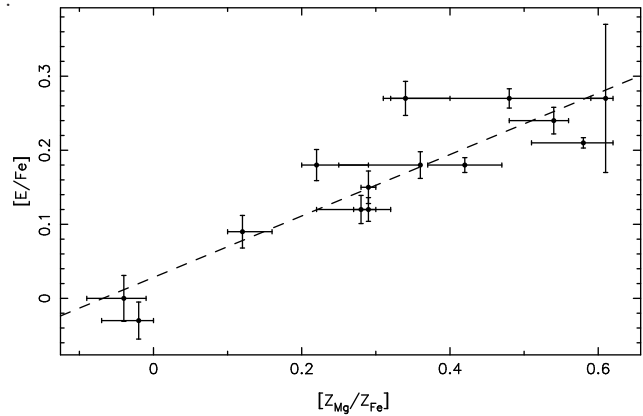}
\caption{[Z$_{\rm Mg}$/Z$_{\rm Fe}$] vs. \mbox{$\mbox{[Mg/Fe]}$} abundance ratio estimates
for the Virgo elliptical galaxies of Yamada et al. (2006). The proxy for the abundance
ratio is the metallicity difference derived from plotting the H$\gamma_{\sigma}$ age
indicator of Vazdekis \& Arimoto (1999) vs. the magnesium-dominated index, Mg$b$, and
H$\gamma_{\sigma}$ vs. the iron-dominated index, Fe3. We plot the $[Z_{\rm Mg}/Z_{\rm
Fe}]$ values listed in Table~5 of Yamada et al. To estimate the \mbox{$\mbox{[Mg/Fe]}$}
abundance ratio we use the Lick/IDS indices and the models of Thomas et al. (2003).}
\label{fig:heuristic}
\end{figure}

We show here how our base models can be safely used for studying galaxies with
non-solar element mixtures. A good example of such use is shown in the
line-strength analysis of Yamada et al. (2006). A proxy for the
\mbox{$\mbox{[Mg/Fe]}$} abundance ratio is obtained with the aid of the base
model grids and an age indicator that is almost insensitive to metallicity. If a
galaxy spectrum is enhanced in Mg/Fe, we obtain a higher metallicity when this
age indicator is plotted vs. a Mg-dominated index than when the index is
Fe-dominated. Yamada et al. use the H$\gamma_{\sigma}$ index of Vazdekis \&
Arimoto (1999), which requires spectra with extremely high S/N, and the
metallicity indicators Mg$b$ and Fe3 ($\equiv\frac{{\rm Fe4383}+{\rm
Fe5270}+{\rm Fe5335}}{3}$ defined in Kuntschner, 2000). Irrespective of the
metallicity indicator in use, such optimized age indicators provide orthogonal
model grids, and the lines of constant age are essentially horizontal, which
means that a given measurement for the age indicator corresponds to a unique age
(see Cervantes \& Vazdekis 2009).  The metallicity difference obtained in this
way, [Z$_{\rm Mg}$/Z$_{\rm Fe}$], is a good proxy for the abundance ratio
\mbox{$\mbox{[Mg/Fe]}$}, as was shown in, e.g., S\'anchez-Bl\'azquez et al.
(2006b), de la Rosa et al. (2007) and Michielsen et al. (2008). In
Fig.~\ref{fig:heuristic} we plot the values obtained by Yamada et al. (2006,
Table~5) for this proxy vs. the \mbox{$\mbox{[Mg/Fe]}$} estimates derived with
the aid of the models of Thomas et al. (2003), which specifically account for
non scaled-solar element partitions. To apply these models we use the Lick/IDS
version of the H$\beta$, Mg$b$ and Fe3 indices of Yamada et al. (2006).  As an
example, Table~5 of Yamada et al. lists [Z$_{\rm Mg}$/Z$_{\rm Fe}$]$\sim$0.6 for
NGC\,4473 and NGC\,4621. Following Fig.~\ref{fig:heuristic} this translates to
\mbox{$\mbox{[Mg/Fe]}$}$\sim+0.25$ on the abundance ratio scale of Thomas et al.
(2003). Note that this [Mg/Fe] abundance scale is also a considerable
approximation because {\it i)} the use of scaled-solar isochrones, {\it ii)} it
is based on three synthetic spectra to model the SSPs and {\it iii)} a fixed
element ratio pattern is adopted.  Note also that in Fig.~\ref{fig:heuristic} we
combine the results from two different model predictions, and that our method
requires extrapolating to higher metallicities to obtain $Z_{Mg}$ (see the
corresponding plots in Yamada et al. 2006). This explains in part the larger
dispersion obtained for the higher abundance ratios. 

It is worth noting that the newly defined H$\beta_o$ index of Cervantes \&
Vazdekis (2009) provides similar age disentangling power as the
H$\gamma_{\sigma}$ index, but with the advantage that H$\beta_o$ requires
spectra of similar qualities as usually employed for measuring the standard
H$\beta$ index (see Cervantes \& Vazdekis 2009). H$\beta$ can also be used for
this purpose, but the resulting model grids are significantly less orthogonal,
increasing the error bars of the [Z$_{\rm Mg}$/Z$_{\rm Fe}$] determinations.

Finally, to obtain the total metallicity, these optimized age indicators can be
plotted versus the combined indices [MgFe] (Gonz\'alez 1993) or [MgFe]' (Thomas
et al. 2003), which have been shown to be virtually insensitive 
to the \mbox{$\mbox{[Mg/Fe]}$}
abundance ratio.

In the second paper of this series we will present an updated version of
Fig.~\ref{fig:heuristic}, where we use both the scaled-solar and the
$\alpha$-enhanced version of our MILES models, in order to be fully consistent.
In addition, these models reach higher metallicities, avoiding in this way the
extrapolation that might be required to obtain [Z$_{\rm Mg}$/Z$_{\rm Fe}$]. 
However, we have shown here that  our base models, which combine scaled-solar isochrones with an
empirical stellar library, can be used for studying galaxies with
$\alpha$-enhanced element partitions via the line-strength analyses.

\subsubsection{Cautionary remark on the H$\beta$ model predictions}
\label{sec:jones}

\begin{figure}
\includegraphics[angle=0,width=3.45in.]{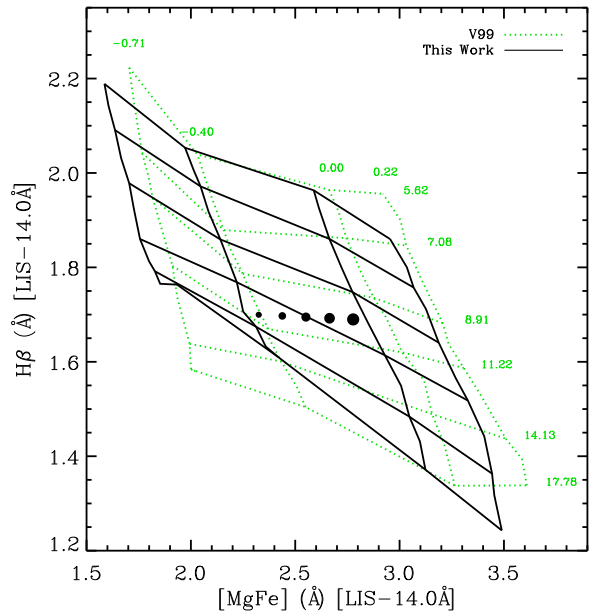}
\caption{H$\beta$ vs. [MgFe] diagnostic diagram obtained from our MILES base
model SEDs (solid line) and V99 (dotted line), which uses the Jones (1999)
stellar library. The indices were measured on the SSP spectra once smoothed to
match the LIS-14.0\AA\, resolution. We plot the radial line-strength profile for the
Virgo galaxy NGC\,4387, which has been obtained from low resolution Keck/LRIS
spectra by S\'anchez-Bl\'azquez et al. (2007). The size of the filled circles
decrease with increasing galactocentric distance: 1.25, 2.5, 5, 10, 20\,arcsec.}
\label{fig:hbeta_Jones_MILES}
\end{figure}

Figure \ref{fig:hbeta_Jones_MILES} shows the very popular diagnostic diagram 
H$\beta$ vs. [MgFe], which is obtained from our MILES SSP SEDs once smoothed to
match the LIS-14.0\AA\ resolution. For comparison we show the equivalent model grid
from the V99 models, as updated in this work, which use the Jones (1999) stellar
spectral library. 
Fig.~\ref{fig:hbeta_Jones_MILES} shows that the V99 model grid is significantly
more orthogonal than the grid obtained with the MILES SSP SEDs, mainly due to the
lower sensitivity of H$\beta$ index to metallicity variations. The MILES SSP grid looks,
in fact, more similar to the equivalent
standard grid obtained on the basis of the empirical fitting functions of Worthey
et al. (1994) on the Lick/IDS system. 

We plot the radial line-strength profile for the Virgo galaxy
NGC\,4387, which has been obtained from low resolution Keck/LRIS spectra by
S\'anchez-Bl\'azquez et al. (2007). We deliberately do not plot the
observational errors, as we are only interested in illustrating the net effect on
the trend. Note that we also plotted in Fig.~\ref{fig:galaxies} the higher
resolution spectrum obtained at the William Herschel Telescope for the central
aperture r$_{e}/10$, i.e., 1.6\,arcsec (Yamada et al. 2006). 
According to the V99 model grid, the observed gradient can be attributed to a
decreasing metallicity with increasing galactocentric distance. Using the MILES
SSP grid, the gradient can be explained by a combination of increasing age and
decreasing metallicity.

We have investigated the possible cause for the  differences between the two
model grids. As we use here an updated version for the V99 models, which is
now based on the same isochrones and transformations to the observational
plane that we employ for the MILES models, these two ingredients cannot be
the  the ones we are looking for. We identified three other possible sources:
{\it i)} intrinsic differences in the spectra of MILES and Jones (1999)
stellar libraries, mostly around the H$\beta$ feature (such differences have
been quoted before by Worthey \& Ottaviani (1997); {\it ii)} differences in
the stellar parameters adopted here (i.e. Paper~II) and in V99 and {\it iii)}
differences in the algorithms employed for computing/assigning a stellar
spectrum to a given set of atmospheric parameters. In order to understand this
issue, we synthetized SEDs using the  245 stars, both in the Jones and MILES
libraries, with the following  characteristics: (a) Models using both stellar
libraries.  MILES and Jones.(b) Models using MILES spectra assigning them the
two set of stellar parameters, those used in V99 and in MILES (c) Models using
MILES stellar spectra and with both algorthims,  the one used in V99 and the
new algortihm described in V03.

For the sake of brevity we do not show here this detailed study but we summarize
the main results. We found that these model differences are mostly originated
by the stellar spectrum assignation scheme employed in V99, which works at a
significantly more coarse stellar parameter resolution. The other aspects
described above also have a significant impact, but to a lesser degree.
Furthermore the H$\beta$ index measurements on the SSP SEDs synthesized
with the same algorithm but exchanging the stellar spectral libraries are
consistent with the H$\beta$ index measurements obtained on a star by star basis.
Unfortunately, this problem might have some
impact on certain studies that make use of the  H$\beta$ index derived from V99 models.

Given the improvement on the stellar spectrum assignation scheme used
for the  MILES SSPs, we will provide an updated version of the V99 predictions
in our website using the new algorithms.

\section{The web tool}
\label{sec:web}

The exploitation of stellar population synthesis models has depended in part on
the development of tools by users. Some authors offer software to either help to
interpret or aid the user to handle the provided predictions. A few examples
include LECTOR (V99), GALAXEV (Bruzual \& Charlot 2003), P\'{E}GASE
(Fioc \& Rocca-Volmerange 1997,1999), or EZ Ages (Graves \& Schiavon 2008). 
In an effort to
continue and extend these initiatives, we have created a new website
(http://miles.iac.es) to provide support for using the spectra of both the
MILES stellar library and the model predictions described in this paper,
but also in V99 at 1.8\,\AA\ resolution. The
webpage also handles the near-IR stellar spectral library CaT and models that we
published in Cenarro et al. (2001a), Cenarro et al. (2001b), Cenarro et al.
(2002) and V03. Line-strength predictions based on empirical
fitting functions (including the ones of the Lick/IDS system) are also included.
Furthermore the webpage includes photometric quantities such as fluxes, colours,
mass-to-light ratios and surface brightness fluctuations, which have been
published in Vazdekis et al. (1996) and Blakeslee, Vazdekis \& Ajhar (2001). It
is very important to note that all these predictions will be continuously updated
with the latest developments and with the feedback from users. These comply with
the standards of the Virtual Observatory.

This webpage is intended for users not only to retrieve both the stellar spectra
and the synthetic SSP SED predictions in this project, but also to provide
webtools to facilitate the handling and transformation of the spectra. Beyond
the basic tools for the exploration and visualization of the whole dataset, at a
first stage, we also provide advanced webtools to: 

\begin{itemize} 

\item Generate synthetic stellar spectra for a given set of stellar atmosphere
parameters. These spectra are computed on the basis of our stellar libraries
according to the mathematical algorithm described in Appendix B of V03 (see also
Section~\ref{sec:synthesis}). 

\item Transform the stellar spectra or the SSP spectra to the instrumental setup
of the data observed by the user. This tool will allow the user to input the
details of the instrumental configuration of the observed data (spectral
resolution and sampling) and generate a set of spectra matching that setup. The
model/library spectra will be convolved with the observational PSF 
(typically Gaussian) to match
the spectral resolution of the observed data. During this step, particular
attention will be paid as to whether the user requires the FWHM to be constant
as a function of wavelength (in which case convolution will be performed in 
linear space),
or constant in velocity  (in which case convolution will be performed
in logarithmic space). Finally, the data will be rebinned (i.e. spectrally
sampled, and if desired, in log scale). 

\item Measure a variety of line-strength indices and colors on the
spectra. We make use of LECTOR (V99) and INDEXF (Cardiel 2007) for this purpose. This tool allows
to input newly defined indices by the user.

\item Convert Lick/IDS line-strength indices to the new Line Index System
proposed here. We use for this purpose the transformations described in Appendix
A. 

\item Estimate the best luminosity weighted age and metallicity, and
alpha-enhancement for a given set of line indices using RMODEL (Cardiel et al.
2003). This program determines stellar population parameters (e.g. age,
metallicity, IMF type and slope), using as input line-strength indices by 
interpolating SSP model predictions. Both linear and bivariate fits are
computed to perform the interpolation. 

\item Provide spectra for a given star formation history.

\end{itemize} 

These tools are intended to be the first of many more sophisticated tools, that
we will incorporate in the website over time, to facilitate the comparison between 
the observed data and our stellar library and models, and in this way make the 
analysis easier.


\section{Summary and Conclusions}
\label{sec:conclusions}

Here we present stellar population model SEDs based on the empirical
stellar spectral library MILES . In Paper~I we have presented and described this library, 
whereas in Paper~II we have published the stellar parameter determinations. Here we
present synthetic SEDs for single-age, single-metallicity stellar populations
(SSPs) covering the full optical spectral range, 3540.5--7409.6\,\AA, at
moderately high resolution, FWHM$=$2.3\,\AA, which is virtually
constant as a function of wavelength (see Fig.~4 of Paper~I). 
The SSP SEDs have a reliable, accurate flux
calibration, which is one of the major advantages of the MILES library.
Furthermore the redder part of the SSP spectra do not show telluric residuals,
as these were carefully cleaned from the stellar spectra as shown in Paper~I.

We consider the SEDs presented here as our base models, as they combine an
empirical library, which is imprinted with the chemical composition of the solar
neighbourhood, with scaled-solar stellar isochrones (Girardi et al. 2000).
Therefore, whereas our models are self-consistent, and scaled-solar for solar
metallicity, this is not the case for the low metallicity regime, as the
observed stars there do not show this abundance pattern (see Schiavon 2007). In a
second paper we will present self-consistent models, both scaled-solar and
$\alpha$-enhanced, for a range in metallicities. For this purpose we have used MILES, 
together with
theoretical model atmospheres, which are coupled to the appropriate stellar isochrones. 

An interesting feature of our base models is that they rely, as much as
possible, on empirical ingredients. Apart from using MILES stellar spectra, we 
also use transformations of the theoretical
parameters of the isochrones to observational, measurable, quantities that are
based on extensive empirical, photometric libraries, with
marginal dependence on model atmospheres.
The implementation of MILES in the
population synthesis code has been done very carefully: each star of the library
has been compared with other MILES stars with similar atmospheric
parameters, using the same interpolation algorithm used in the synthesis
code. As a result of these tests and other considerations, we have removed 
or decreased the weight with which a given star
contributes to the synthesis of a stellar spectrum for 135 (out of 985) stars of MILES. 

The unprecedented stellar parameter coverage of the MILES library has allowed us
to safely extend our optical SSP SEDs from intermediate to very old age regimes
(0.06--18\,Gyr) and the metallicity coverage of the SSPs from supersolar
(\mbox{$\mbox{[M/H]}=+0.22$}) to \mbox{$\mbox{[M/H]}=-2.32$}. The very low
metallicity SSPs should be of particular interest for globular cluster studies.
In addition, we have computed SSP SEDs for a suite of IMF shapes and slopes:
unimodal and bimodal IMF shapes as defined in Vazdekis et al. (1996) with slopes
varying in the range 0.3-3.3 (Salpeter: unimodal with slope 1.3), and the
segmented IMFs of Kroupa (2001) (Universal and Revised cases).  In addition
to our reference models, which adopt the MILES temperatures published in
Paper~II, we provide an alternative set of models that adopt the recently
published temperature scale of Gonz\'alez-Hern\'andez \& Bonifacio (2009) (i.e.,
about $\sim$51\,K hotter). The most relevant feature of these models is that the
well known model zero-point problem affecting the age estimates obtained from
the Balmer lines, i.e. the obtained ages are larger than the CMD-derived ages,
is partially alleviated.

To constrain the range of ages, metallicities and IMFs where the use of these
SEDs are safe, we provide a quantitative analysis. This analysis, which
basically takes into account the parameter coverage of the stellar library
feeding the models, clearly shows that our SSP SEDs can be safely used in the
age range 0.06--18\,Gyr, and for all the metallicities, except for
\mbox{$\mbox{[M/H]}=-2.32$}, for which only stellar populations with ages above
$\sim$10\,Gyr can be considered safe. This applies to the standard IMF shapes.
However our quantitative analysis has shown that the quality of the SSP SEDs
decreases as we increase the IMF slope, especially for the Unimodal IMF. For
example, for slopes above 1.8, the synthesized SSP SEDs for
\mbox{$\mbox{[M/H]}=-2.32$} are no longer safe for a Unimodal IMF. According to
this analysis our predictions are of significantly higher quality than any
previous model predictions in the literature  for all the ages and metallicities
that have been computed in this work. For example, any predictions based on the
standard fitting functions of the Lick/IDS system (Worthey et al. 1994), e.g.
line-strength indices, with which most stellar population studies have been
performed so far, should not be considered safe for \mbox{$\mbox{[M/H]}<-1.3$}
(for standard IMF shapes). This limit is set at higher metallicity,
\mbox{$\mbox{[M/H]}\sim-0.7$}, for predictions based on STELIB (e.g., Bruzual \&
Charlot 2003). The comparison of our new models, based on MILES, with the SSP
SEDs published in V99, reveals systematic differences due to limitations in the
flux calibration of the Jones (1999) stellar library that feeds the latter
models. As for the line-strengths the most important difference between these
two models is found for the H$\beta$ index, which shows a lower metallicity
sensitivity in the V99 models. We show an example where
this effect has a non-negligible impact on our interpretation of galaxy
age/metallicity gradients. 

The SSP SEDs increase our ability to perform the stellar populations analysis,
which can be done in a very flexible manner. Observed spectra can be studied by
fitting the full spectrum or by focussing on selected line indices, 
either standard or newly defined. An example of the previous method can be found
in Koleva et al. (2008), where various stellar populations models were compared
following this approach. Our favourite approach is to analyze the spectrum of a
galaxy at the resolution imposed by its velocity dispersion (and its
instrumental resolution),  which require us to smooth the SSP SEDs to match the
observed total resolution. Therefore, the line index analysis to obtain relevant 
stellar population parameters is performed in the system of the galaxy, 
either through full spectrum fitting or line index approaches.

As many authors prefer to plot the line-strength measurements of their whole
galaxy sample in a single index-index diagnostic diagram, or to compare their
data to tabulated indices in the literature, we propose a new Line Index System
(LIS), which makes this method straightforward to apply. With LIS we avoid the
well known uncertainties inherent to the widely employed Lick/IDS system. Unlike
the latter, LIS has two main advantages, i.e., a constant resolution as function
of wavelength and a universal flux-calibrated spectral response. Data can be
analyzed in this system at three different resolutions: 5\,\AA, 8.4\,\AA\ and
14\,\AA\ (FWHM), i.e. $\sigma=$127, 214 and 357\,${\rm km\,s}^{-1}$ at
5000\,\AA, respectively. These resolutions are appropriate for studying globular
cluster, low and intermediate-mass galaxies, and massive galaxies, respectively.
Line-strength measurements given at a higher LIS resolution can be smoothed to
match a lower LIS resolution.  Furthermore we provide polynomials to transform
current Lick/IDS line-index measurements in the literature to the new system. 
We provide LIS line-index tables for various popular samples of Galactic
globular clusters and galaxies. We also show various popular index-index
diagnostic diagrams for these samples in the LIS system.

As an application, we have fitted a number of representative stellar clusters of
varying ages and metallicities with our models, obtaining good agreement with
CMD determinations. Unlike for the open cluster M\,67, our SED fits for a
sample of Galactic globular clusters show non negligible residuals blueward of
4300\,\AA, which mostly reflect the characteristic CN-strong features of these
clusters among other deviations from the scaled-solar pattern. We also applied
our models to representative galaxies with high quality spectra, for which
independent studies are available, obtaining good results. We show that our base
models can be used for studying line-strength indices of galaxies with
$\alpha$-enhanced element partitions. Examples of such a use can be found in,
e.g., Vazdekis et al. (2001b), Kuntschner et al. (2002), Carretero et al.
(2004), Yamada et al. (2006). The method consists in plotting a highly sensitive
age indicator, such as those of Vazdekis \& Arimoto (1999), or the recently
defined H${\beta_o}$ index of Cervantes \& Vazdekis (2009) versus Mg (e.g.,
Mg$b$) and versus Fe (e.g., Fe4383, $\langle\mbox{Fe}\rangle$), obtaining a
virtually orthogonal model grid where the estimated age does not depend on the
metallicity indicator in use. Unlike the age, for a \mbox{$\mbox{[Mg/Fe]}$}
enhanced galaxy the obtained metallicities differ when plotted against, for
example,  Mg$b$ and $\langle\mbox{Fe}\rangle$ indices. This metallicity
difference, [Z$_{{\rm Mg}b}$/Z$_{<{\rm Fe}>}$], can be used as a good proxy for
the abundance ratio determined with the aid of stellar population models that
specifically take into account non-solar element partitions. Although this proxy
yields larger \mbox{$\mbox{[Mg/Fe]}$} values, there is a linear relation between
these two ways of estimating the abundance ratios (e.g., S\'anchez-Bl\'azquez et
al. 2006b; de la Rosa et al. 2007; Michielsen et al. 2008). In the second paper
of this series we
will present both $\alpha$-enhanced and scaled-solar self consistent SSP SEDs,
all based on MILES, but modified with the aid of model atmospheres.  By applying
these models to galaxies with varying abundance ratios, for which high S/N
spectra are available, we confirm the linear relation mentioned above. There is
one advantage  in using this proxy: the abundance ratios determined in this way
do not depend  on the specific details of modeling of $\alpha$-enhanced stellar
populations such as those that are implicit to the computations of the
atmosphere, or the adopted element mixtures.

Our fits to the stellar cluster and galaxy spectra of high quality have
confirmed the high precision of the flux calibration of our SSP SEDs.
Furthermore the colours that we obtained from our SSP SEDs are consistent with
the ones that we derived via the employed photometric stellar libraries, 
within typical zero point uncertaintes. Such agreement shows the reliability of
the temperatures adopted for the MILES stars (Paper~II), which are consistent
with the temperature scales of Alonso et al. (1996,1999). Such accuracy opens
new applications for these models. For example, the model SEDs can be used as
templates for codes aiming at determining spectrophotometric redshifts,
particularly for those based on narrow-band filters (e.g. ALHAMBRA, PAU).

Finally we present a webpage from which not only these models and stellar
libraries can be downloaded, but where we also provide a suite of user-friendly
webtools to facilitate the handling and transformation of the spectra. For
example, once the user enters the details of the instrumental setup employed in
the observations (e.g., PSF, sampling), the users can obtain their favourite
line strength indices and diagnostic diagrams, ready to plot their
observational measurements.  Apart of the model SEDs shown here, the webpage
also provides predictions for a suite of observable quantities.


\section*{Acknowledgments}

We are indebted to the Padova group for making available their isochrone
calculations. We are grateful to P. Coelho, J.~L. Cervantes, I.~G. de la Rosa,
M. Koleva, E. Ricciardelli and J. Gonz\'alez-Hern\'andez for very useful discussions. We also
would like to thank J.~A. Perez Prieto for helping us in the construction of the
web page for the models. We thank the referee for pointing out relevant aspects
that were useful for improving the original manuscript. The MILES library was
observed at the INT on the island of La Palma, operated by the Isaac Newton
Group at the Observatorio del Roque de los Muchachos of the Instituto de
Astrof\'{\i}sica de Canarias. This research has made an extensive use of the
SIMBAD data base (operated at CDS, Strasbourg, France), the NASA's Astrophysics
Data System Article Service, and the {\it Hipparcos} Input Catalogue. PSB
acknowledges the support of a Marie Curie Intra-European Fellowship within the
6th European Community Framework Programme. JFB acknowledges support from the
Ram\'on y Cajal Program financed by the Spanish Ministry of Science and
Innovation. AJC is a {\it Ram\'on y Cajal} Fellow of the Spanish Ministry of
Science and Innovation. This work has been supported by the Programa Nacional de
Astronom\'{\i}a y Astrof\'{\i}sica of the Spanish Ministry of Science and
Innovation under grants \emph{AYA2007-67752-C03-01, AYA2007-67752-C03-02 and
AYA2006-15698-C02-02}.

\label{lastpage}

\appendix

\section{Transformations from the Lick/IDS system to the new Line Index System}
\label{ap:Transformations} 

\begin{figure*}
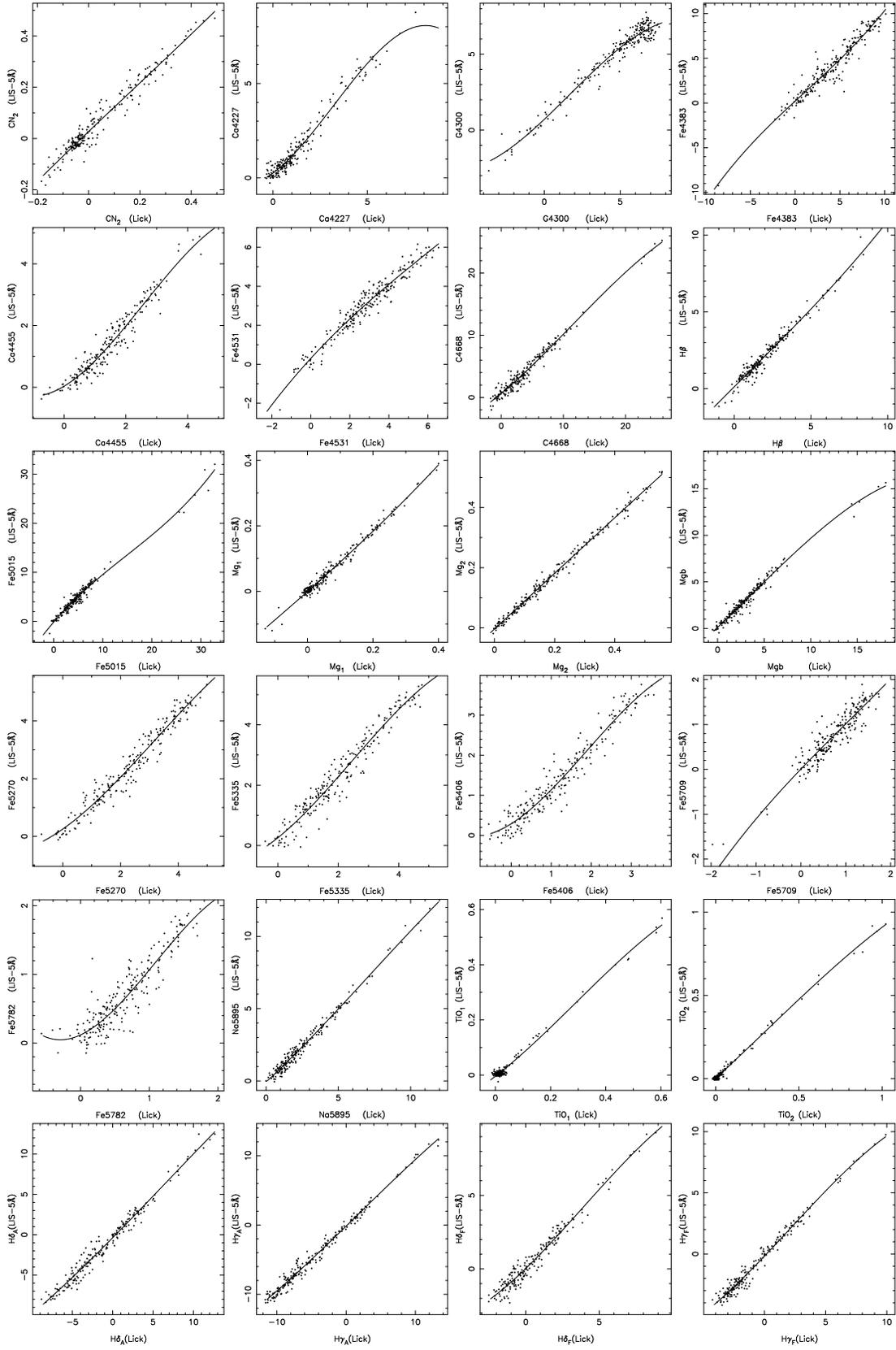

\resizebox{0.2\textwidth}{!}{\includegraphics[angle=-90]{cn2.lis1.ps}}
\resizebox{0.2\textwidth}{!}{\includegraphics[angle=-90]{ca4227.lis1.ps}}
\resizebox{0.2\textwidth}{!}{\includegraphics[angle=-90]{g4300.lis1.ps}}
\resizebox{0.2\textwidth}{!}{\includegraphics[angle=-90]{fe4383.lis1.ps}}
\resizebox{0.2\textwidth}{!}{\includegraphics[angle=-90]{ca4455.lis1.ps}}
\resizebox{0.2\textwidth}{!}{\includegraphics[angle=-90]{fe4531.lis1.ps}}
\resizebox{0.2\textwidth}{!}{\includegraphics[angle=-90]{c4668.lis1.ps}}
\resizebox{0.2\textwidth}{!}{\includegraphics[angle=-90]{hbeta.lis1.ps}}
\resizebox{0.2\textwidth}{!}{\includegraphics[angle=-90]{fe5015.lis1.ps}}
\resizebox{0.2\textwidth}{!}{\includegraphics[angle=-90]{mg1.lis1.ps}}
\resizebox{0.2\textwidth}{!}{\includegraphics[angle=-90]{mg2.lis1.ps}}
\resizebox{0.2\textwidth}{!}{\includegraphics[angle=-90]{mgb.lis1.ps}}
\resizebox{0.2\textwidth}{!}{\includegraphics[angle=-90]{fe5270.lis1.ps}}
\resizebox{0.2\textwidth}{!}{\includegraphics[angle=-90]{fe5335.lis1.ps}}
\resizebox{0.2\textwidth}{!}{\includegraphics[angle=-90]{fe5406.lis1.ps}}
\resizebox{0.2\textwidth}{!}{\includegraphics[angle=-90]{fe5709.lis1.ps}}
\resizebox{0.2\textwidth}{!}{\includegraphics[angle=-90]{fe5782.lis1.ps}}
\resizebox{0.2\textwidth}{!}{\includegraphics[angle=-90]{na5895.lis1.ps}}
\resizebox{0.2\textwidth}{!}{\includegraphics[angle=-90]{tio1.lis1.ps}}
\resizebox{0.2\textwidth}{!}{\includegraphics[angle=-90]{tio2.lis1.ps}}
\resizebox{0.2\textwidth}{!}{\includegraphics[angle=-90]{hda.lis1.ps}}
\resizebox{0.2\textwidth}{!}{\includegraphics[angle=-90]{hga.lis1.ps}}
\resizebox{0.2\textwidth}{!}{\includegraphics[angle=-90]{hdf.lis1.ps}}
\resizebox{0.2\textwidth}{!}{\includegraphics[angle=-90]{hgf.lis1.ps}}

\caption{Comparison of the Lick/IDS indices of the Lick/IDS library 
and those in MILES broadened to a constant resolution FWHM=5\,\AA. Solid lines
show the transformation from one system to the other.}
\label{ plots}
\end{figure*}

\begin{figure*}
\resizebox{0.2\textwidth}{!}{\includegraphics[angle=-90]{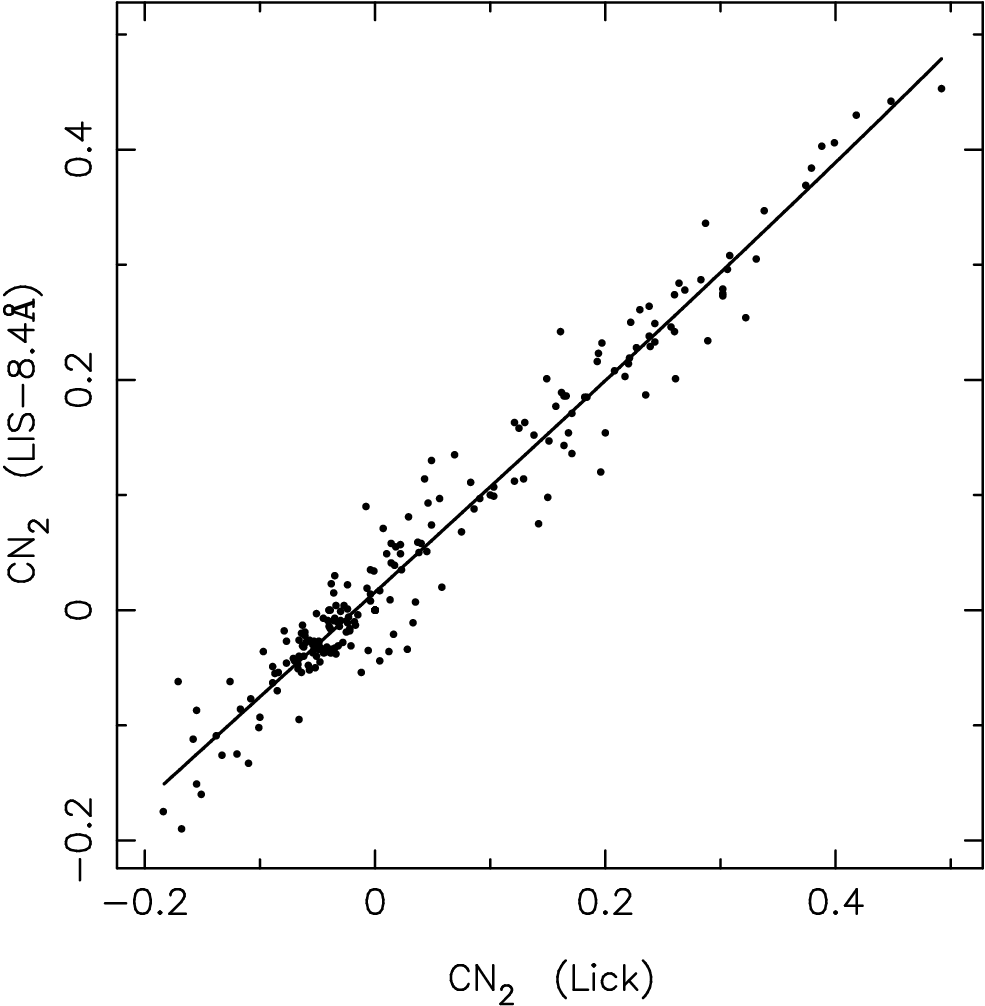}}
\resizebox{0.2\textwidth}{!}{\includegraphics[angle=-90]{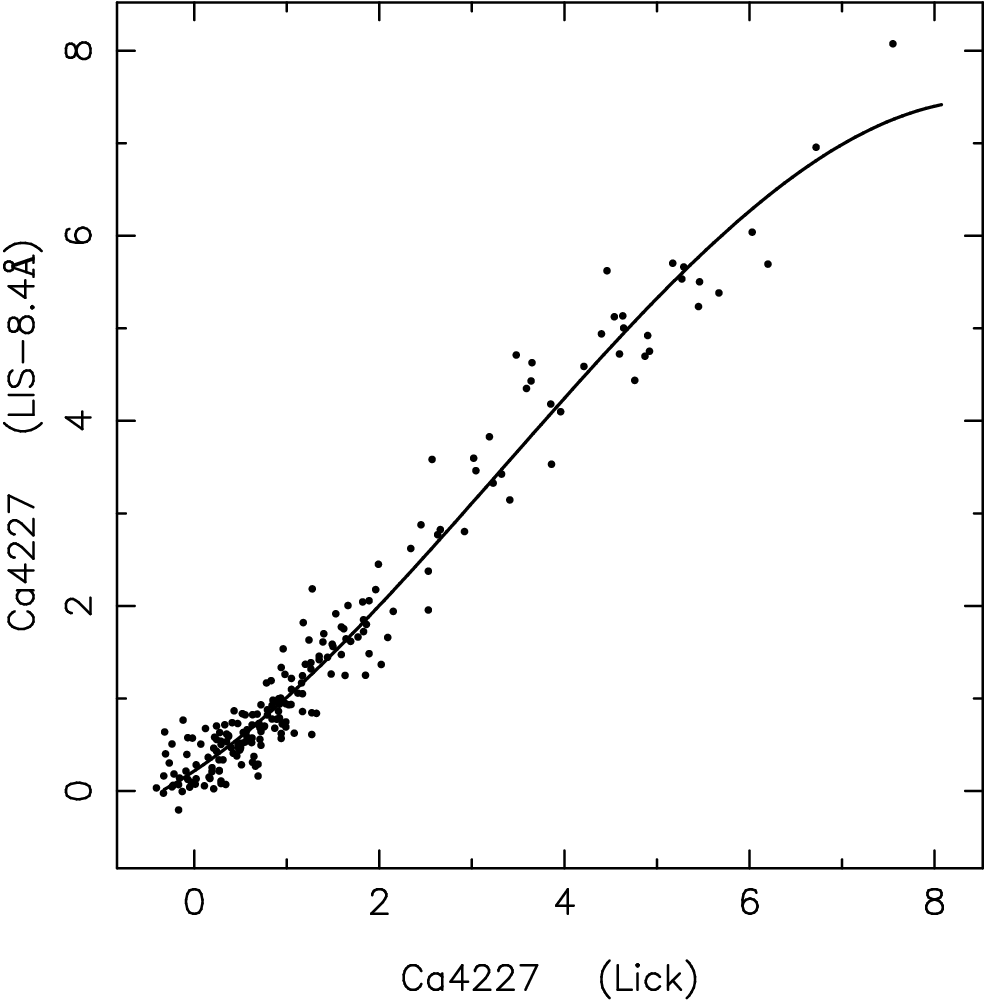}}
\resizebox{0.2\textwidth}{!}{\includegraphics[angle=-90]{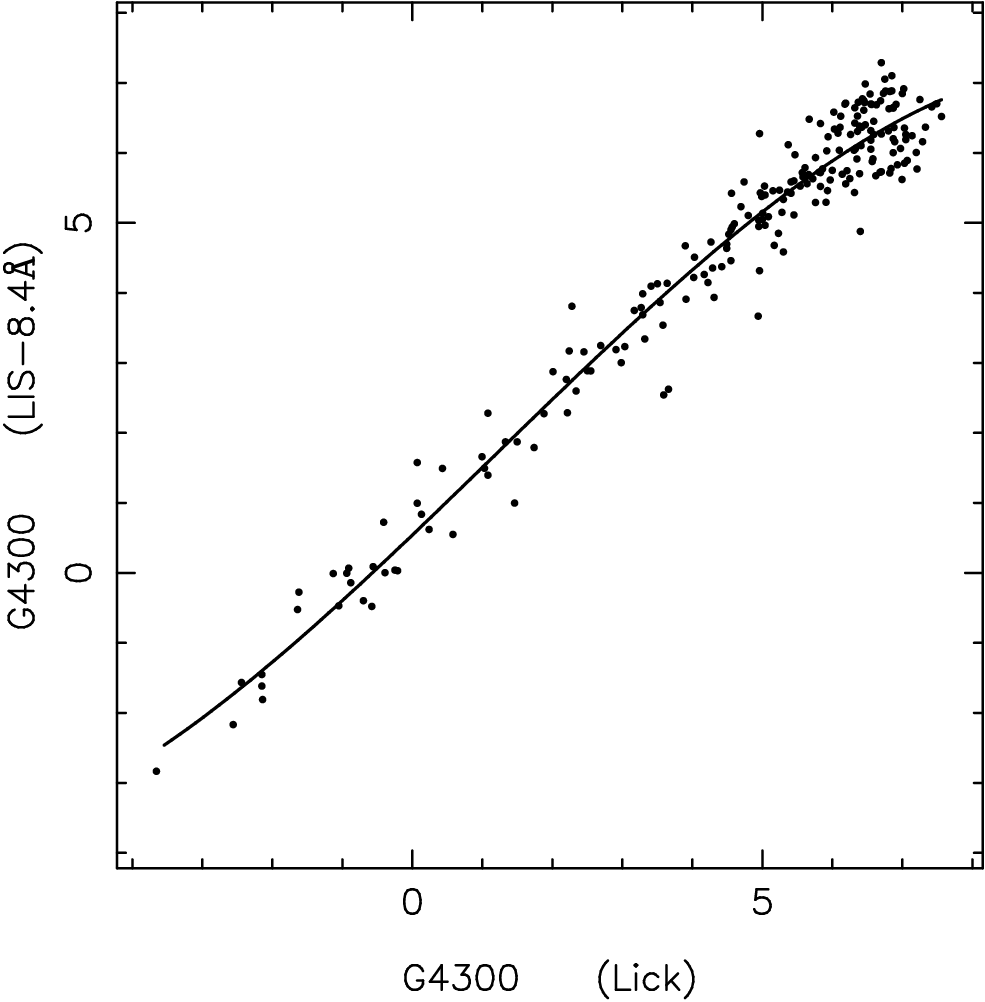}}
\resizebox{0.2\textwidth}{!}{\includegraphics[angle=-90]{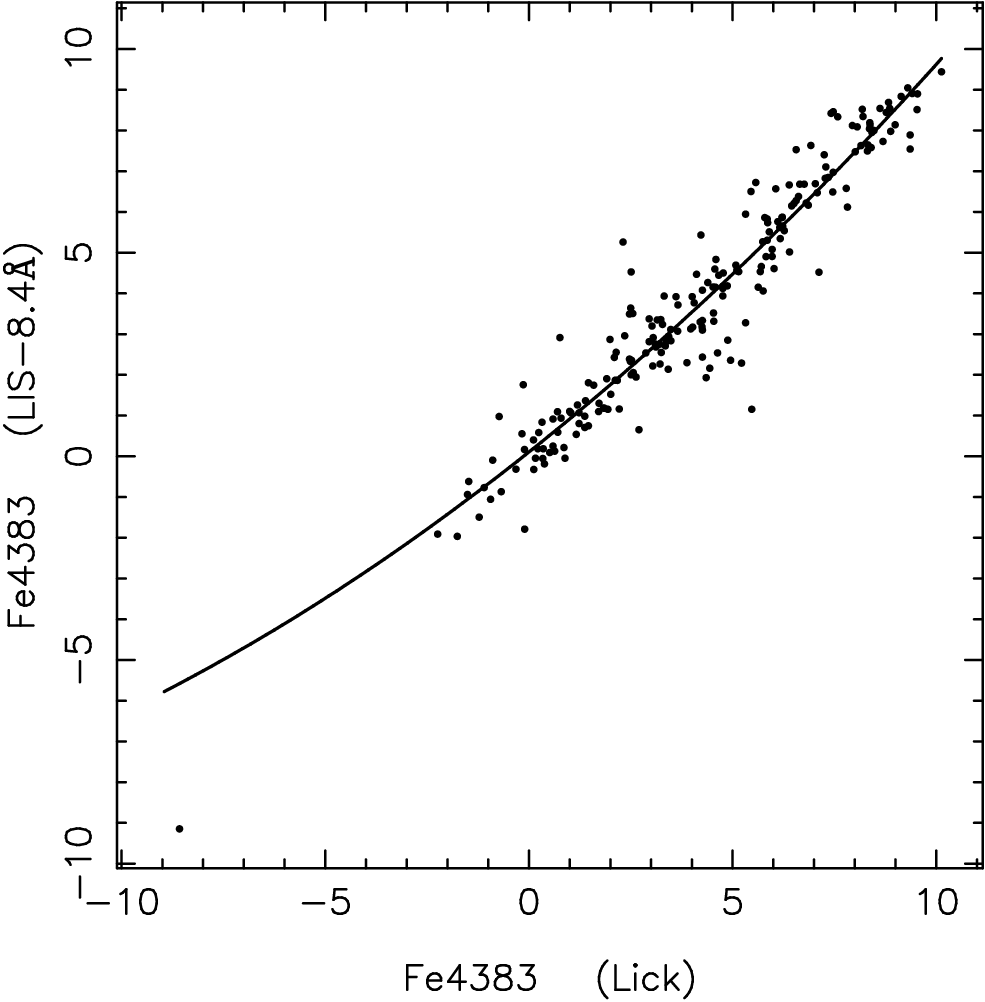}}
\resizebox{0.2\textwidth}{!}{\includegraphics[angle=-90]{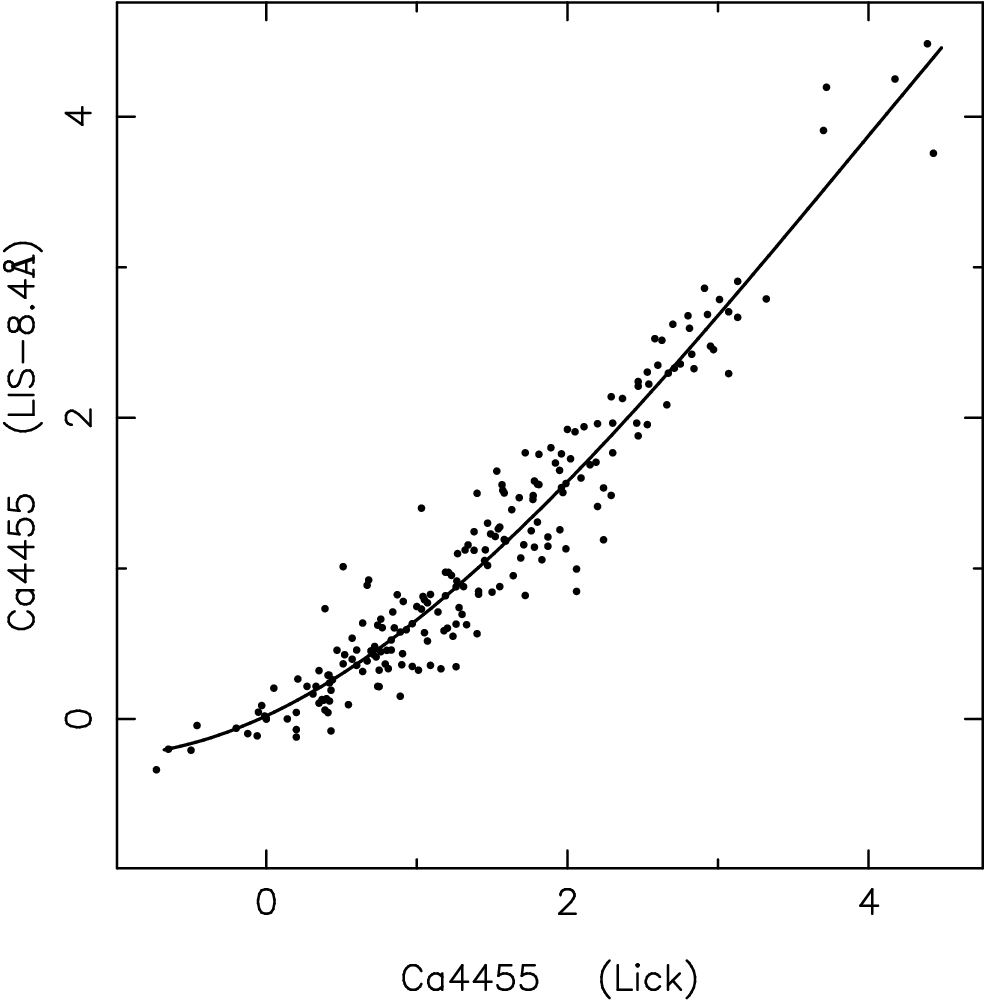}}
\resizebox{0.2\textwidth}{!}{\includegraphics[angle=-90]{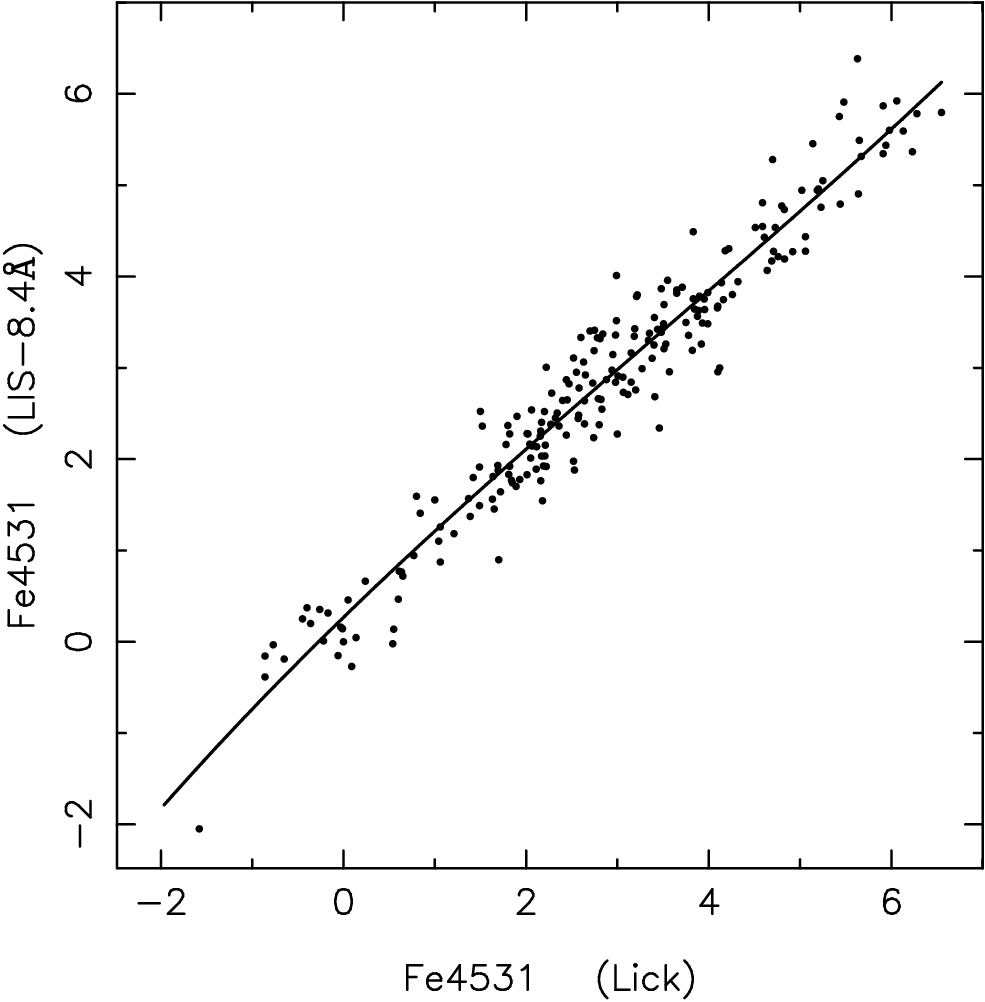}}
\resizebox{0.2\textwidth}{!}{\includegraphics[angle=-90]{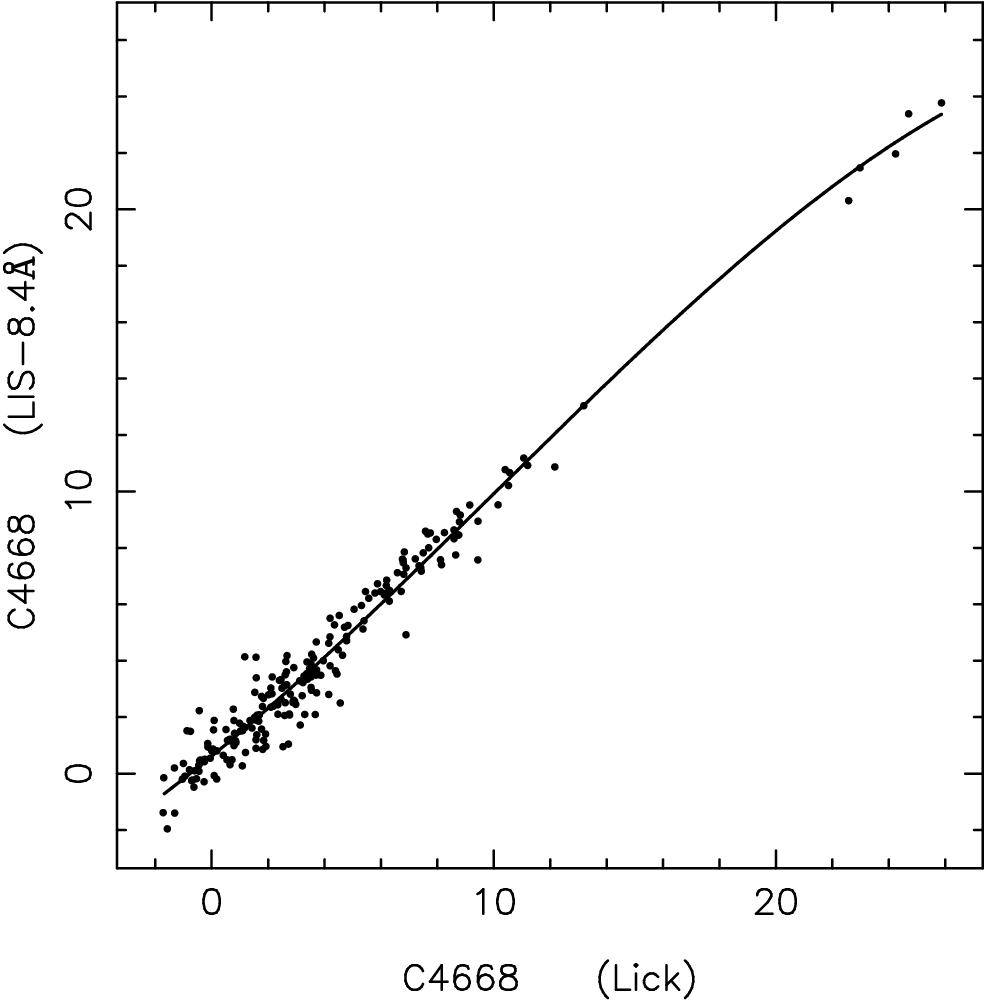}}
\resizebox{0.2\textwidth}{!}{\includegraphics[angle=-90]{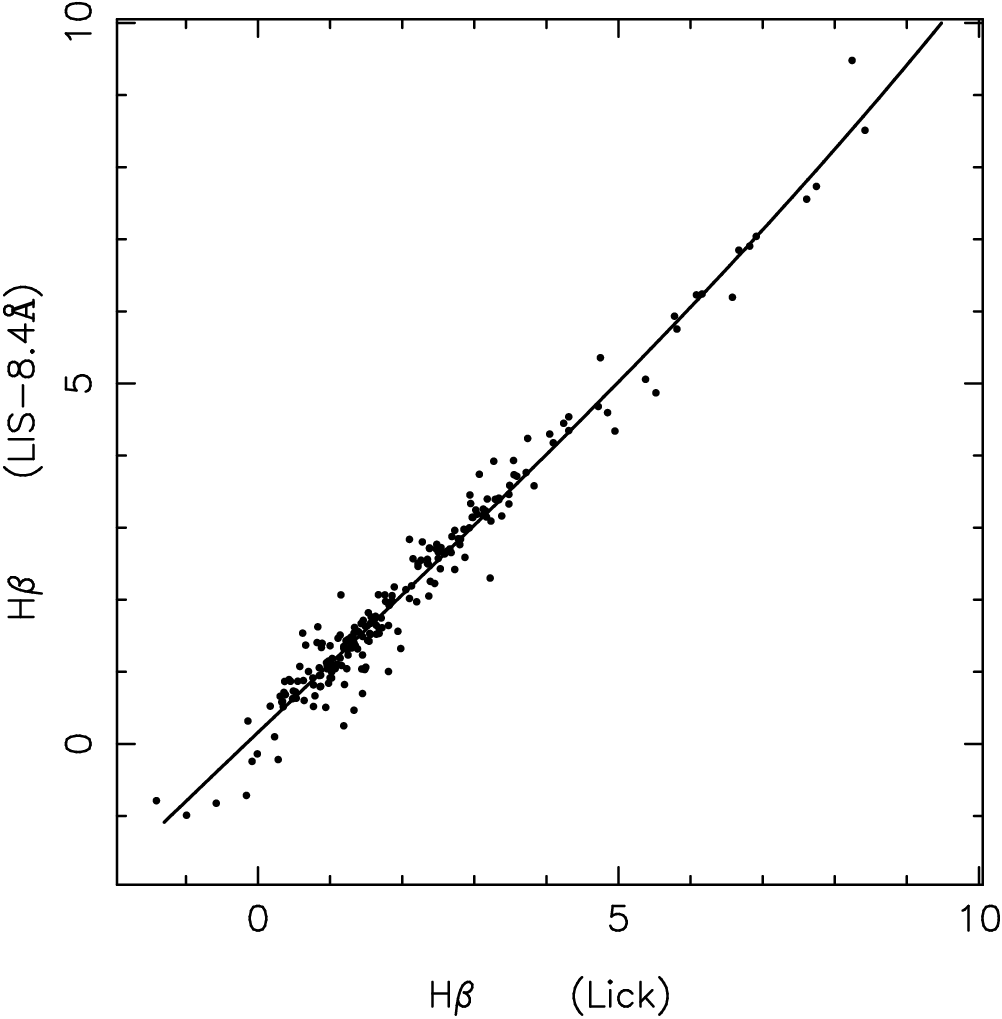}}
\resizebox{0.2\textwidth}{!}{\includegraphics[angle=-90]{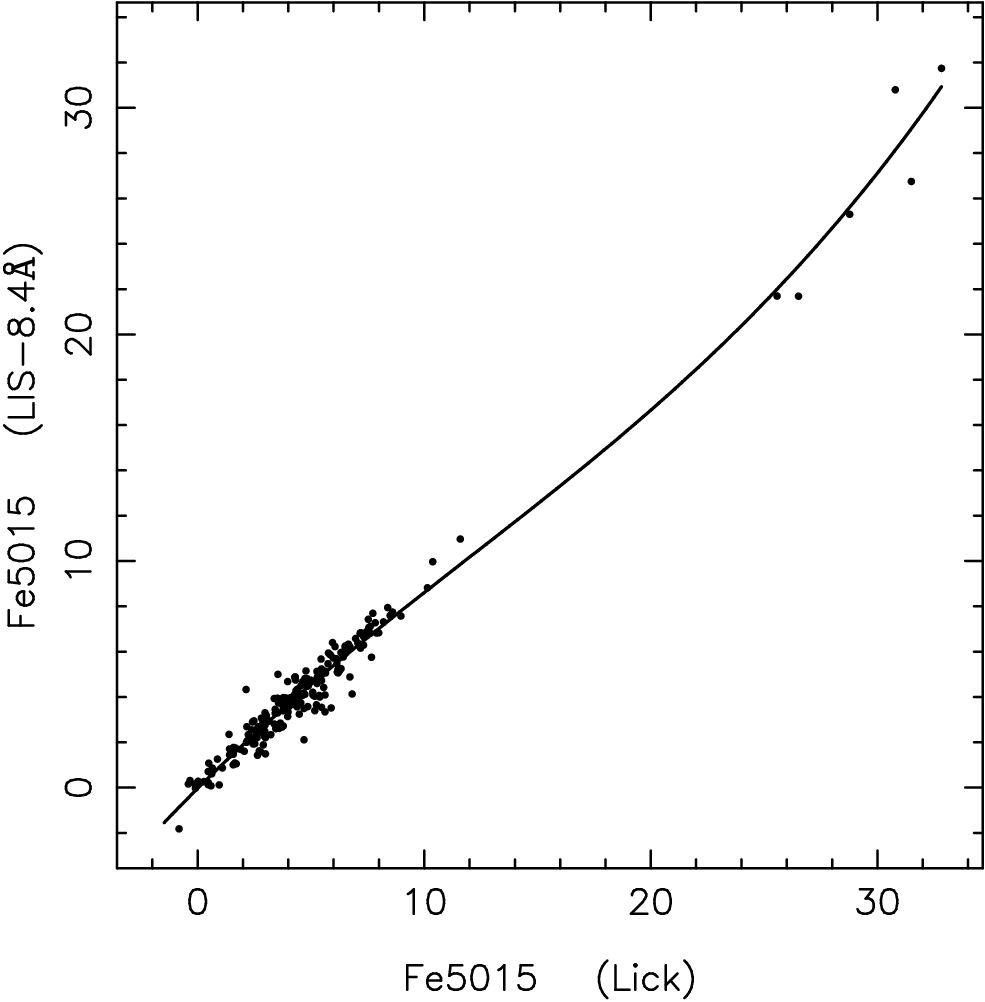}}
\resizebox{0.2\textwidth}{!}{\includegraphics[angle=-90]{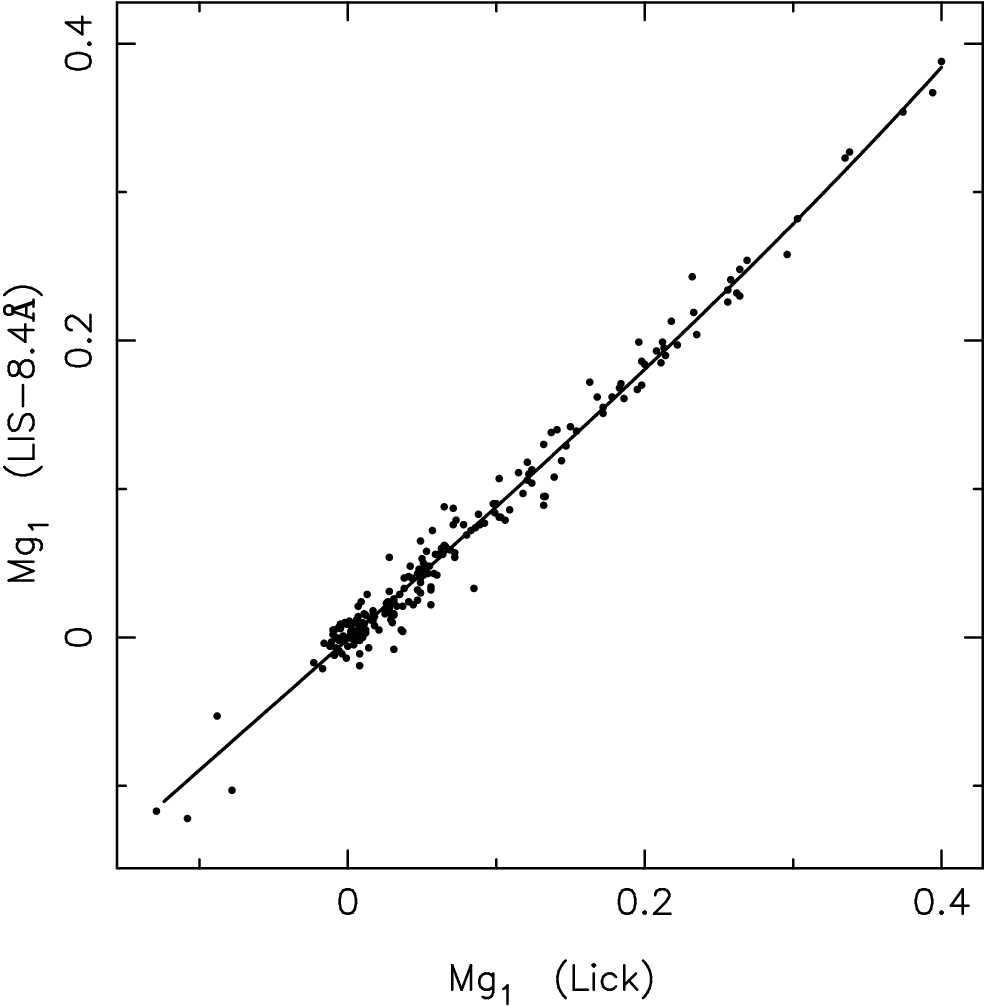}}
\resizebox{0.2\textwidth}{!}{\includegraphics[angle=-90]{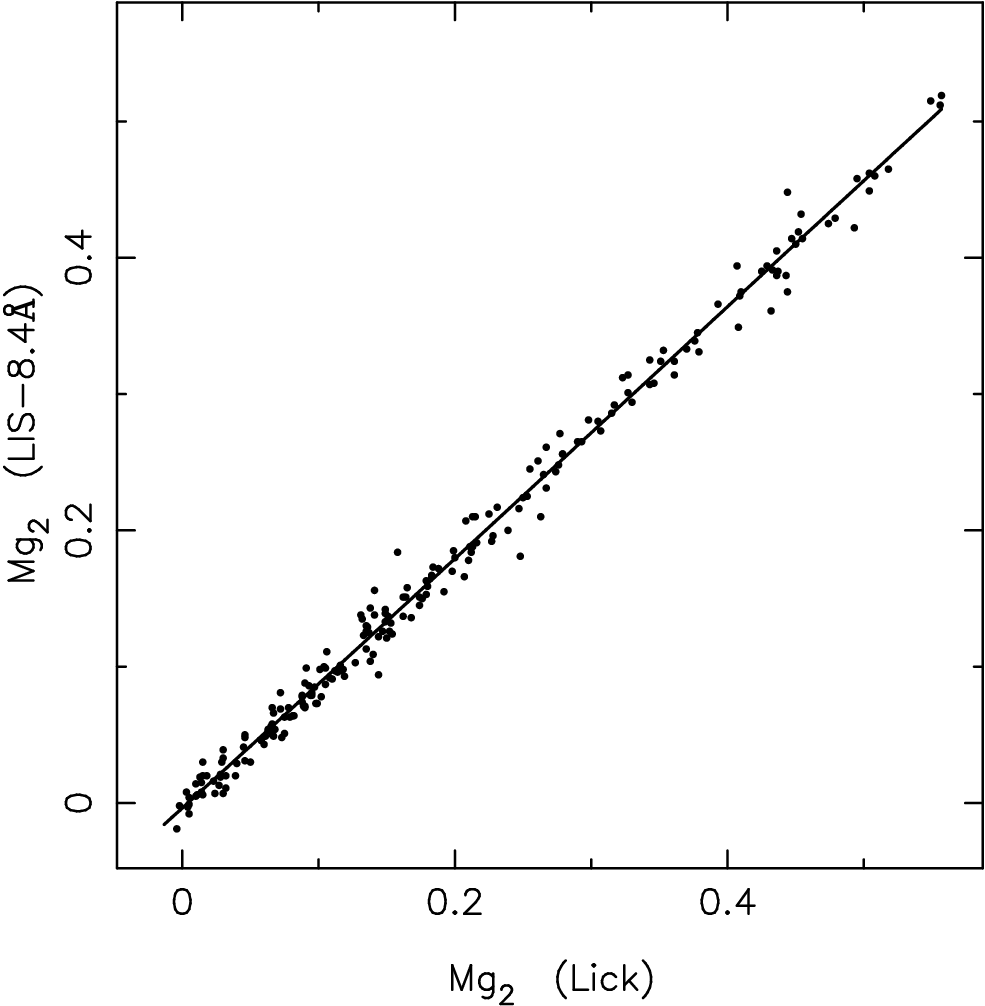}}
\resizebox{0.2\textwidth}{!}{\includegraphics[angle=-90]{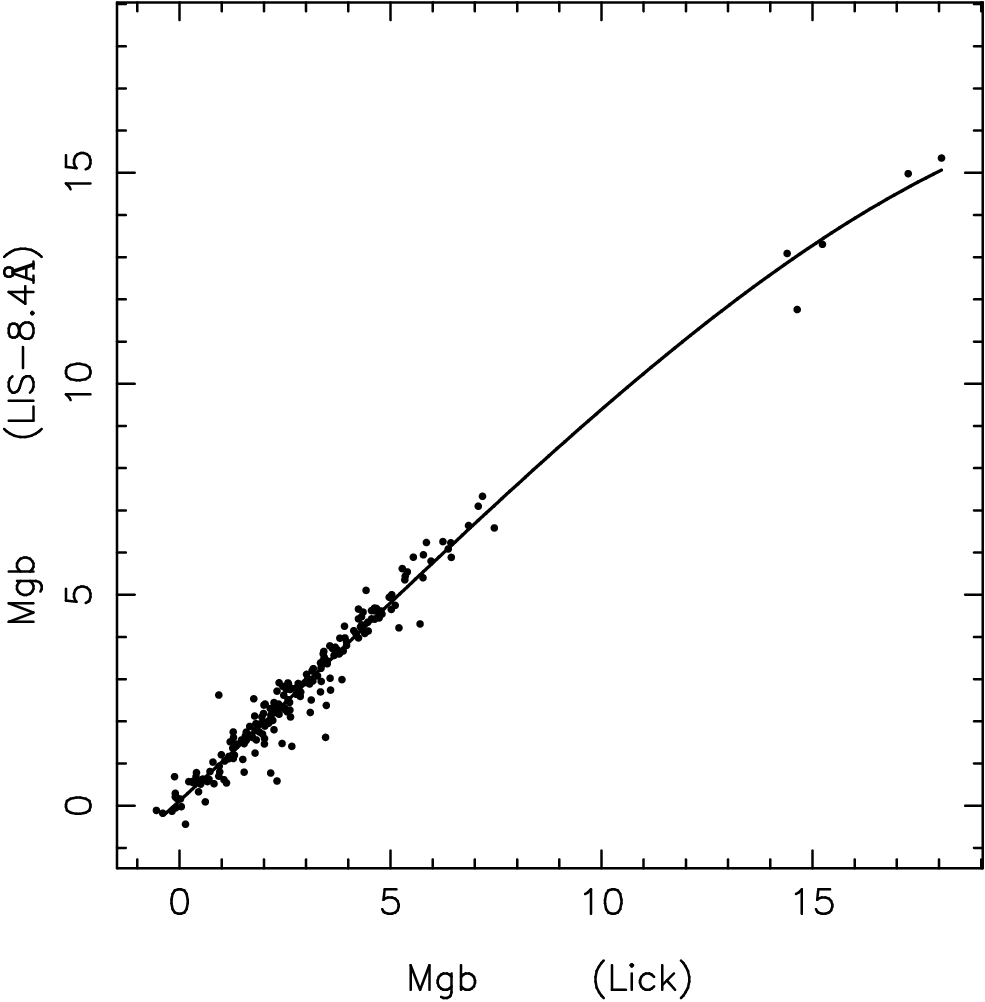}}
\resizebox{0.2\textwidth}{!}{\includegraphics[angle=-90]{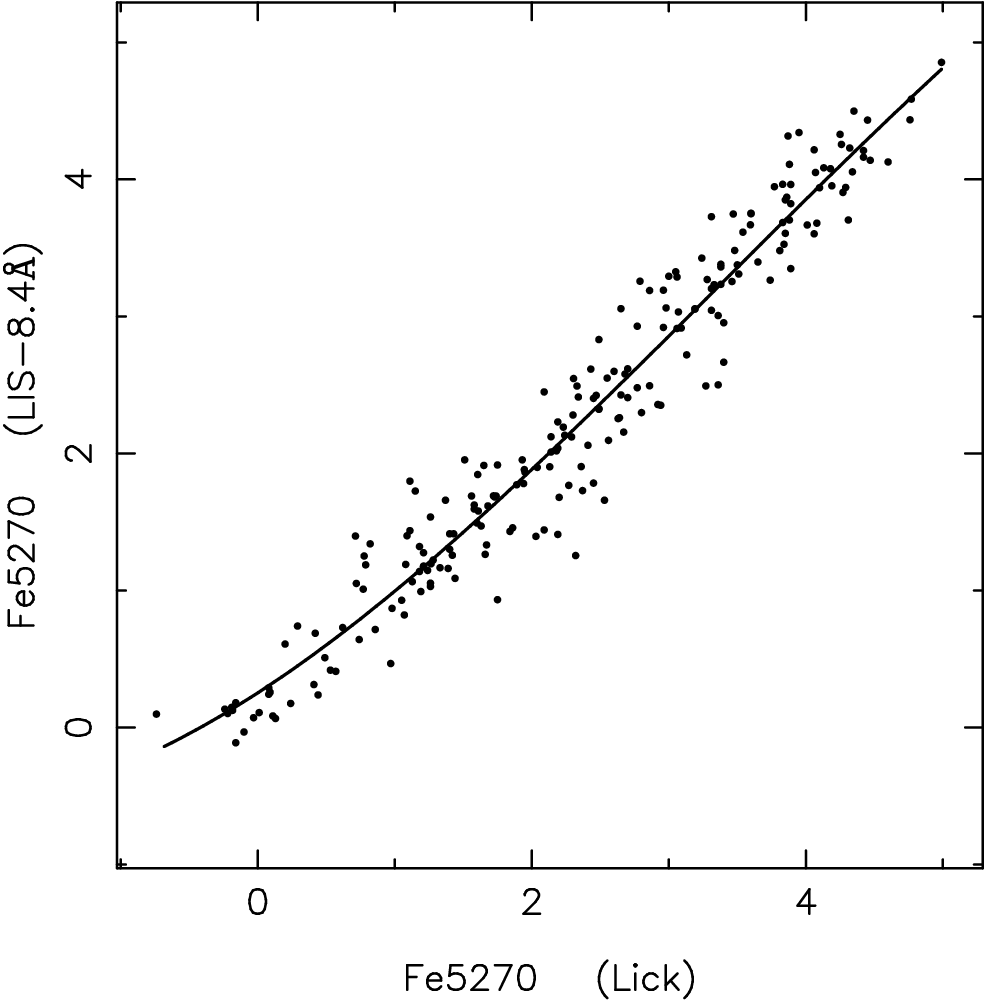}}
\resizebox{0.2\textwidth}{!}{\includegraphics[angle=-90]{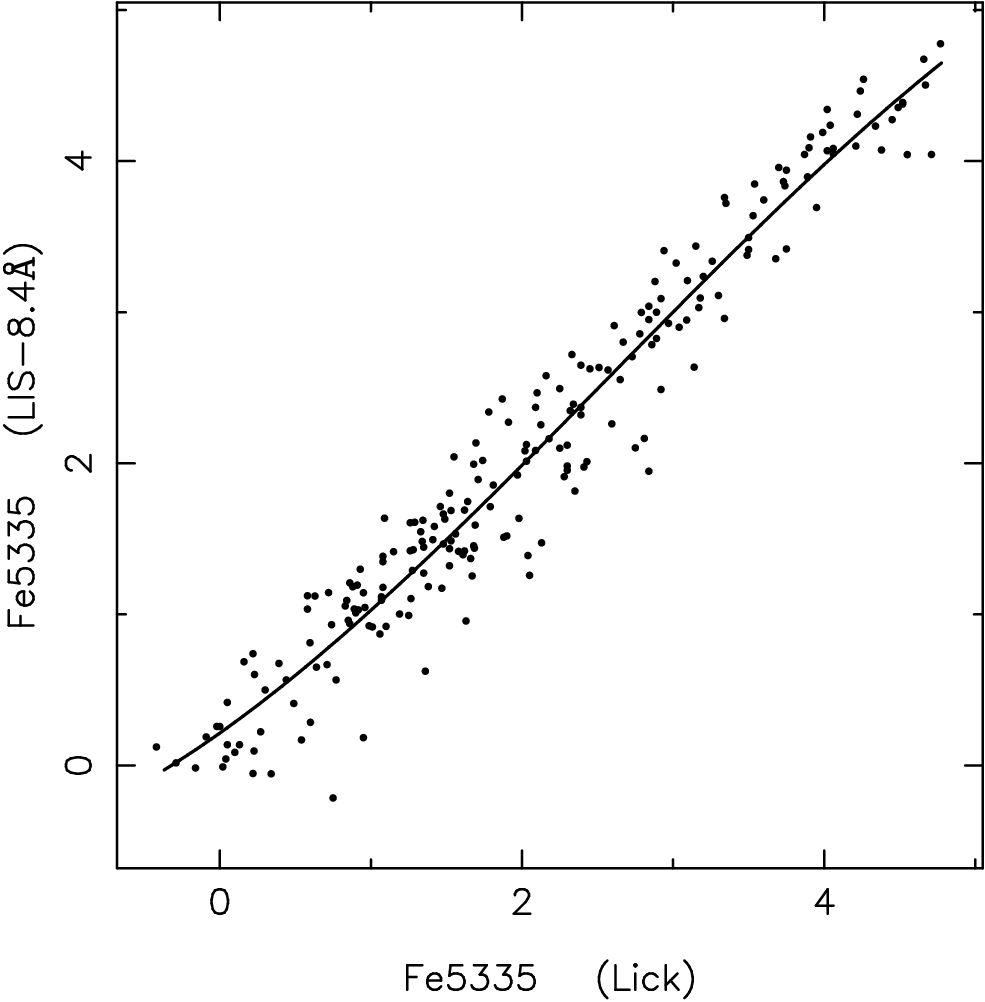}}
\resizebox{0.2\textwidth}{!}{\includegraphics[angle=-90]{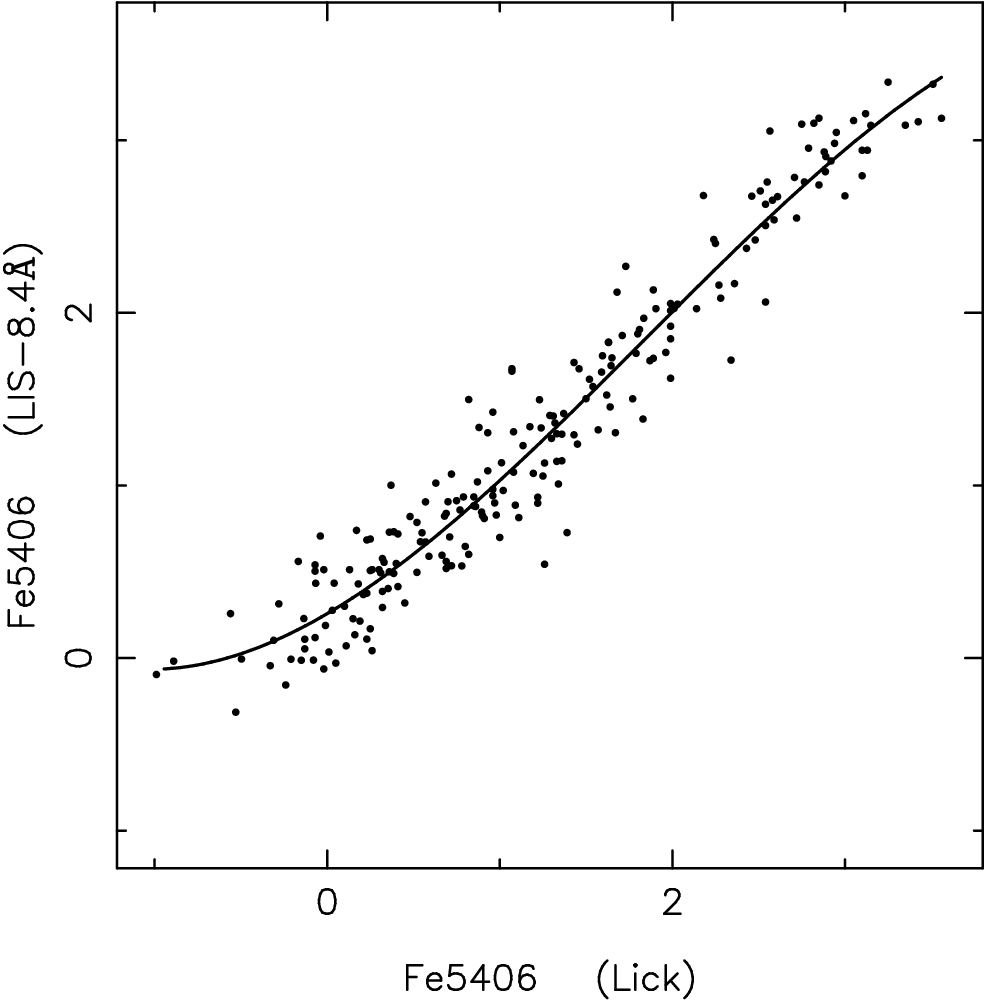}}
\resizebox{0.2\textwidth}{!}{\includegraphics[angle=-90]{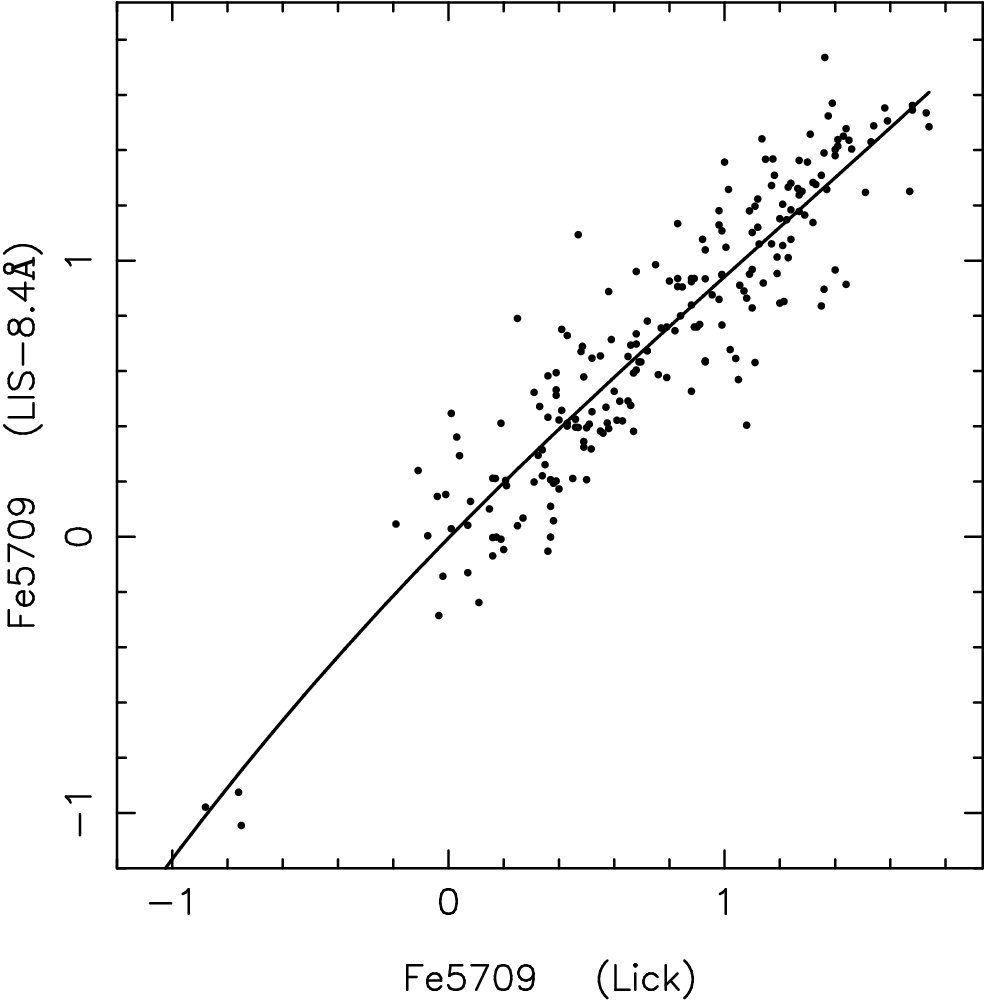}}
\resizebox{0.2\textwidth}{!}{\includegraphics[angle=-90]{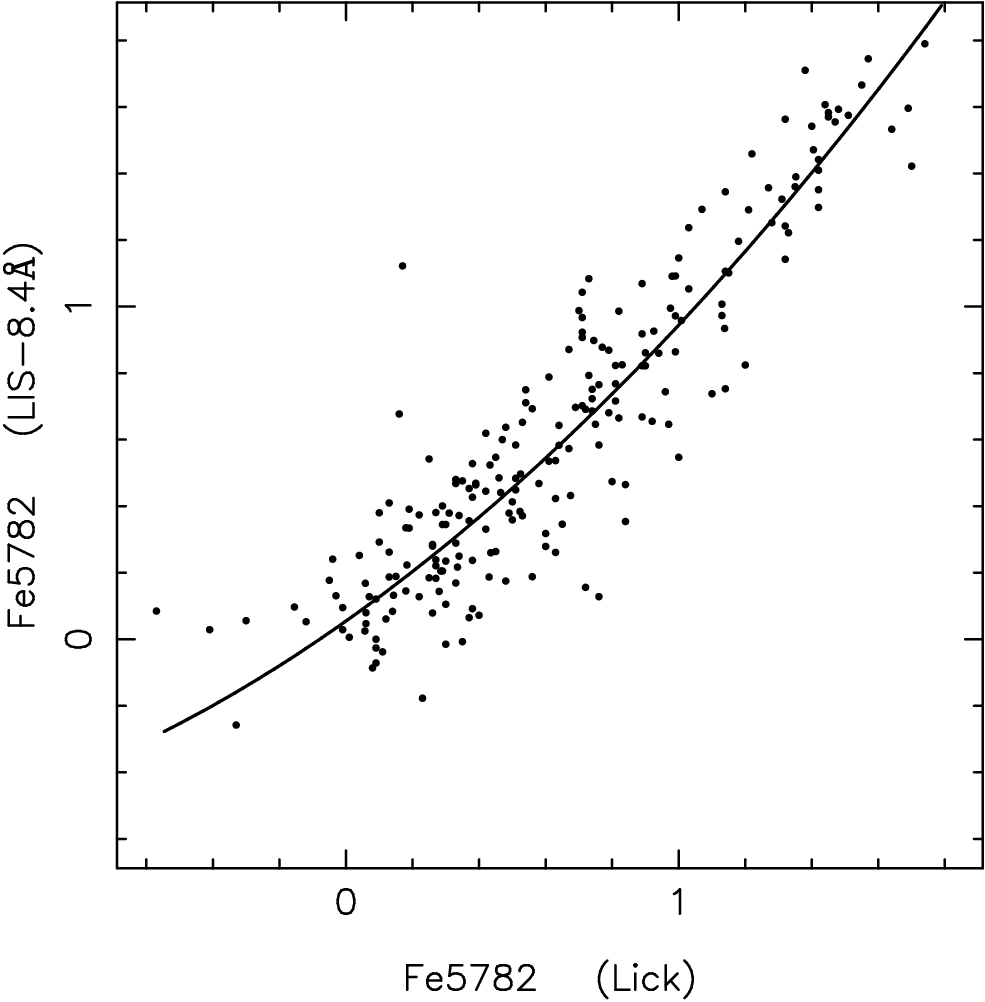}}
\resizebox{0.2\textwidth}{!}{\includegraphics[angle=-90]{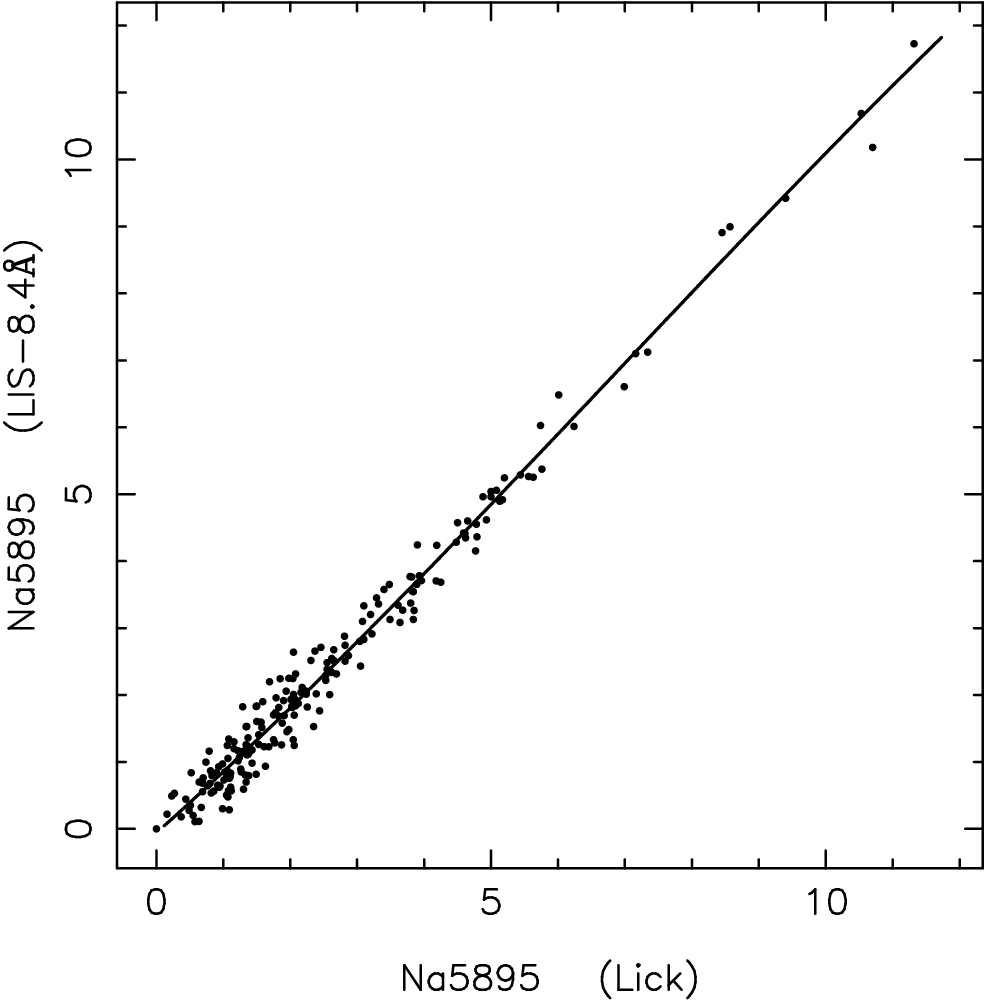}}
\resizebox{0.2\textwidth}{!}{\includegraphics[angle=-90]{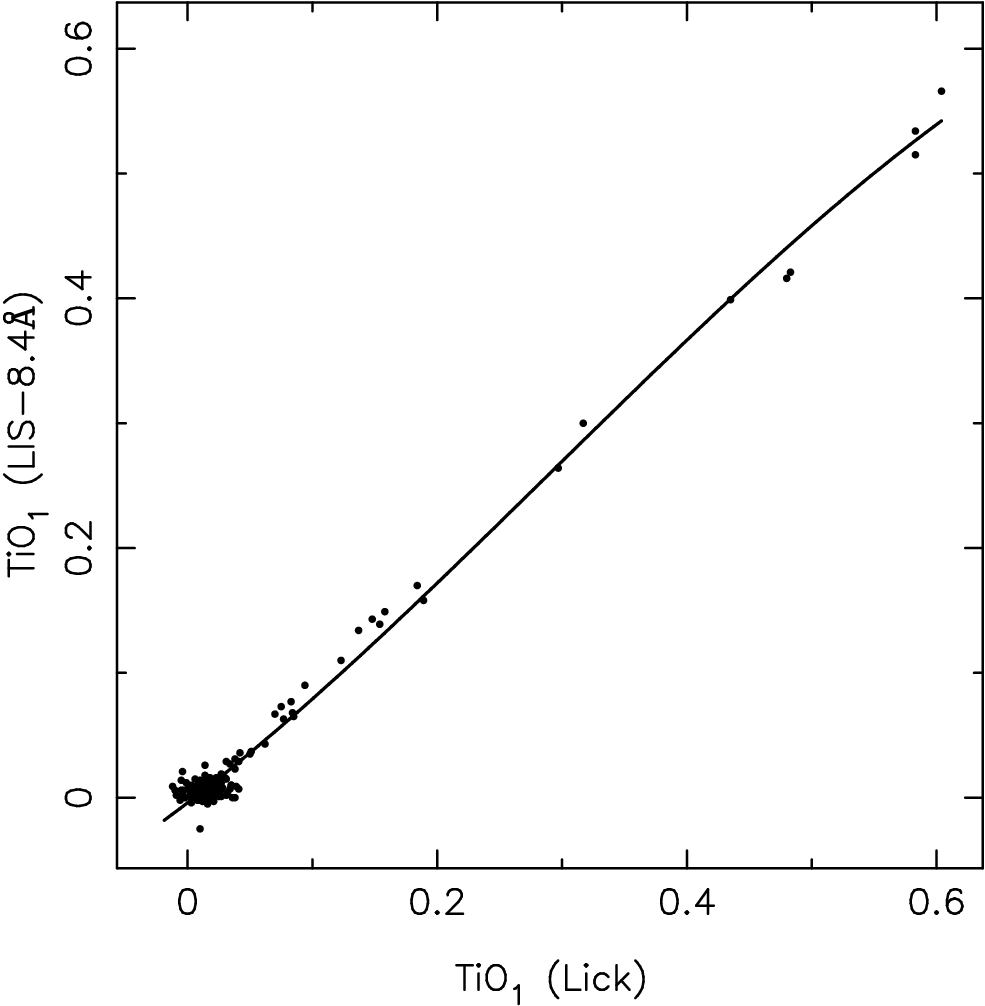}}
\resizebox{0.2\textwidth}{!}{\includegraphics[angle=-90]{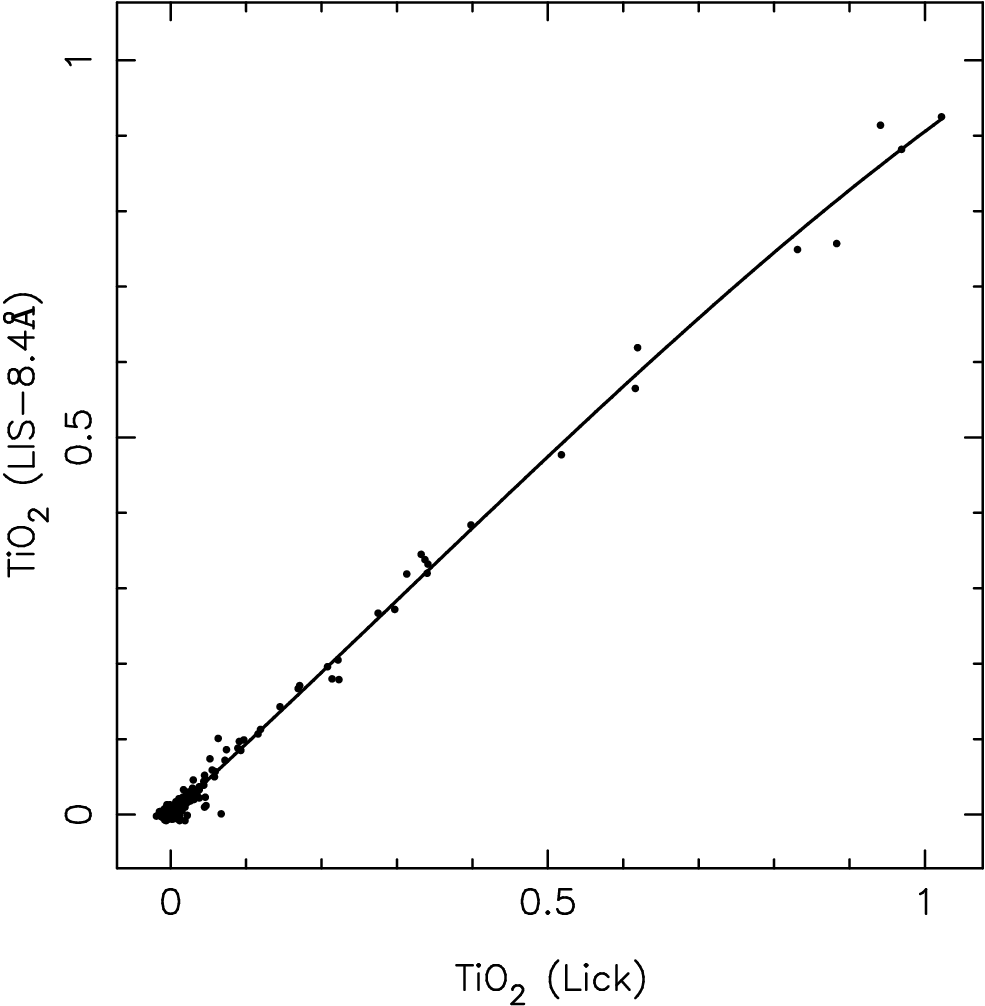}}
\resizebox{0.2\textwidth}{!}{\includegraphics[angle=-90]{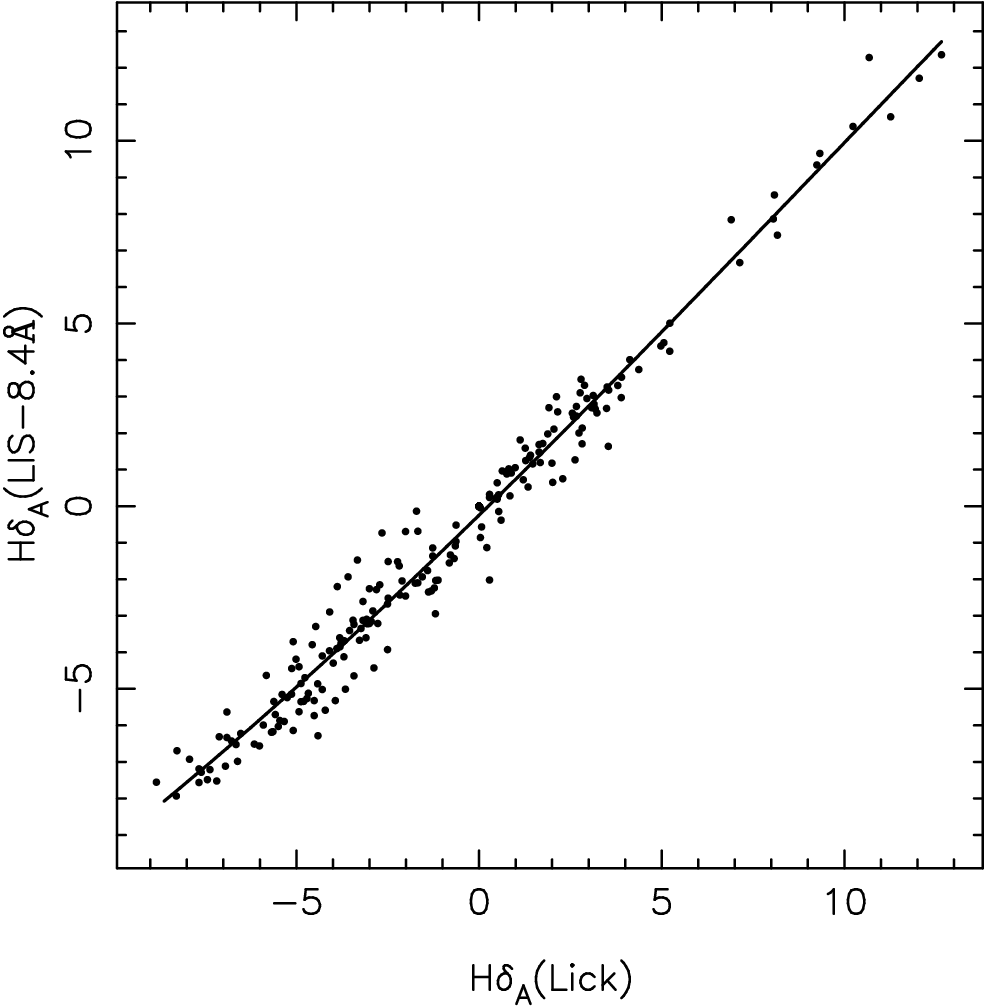}}
\resizebox{0.2\textwidth}{!}{\includegraphics[angle=-90]{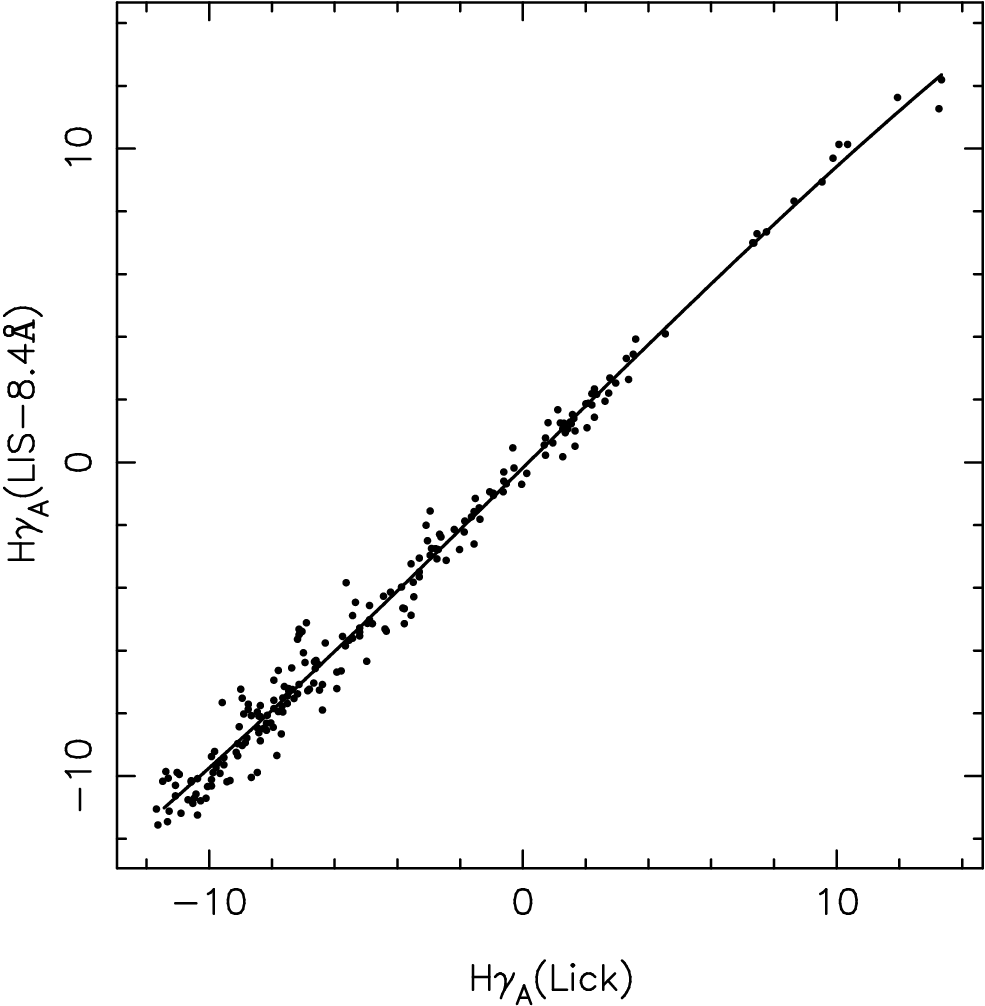}}
\resizebox{0.2\textwidth}{!}{\includegraphics[angle=-90]{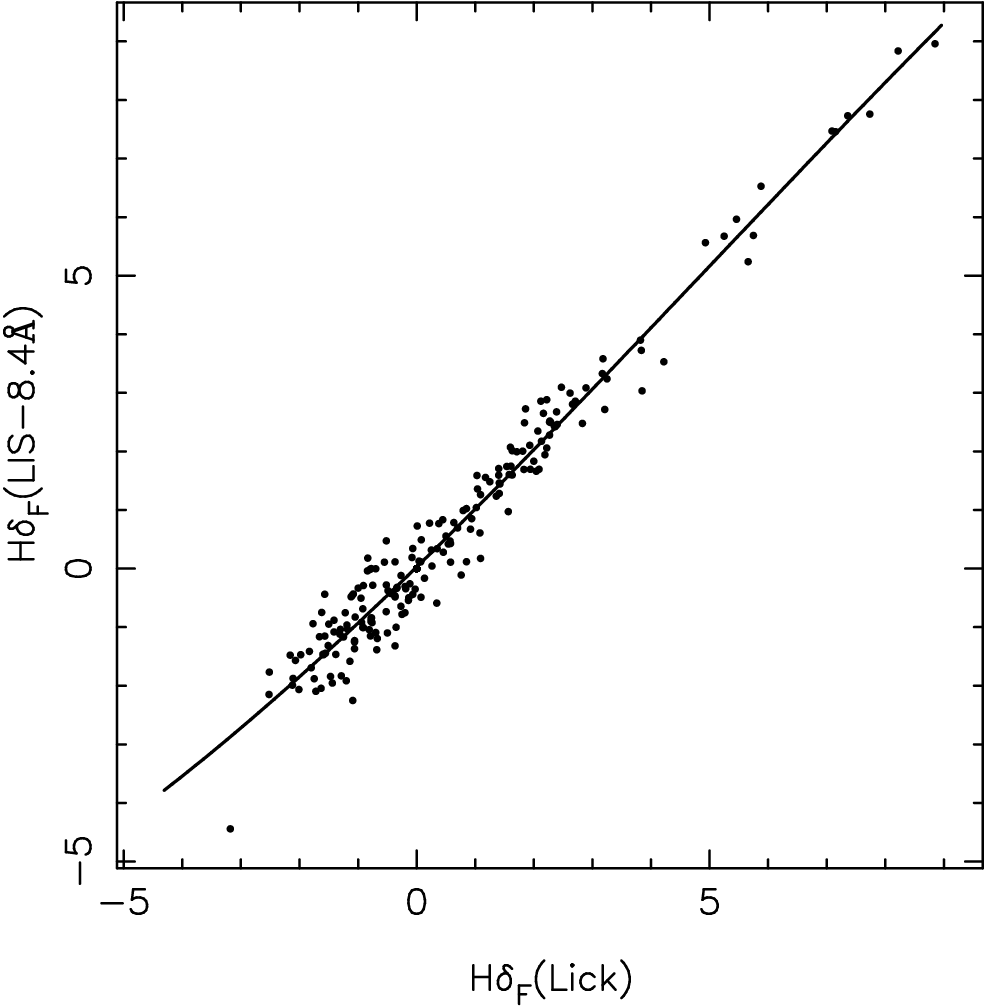}}
\resizebox{0.2\textwidth}{!}{\includegraphics[angle=-90]{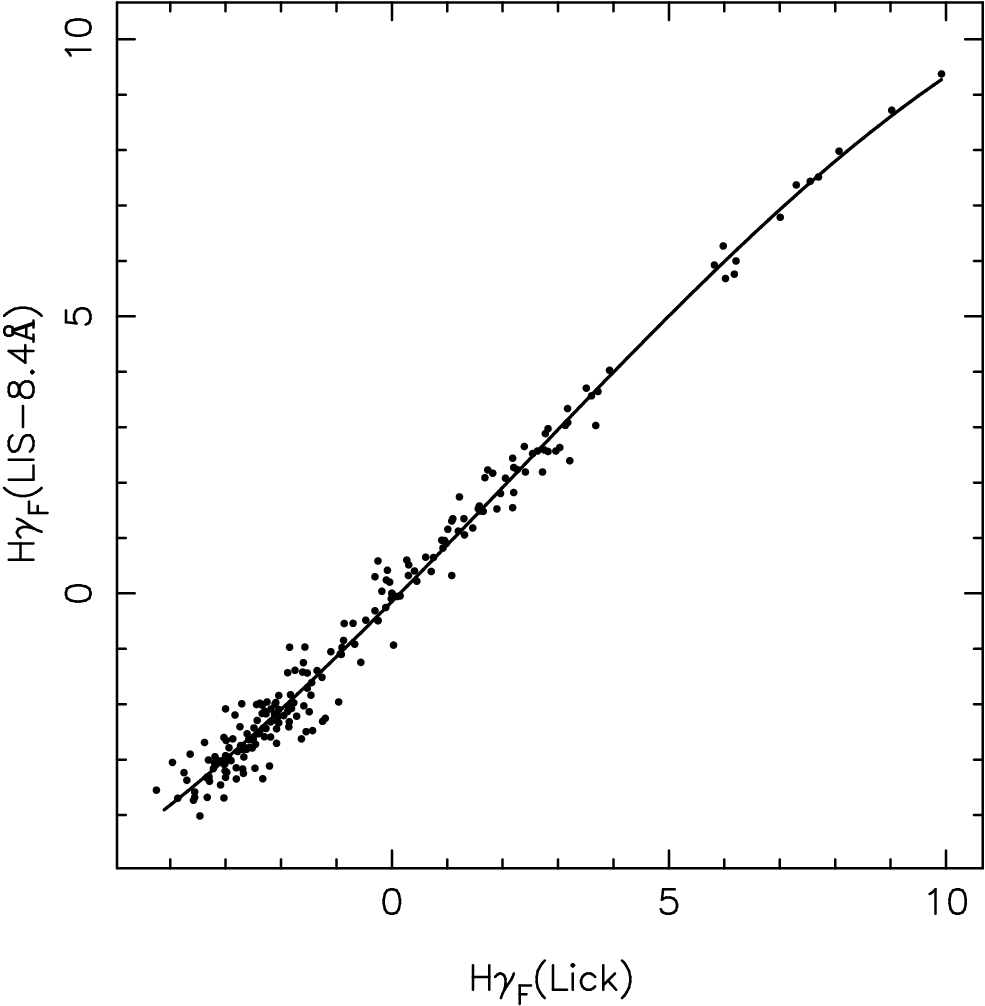}}
\caption{Comparison between the Lick/IDS indices of the Lick library
and those measured on the stars in common with the MILES library broadened to
FWHM=8.4\,\AA. Solid lines represent the transformation from one system to the
other (see Table~\ref{conversion.gal})}
\label{lick.gal}
\end{figure*}

\begin{figure*}
\resizebox{0.2\textwidth}{!}{\includegraphics[angle=-90]{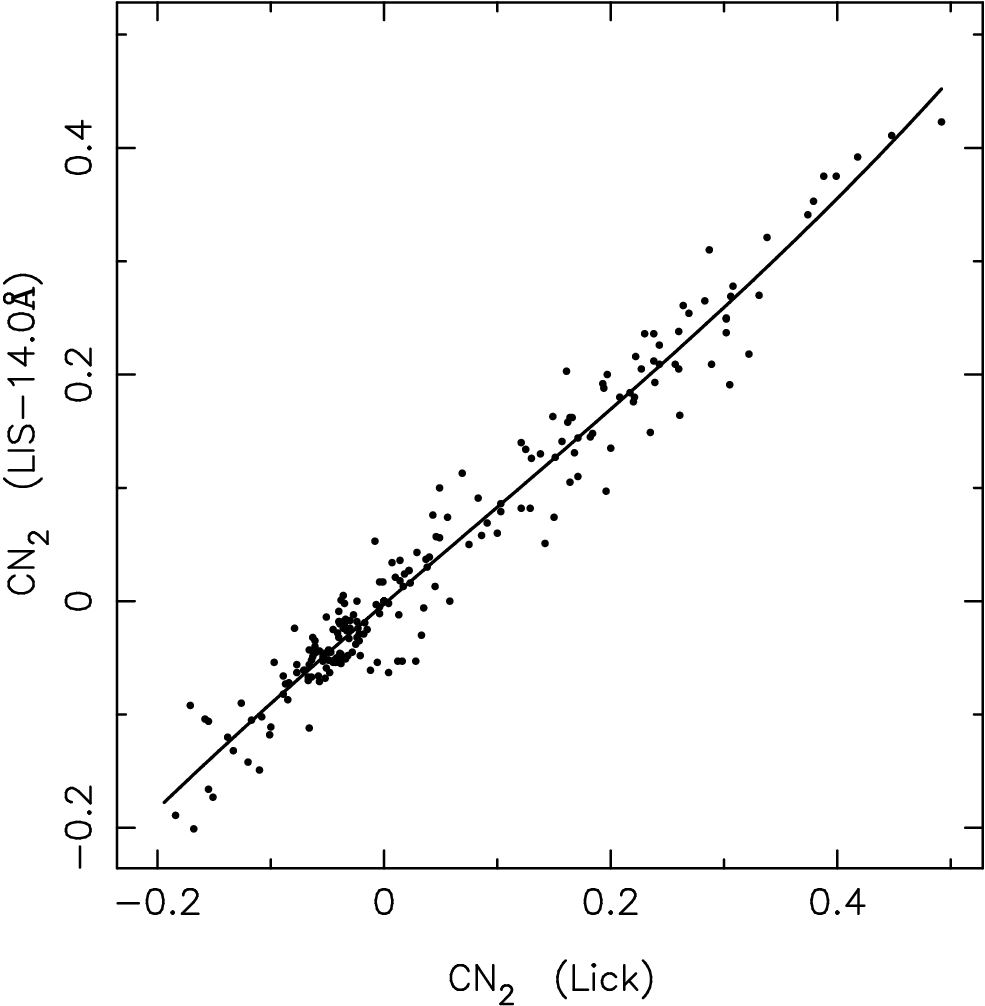}}
\resizebox{0.2\textwidth}{!}{\includegraphics[angle=-90]{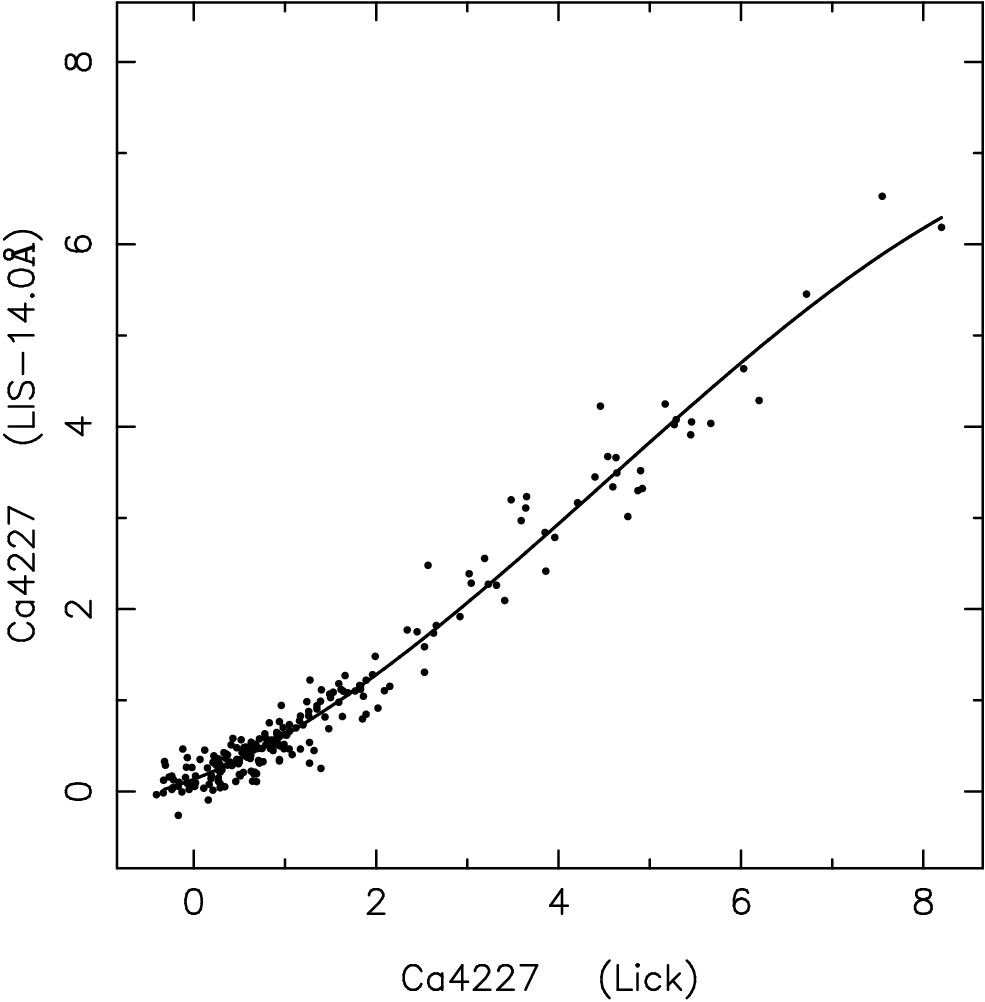}}
\resizebox{0.2\textwidth}{!}{\includegraphics[angle=-90]{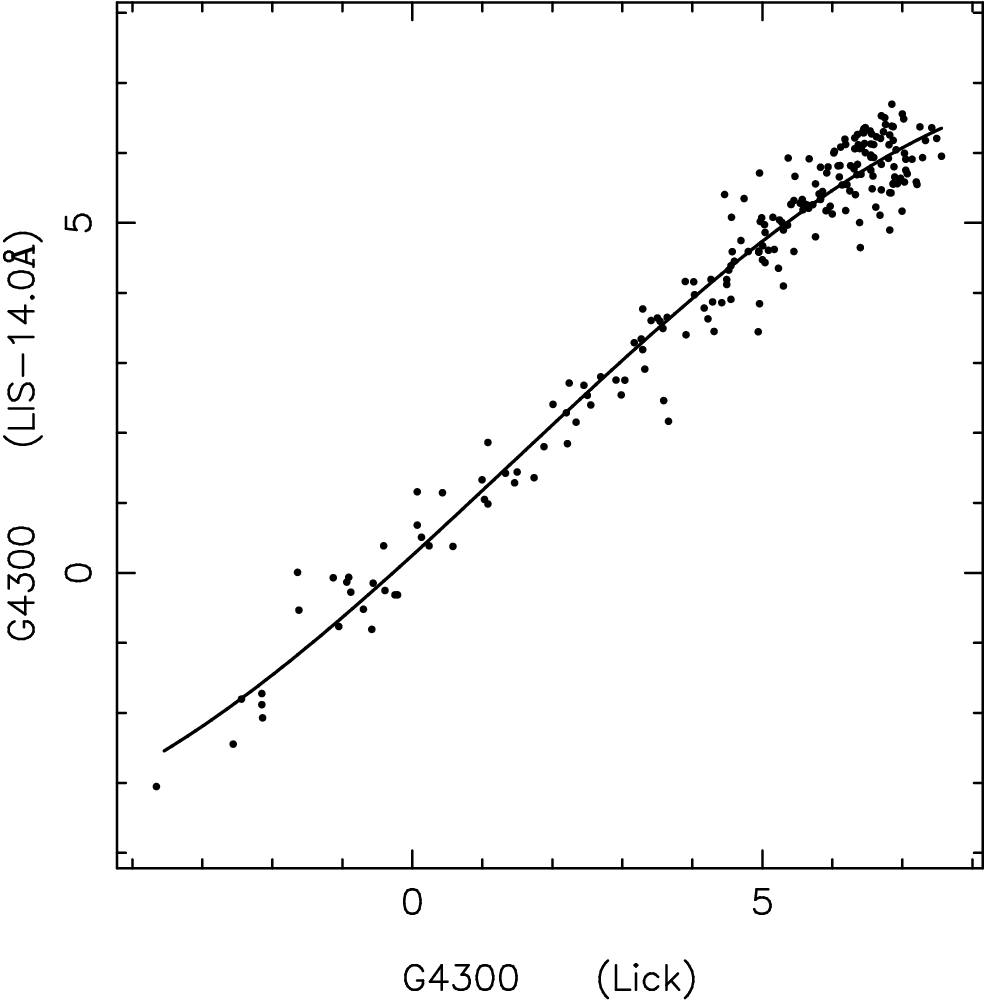}}
\resizebox{0.2\textwidth}{!}{\includegraphics[angle=-90]{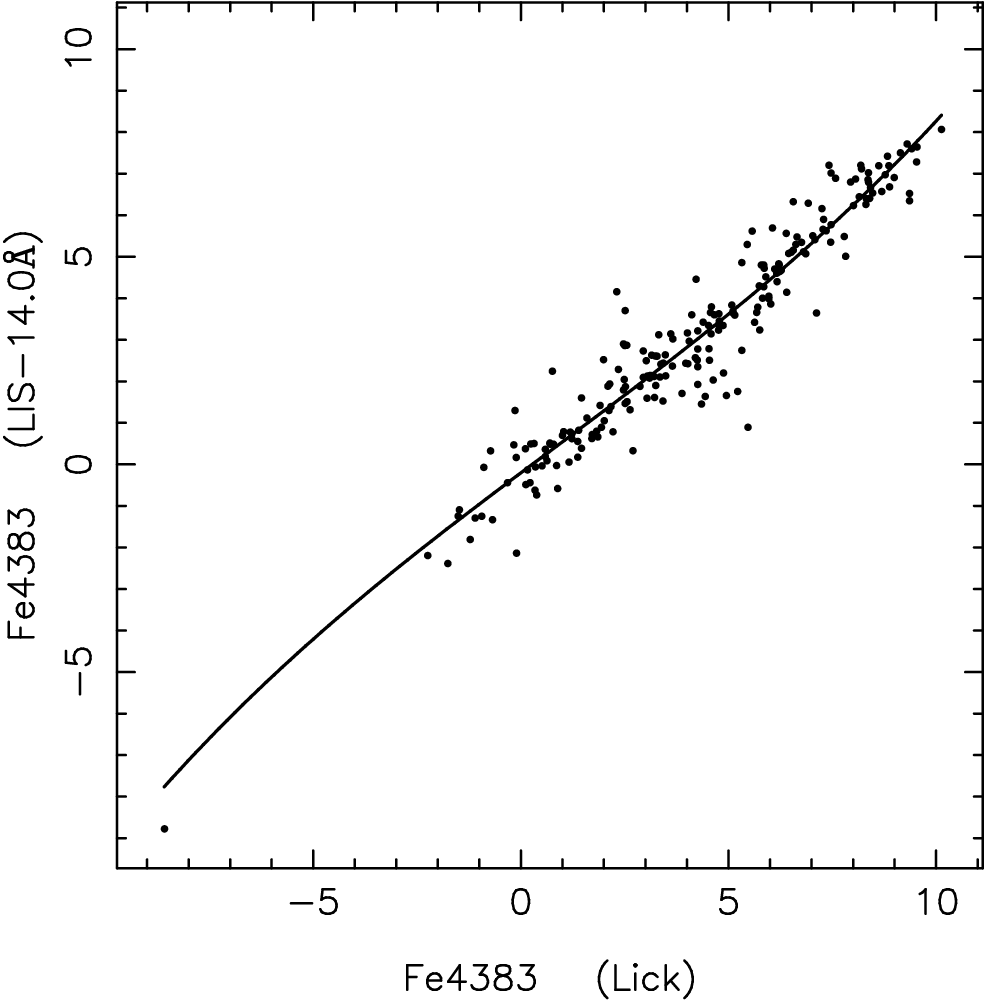}}
\resizebox{0.2\textwidth}{!}{\includegraphics[angle=-90]{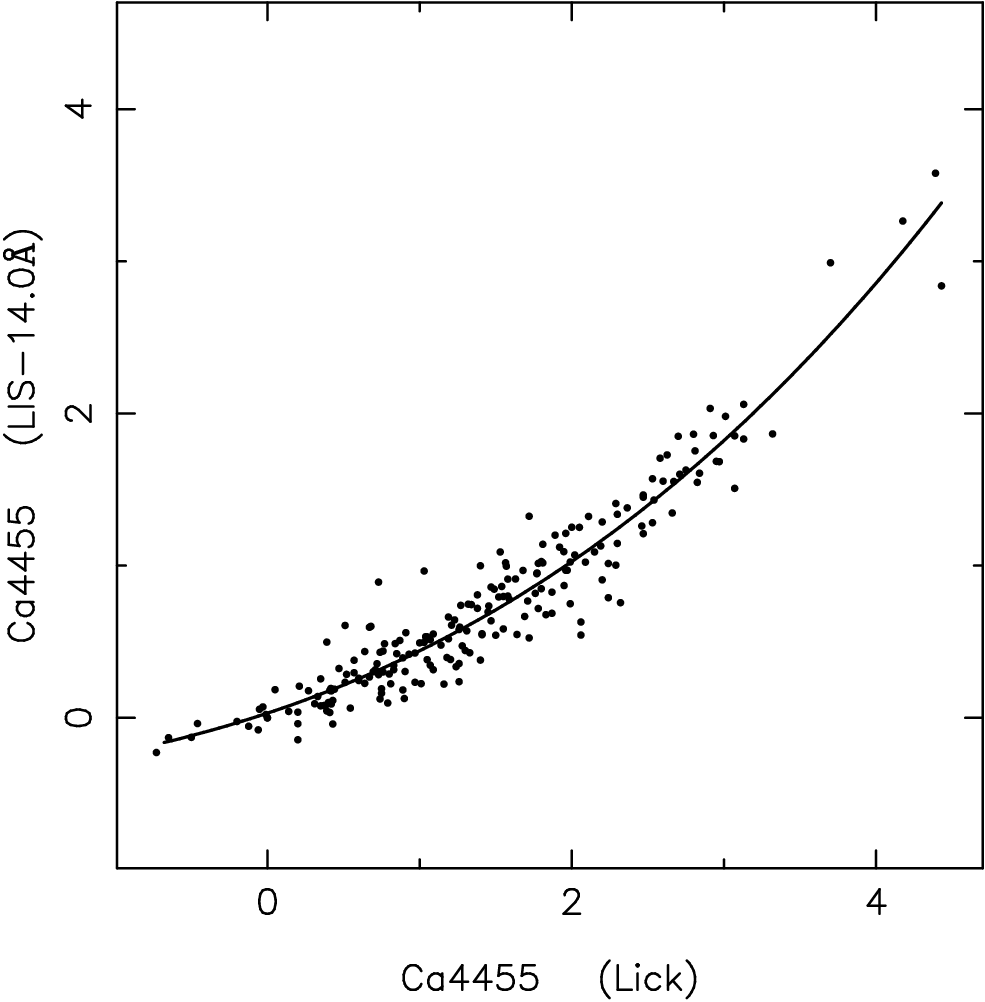}}
\resizebox{0.2\textwidth}{!}{\includegraphics[angle=-90]{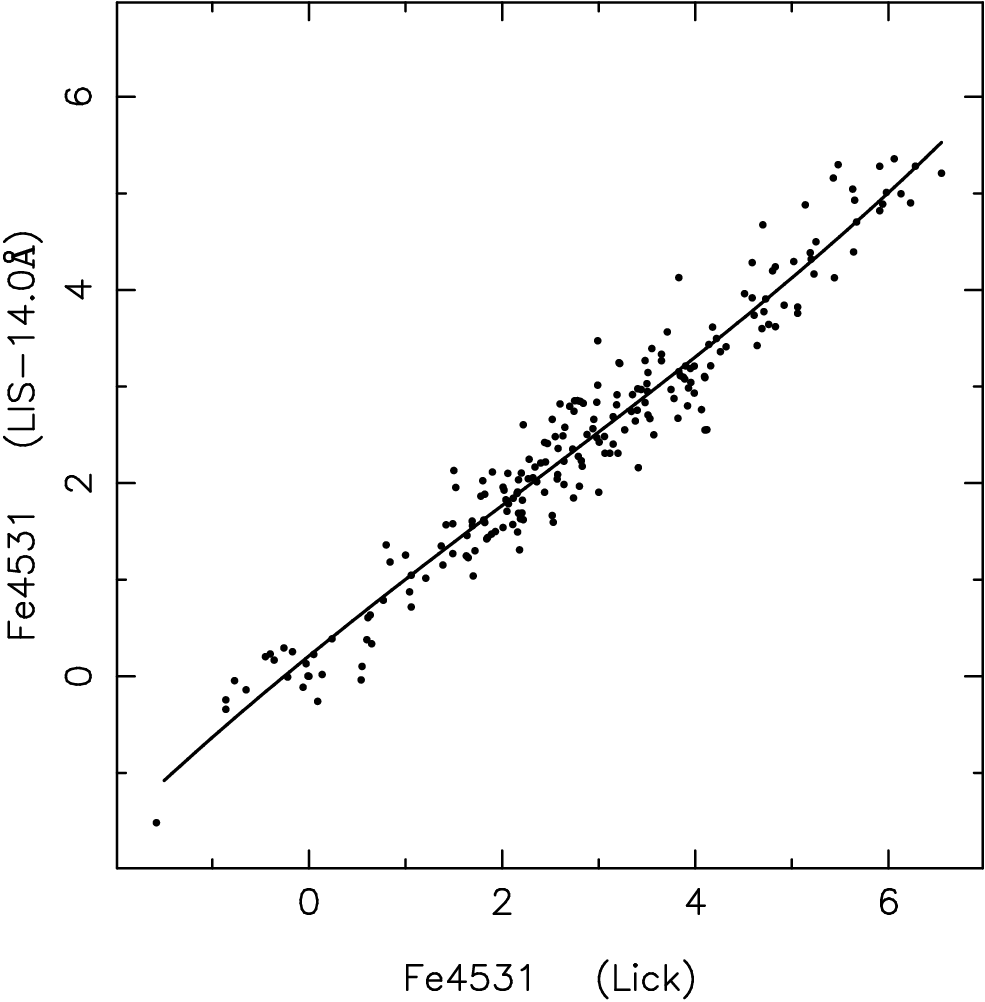}}
\resizebox{0.2\textwidth}{!}{\includegraphics[angle=-90]{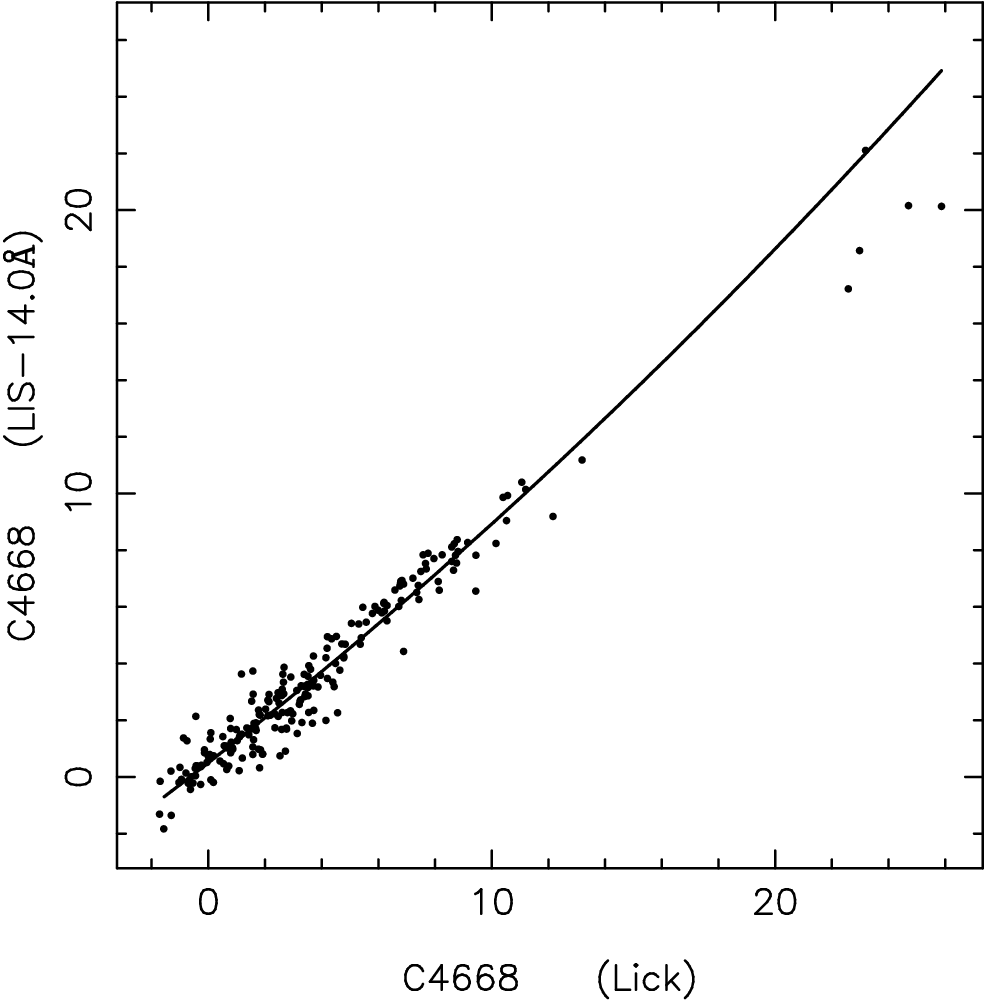}}
\resizebox{0.2\textwidth}{!}{\includegraphics[angle=-90]{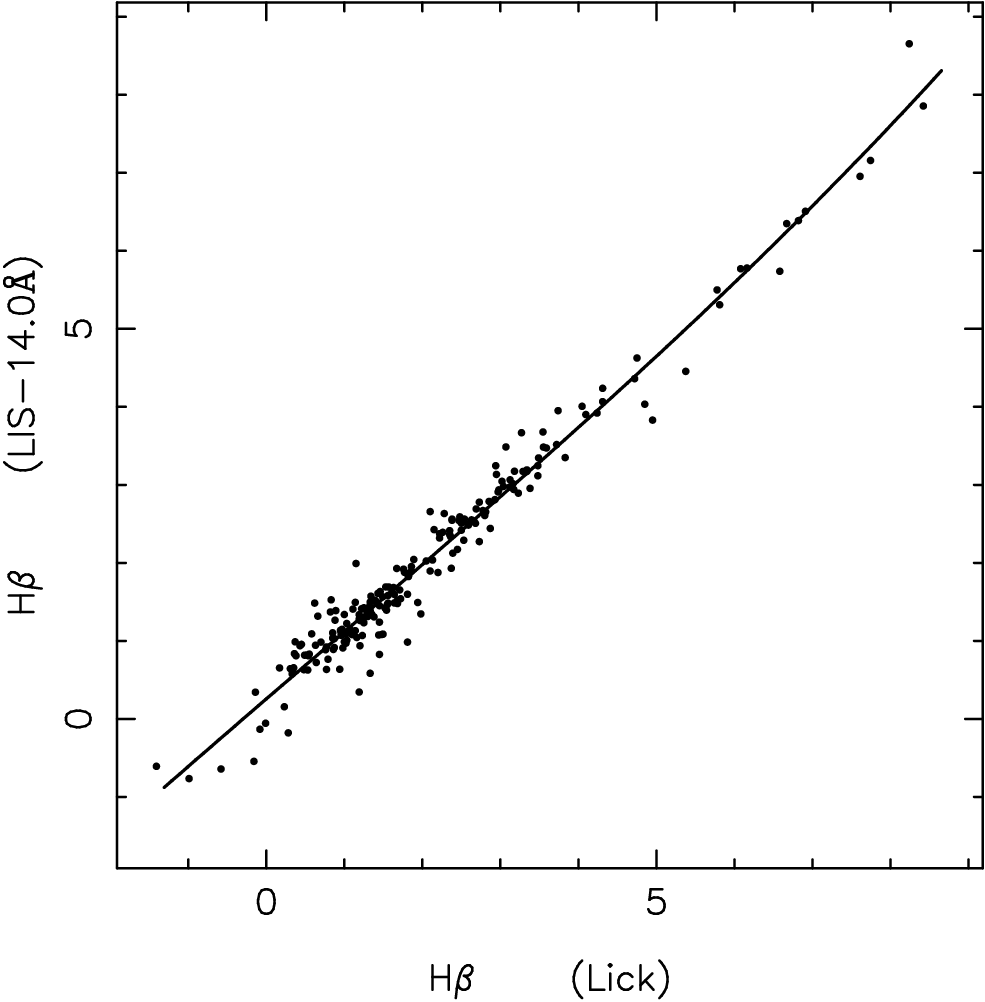}}
\resizebox{0.2\textwidth}{!}{\includegraphics[angle=-90]{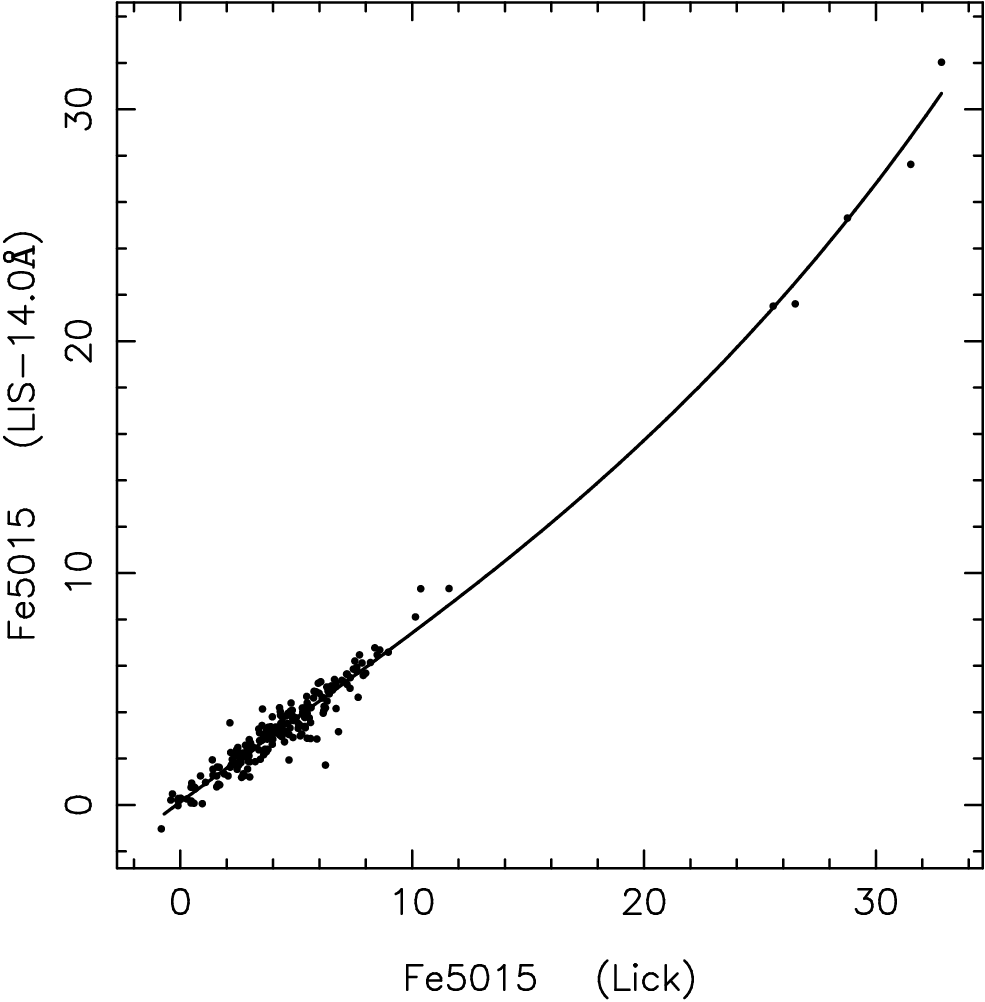}}
\resizebox{0.2\textwidth}{!}{\includegraphics[angle=-90]{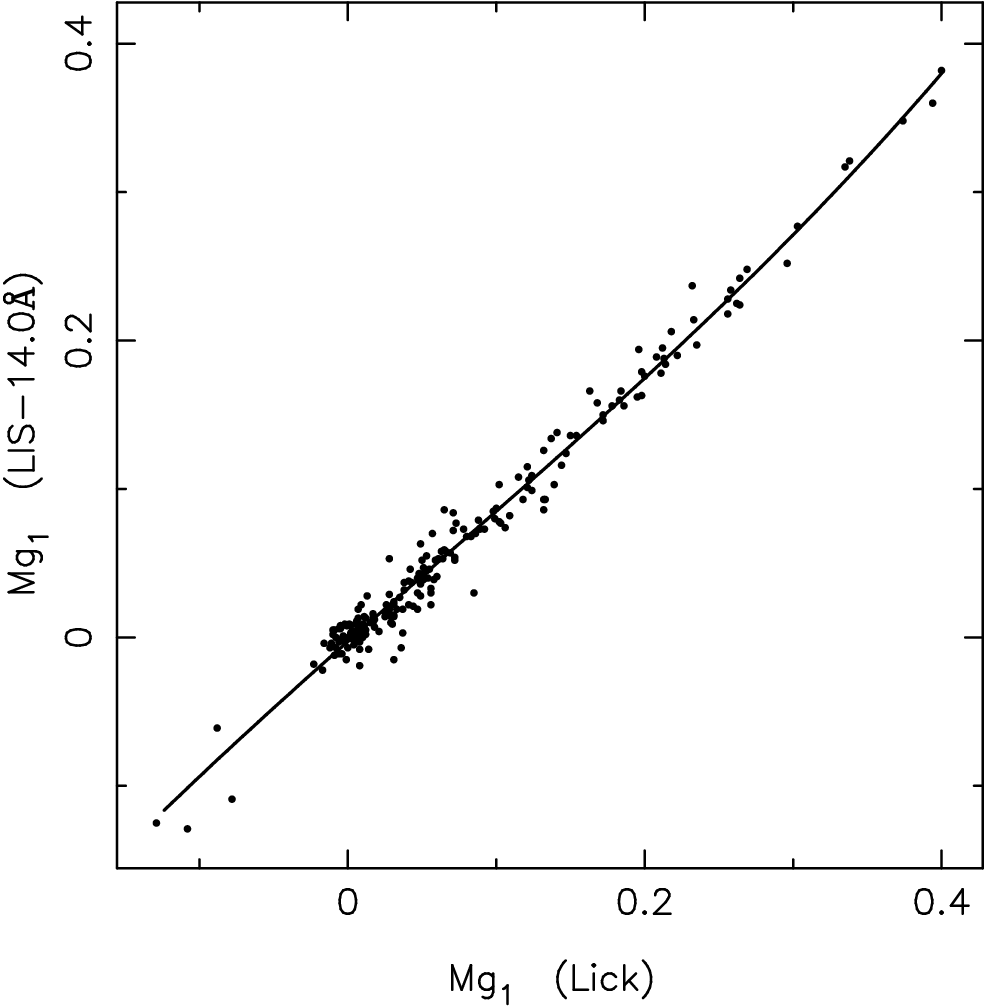}}
\resizebox{0.2\textwidth}{!}{\includegraphics[angle=-90]{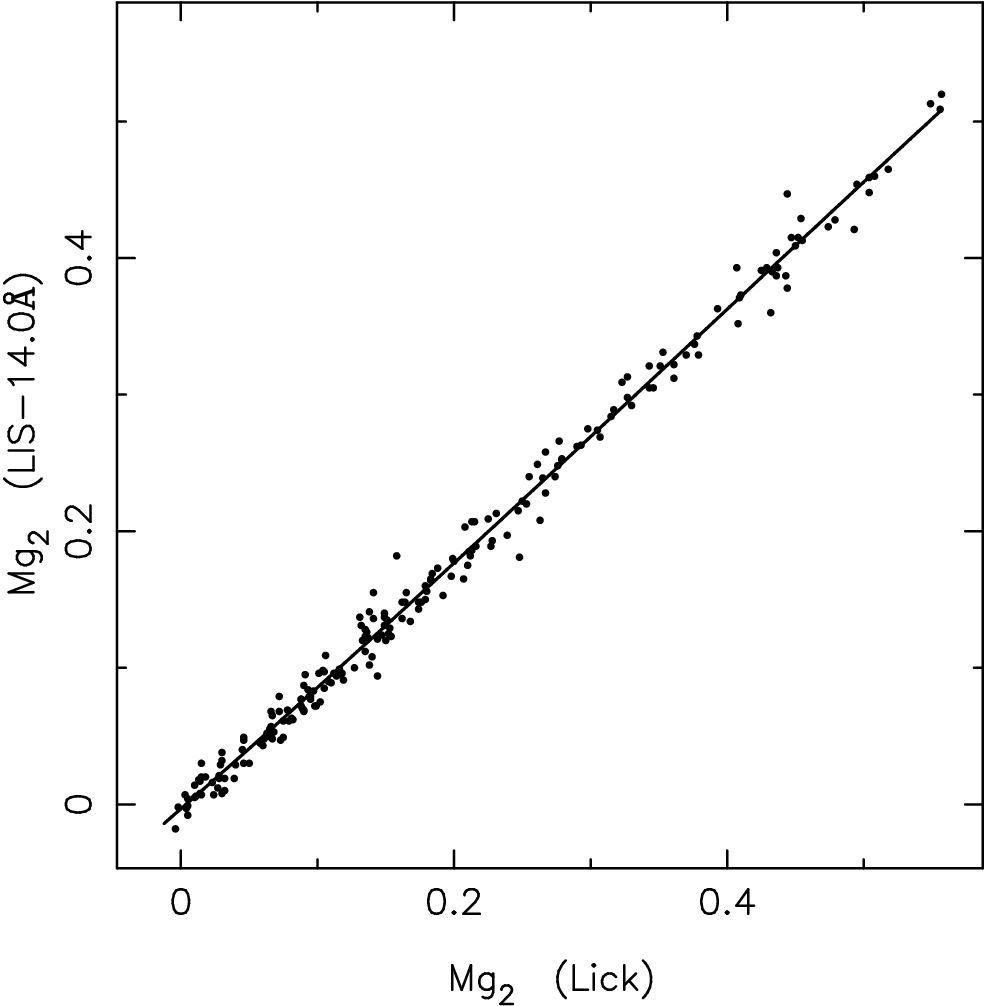}}
\resizebox{0.2\textwidth}{!}{\includegraphics[angle=-90]{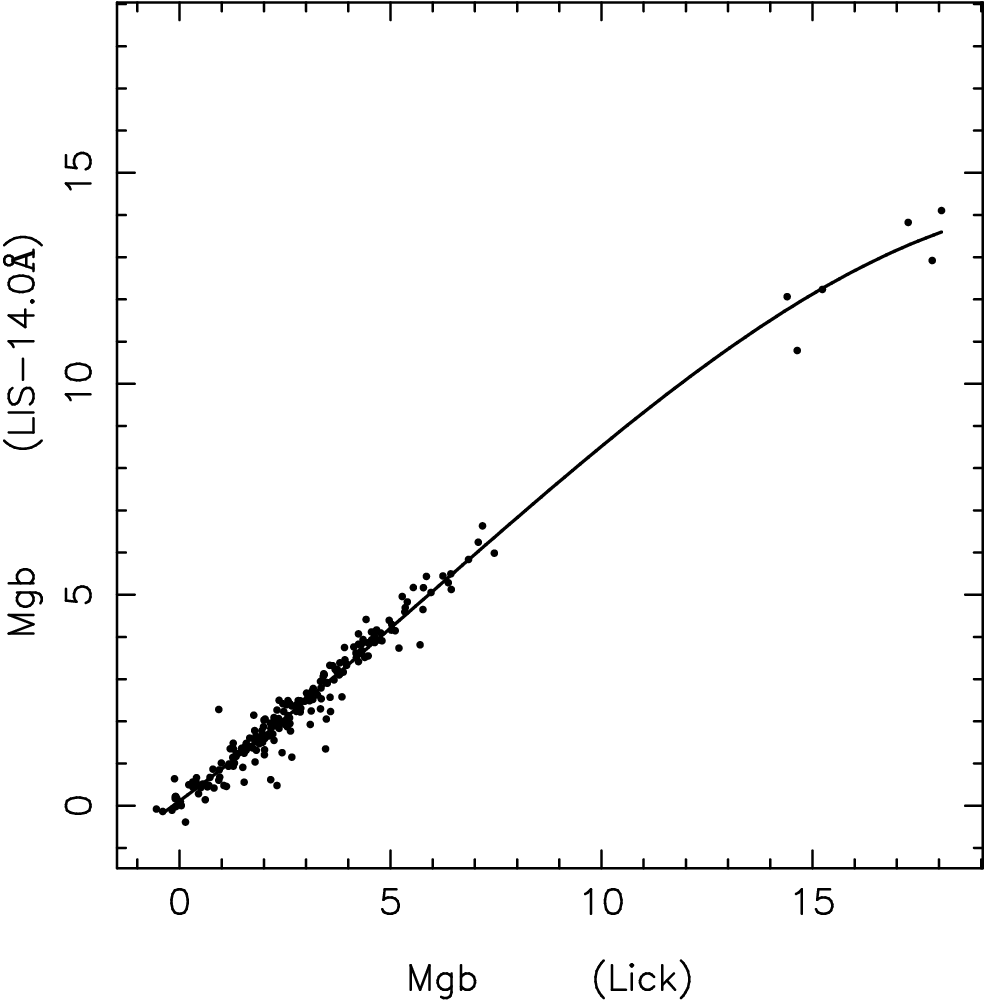}}
\resizebox{0.2\textwidth}{!}{\includegraphics[angle=-90]{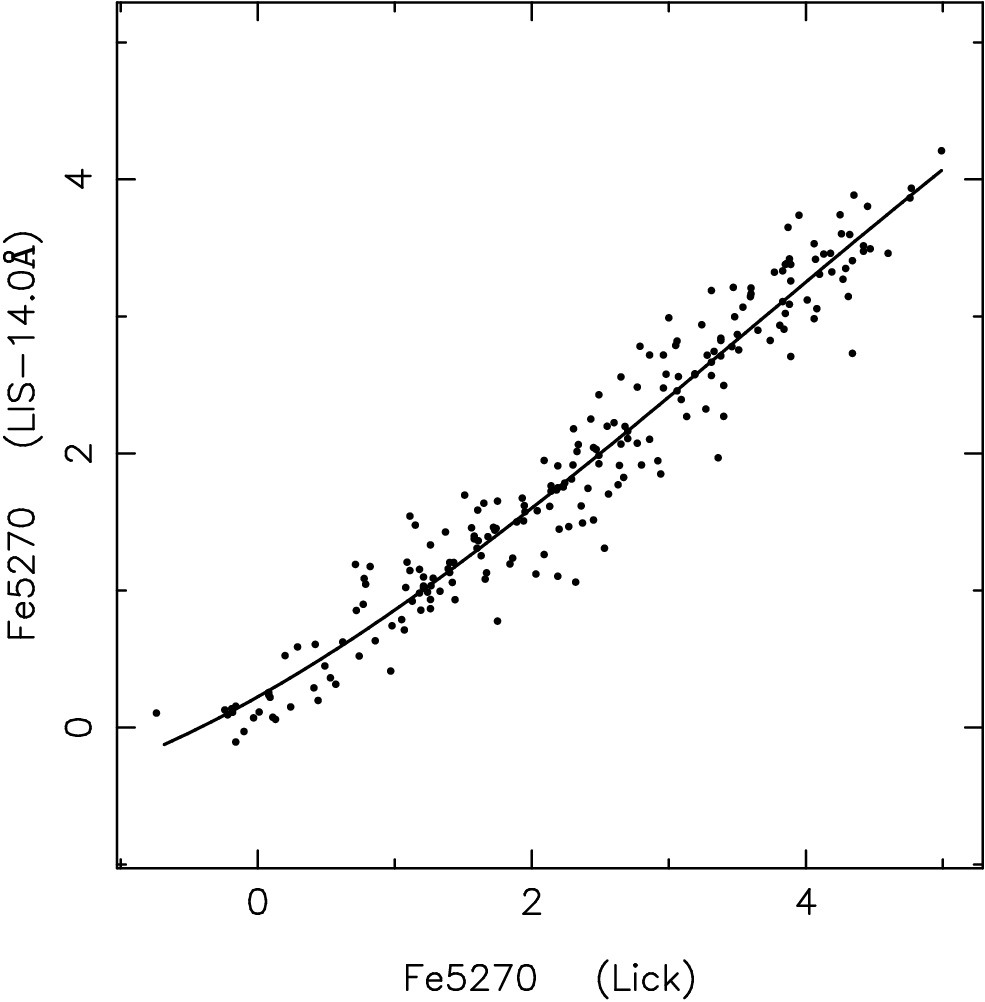}}
\resizebox{0.2\textwidth}{!}{\includegraphics[angle=-90]{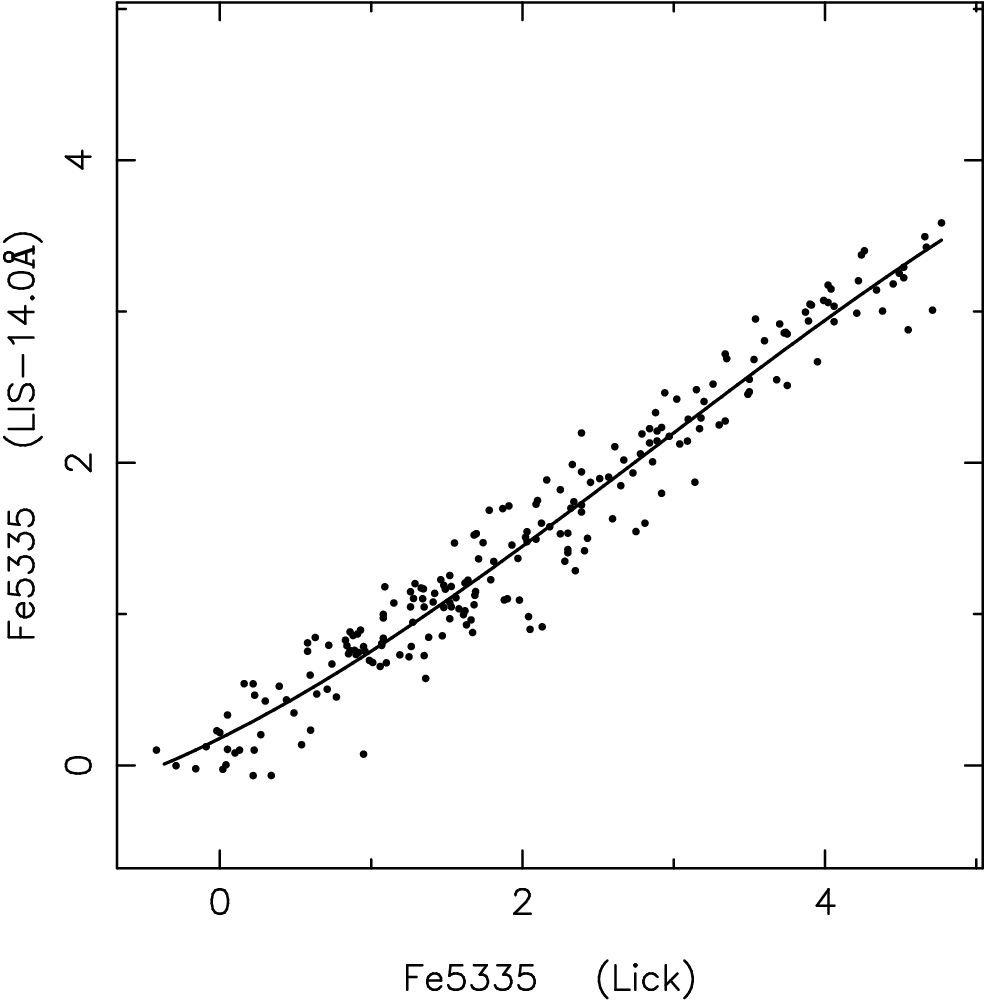}}
\resizebox{0.2\textwidth}{!}{\includegraphics[angle=-90]{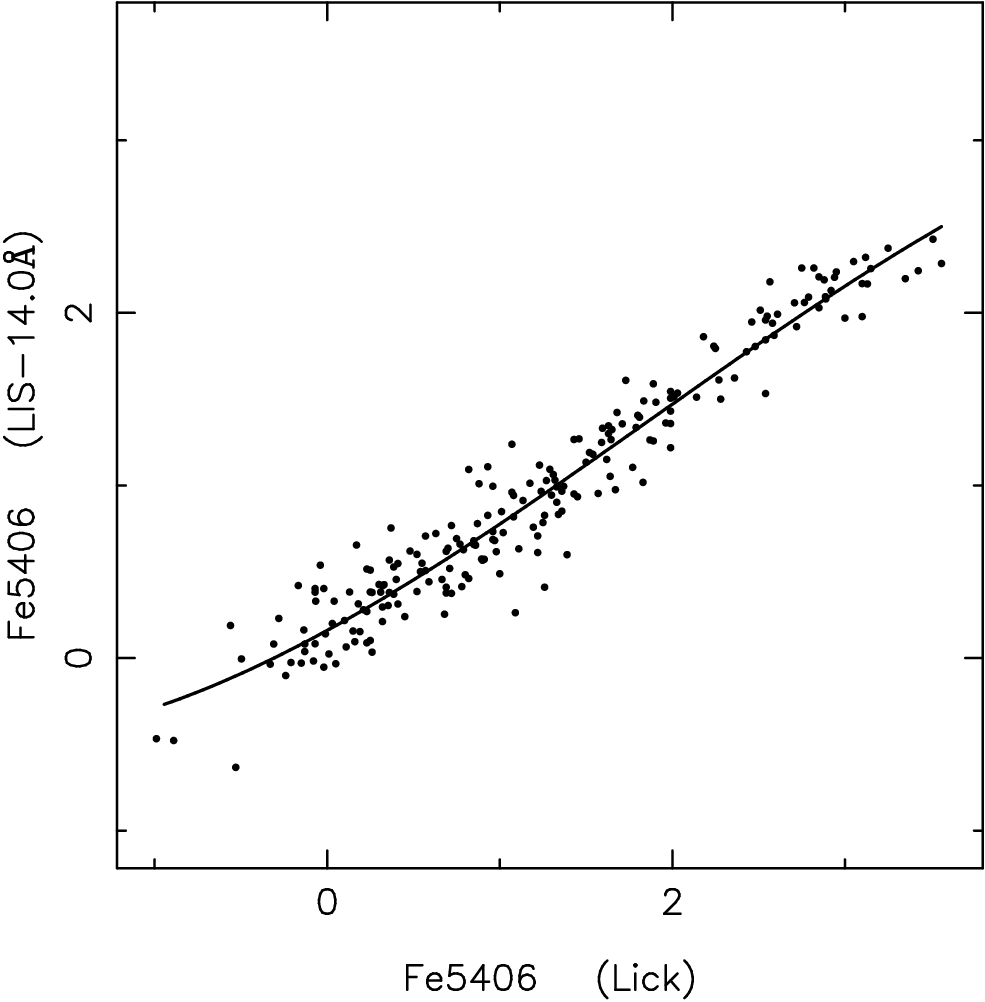}}
\resizebox{0.2\textwidth}{!}{\includegraphics[angle=-90]{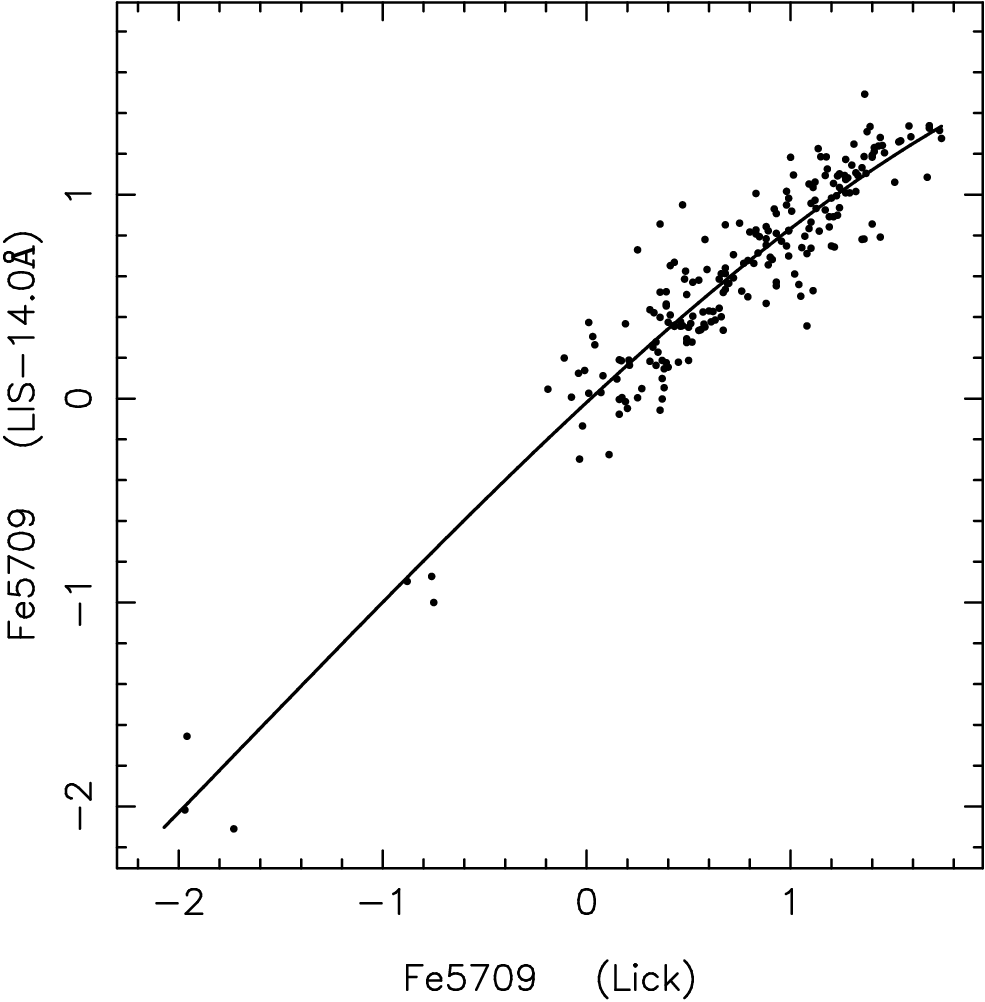}}
\resizebox{0.2\textwidth}{!}{\includegraphics[angle=-90]{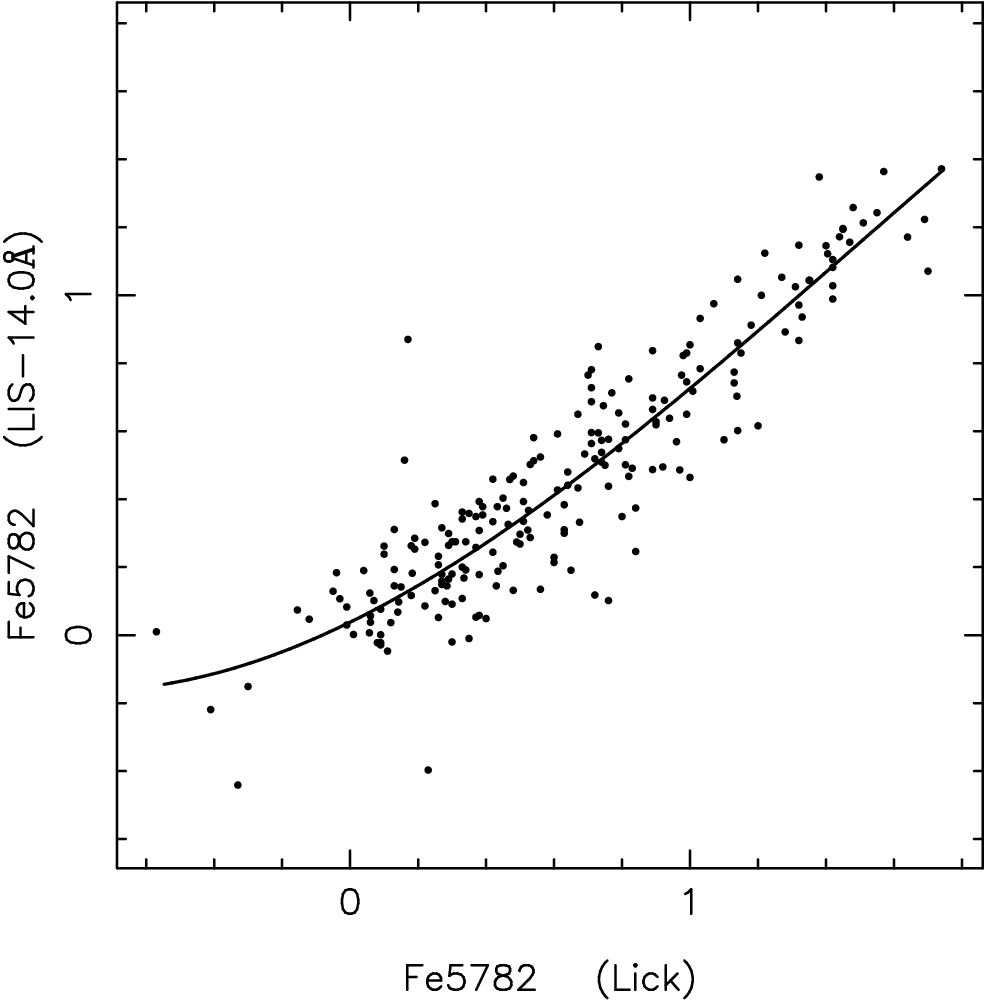}}
\resizebox{0.2\textwidth}{!}{\includegraphics[angle=-90]{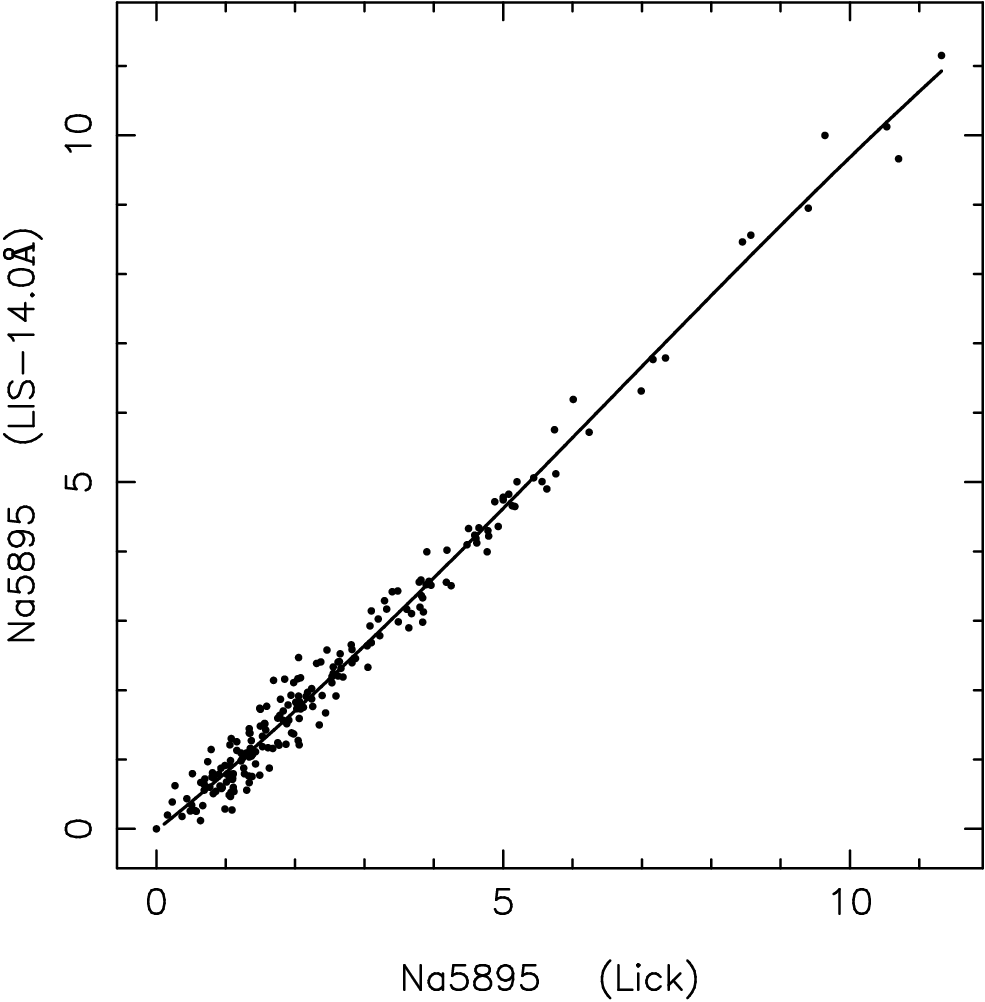}}
\resizebox{0.2\textwidth}{!}{\includegraphics[angle=-90]{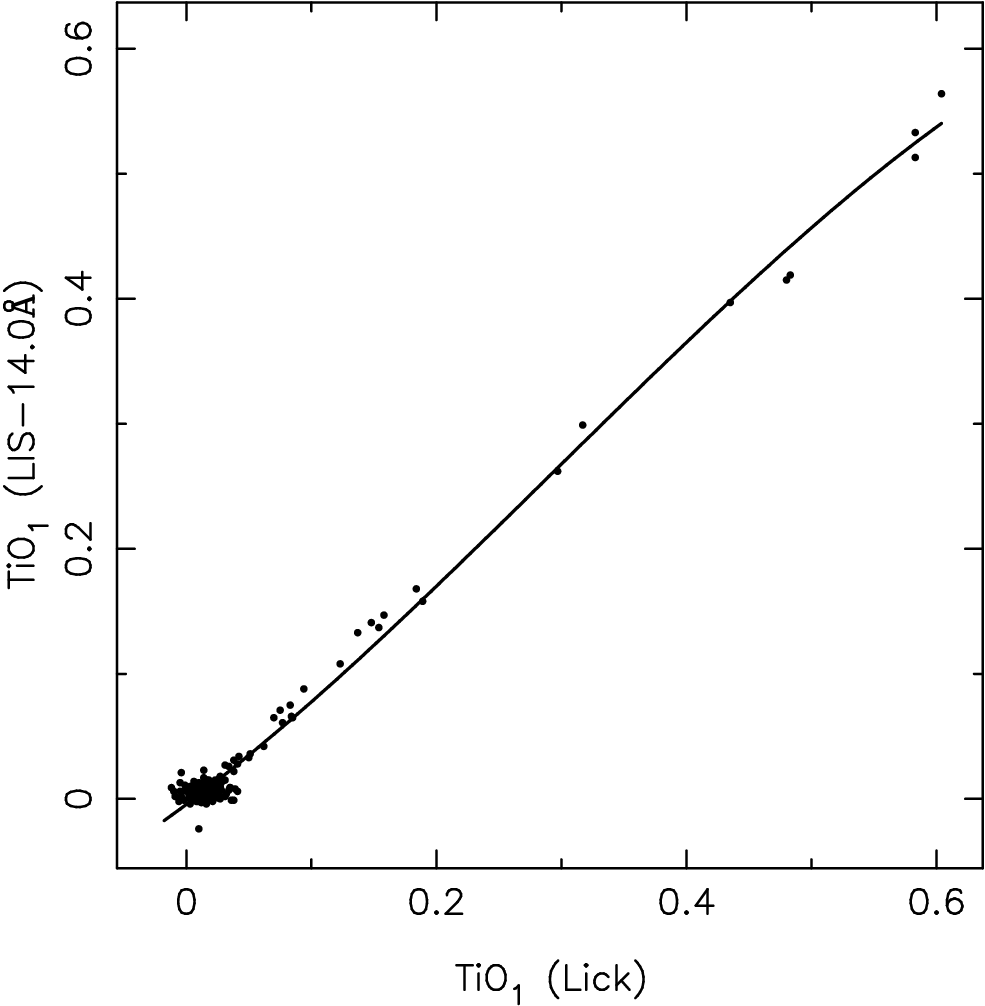}}
\resizebox{0.2\textwidth}{!}{\includegraphics[angle=-90]{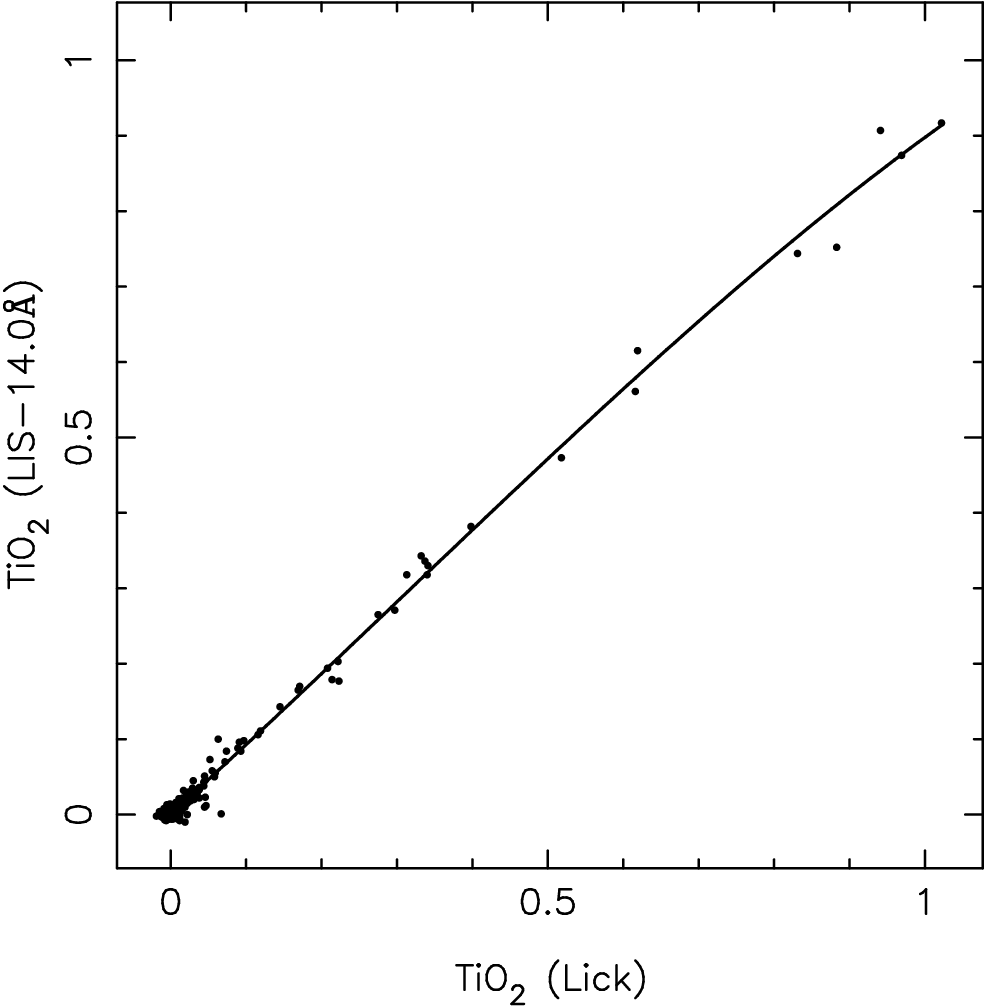}}
\resizebox{0.2\textwidth}{!}{\includegraphics[angle=-90]{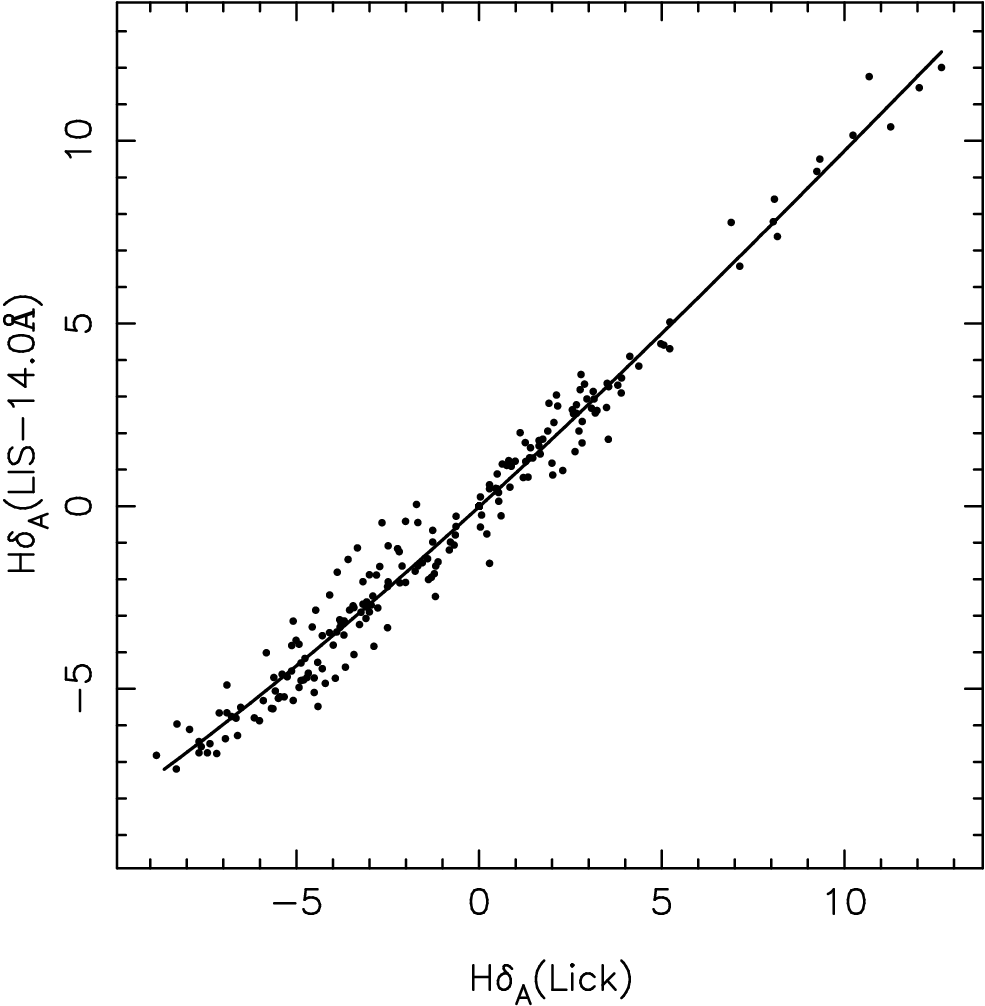}}
\resizebox{0.2\textwidth}{!}{\includegraphics[angle=-90]{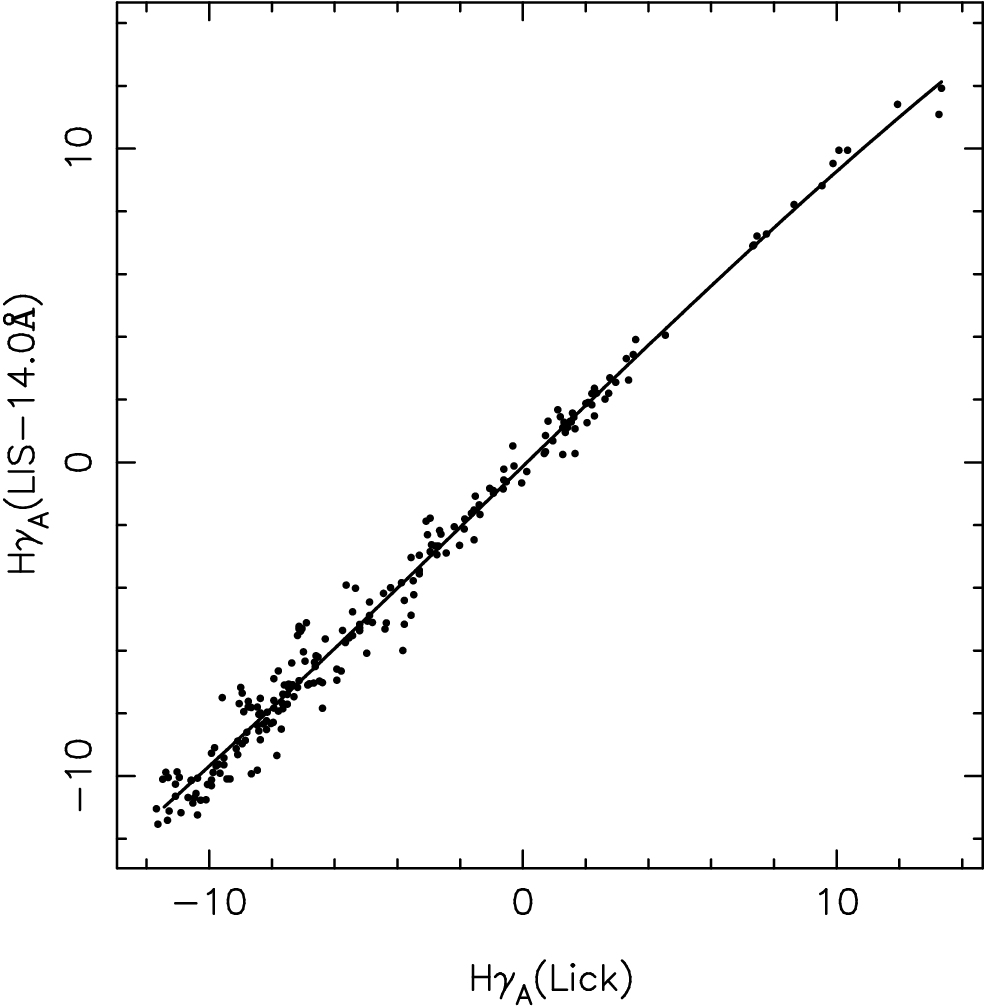}}
\resizebox{0.2\textwidth}{!}{\includegraphics[angle=-90]{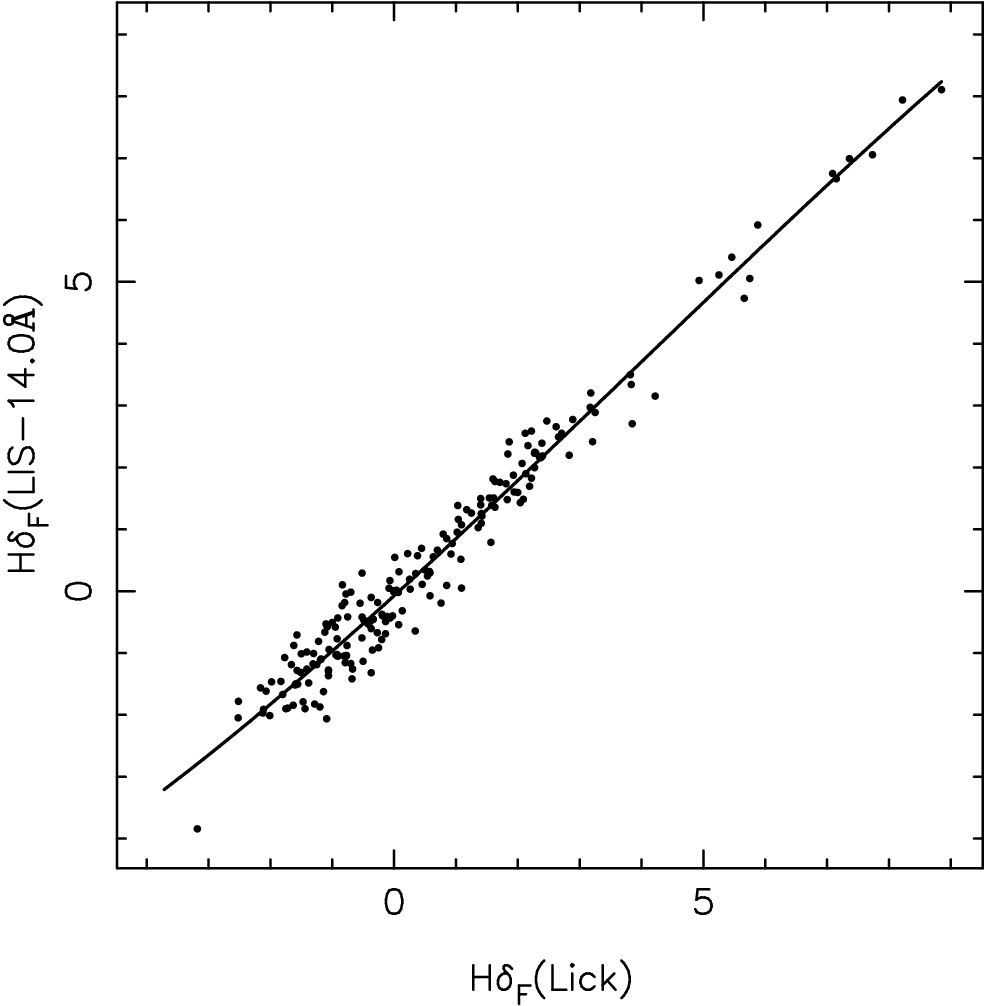}}
\resizebox{0.2\textwidth}{!}{\includegraphics[angle=-90]{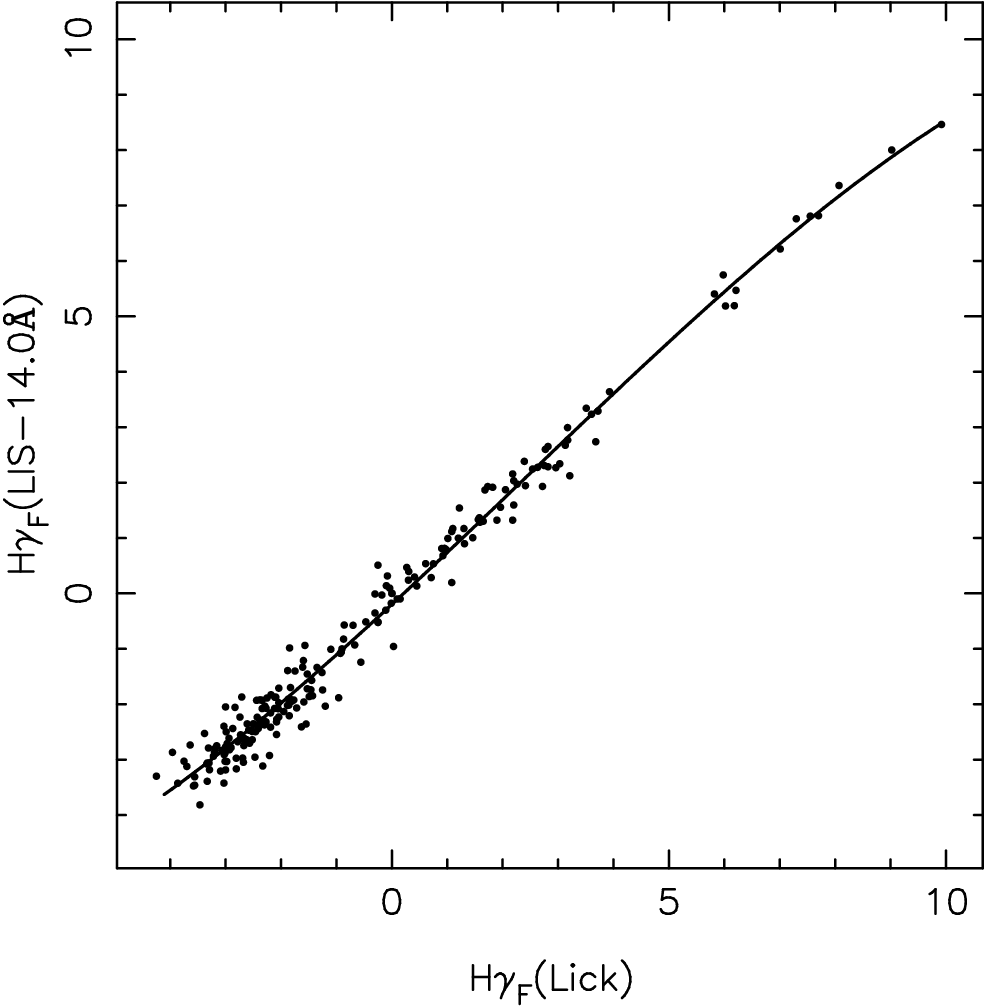}}
\caption{Comparison between the Lick/IDS indices of the Lick library and
those measured on the stars in common with the MILES library broadened to
FWHM=14\,\AA. Solid lines represent the transformation from one system to the other
(see Table~\ref{conversion.gal3})}
\label{lick.gal3}
\end{figure*}

In this Appendix we provide the required transformations that make it possible to convert
Lick/IDS index measurements to the new system proposed here. For this purpose we
have compared the index measurements of the 218 MILES stars in common with the
Lick library. The indices were measured once the MILES stellar spectra were
broadened to 5.0\,\AA, 8.4\,\AA\ and 14.0\,\AA\ to obtain the conversions
corresponding to LIS-5.0\AA, LIS-8.4\AA\ and LIS-14.0\AA, respectively. To
calculate these conversions we fitted 3$^{rd}$ order polynomials in all cases.
These polynomials are defined as follows:

\begin{equation} 
{\rm I_{LIS}={a_0}+{a_1}I_{Lick/IDS}+{a_2}I_{Lick/IDS}^2+{a_3}I_{Lick/IDS}^3}  
\label{eq:transf_Lick_LIS}
\end{equation}  

\noindent where I$_{\rm LIS}$ and I$_{\rm Lick/IDS}$ represent the index values
in the LIS (at either of the three resolutions) and Lick/IDS system,
respectively. Figure\,\ref{ plots} shows the results of the comparison between
the indices of the Lick/IDS library and of MILES at FWHM=5\,\AA\
resolution. Similarly, Fig.~\ref{lick.gal} and Fig.~\ref{lick.gal3} show the
comparison of the indices of the Lick/IDS stars and of the MILES stars
broadened to FWHM=8.4\,\AA\ and FWHM=14\,\AA, respectively. The coefficients of
these polinomia, which convert the Lick/IDS indices to the LIS-5.0\AA,
LIS-8.4\AA\ and LIS-14.0\AA, are given in Table~\ref{conversion.},
Table~\ref{conversion.gal} and Table~\ref{conversion.gal3}, respectively.

\begin{table}
\centering
\begin{tabular}{lrrrr}
\hline\hline
Index        &  a$_0$ & a$_1$ & a$_2$ &  a$_3$ \\
\hline
CN$_1$       &$ 0.018$& $0.941$ & $-0.137$ & $ 0.352$\\
CN$_2$       &$ 0.027$& $0.955$ & $-0.059$ & $ 0.106$\\
Ca4227       &$ 0.253$& $0.738$ & $ 0.177$ & $-0.018$\\
G4300        &$ 0.709$& $0.944$ & $ 0.029$ & $-0.006$\\
Fe4383       &$ 0.211$& $0.923$ & $-0.005$ & $ 0.001$\\
Ca4455       &$ 0.035$& $0.602$ & $ 0.250$ & $-0.032$\\
Fe4531       &$ 0.304$& $1.092$ & $-0.046$ & $ 0.002$\\
C4668        &$ 0.642$& $0.831$ & $ 0.017$ &  \\
H$\beta$     &$ 0.116$& $1.001$ & $-0.003$ & $0.001$\\ 
Fe5015       &$-0.154$& $1.162$ & $-0.025$ &  \\
Mg$_1$       &$-0.001$& $0.886$ & $ 0.146$ & $ 0.140$\\
Mg$_2$       &$-0.004$& $0.924$ & $ 0.013$ & $-0.028$\\
Mgb          &$ 0.116$& $0.950$ & $ 0.009$ & $-0.001$\\
Fe5270       &$ 0.256$& $0.705$ & $ 0.122$ & $-0.013$\\
Fe5335       &$ 0.259$& $0.805$ & $ 0.141$ & $-0.019$\\
Fe5406       &$ 0.289$& $0.629$ & $ 0.266$ & $-0.047$\\
Fe5709       &$ 0.006$& $1.045$ & $-0.052$ & $ 0.016$\\
Fe5782       &$ 0.119$& $0.467$ & $ 0.693$ & $-0.213$\\
Na5895       &$-0.074$& $0.921$ & $ 0.023$ & $-0.001$\\
TiO$_1$      &$-0.004$& $0.773$ & $ 0.723$ & $-0.826$\\
TiO$_2$      &$ 0.001$& $0.924$ & $ 0.116$ & $-0.132$\\ 
H$\delta_A$  &$-0.369$& $1.012$ & $ 1.012$ &      \\
H$\gamma_A$  &$-0.151$& $0.974$ & $ 0.001$ &      \\
H$\delta_F$  &$ 0.028$& $0.970$ & $ 0.037$ & $-0.003$\\
H$\gamma_F$  &$-0.174$& $1.050$ & $ 0.013$ & $-0.002$\\
\hline
\end{tabular}
\caption{Polynomials to transform from the Lick/IDS system to the new system
defined here for GC (LIS-5.0\AA) (see
Eq.~\ref{eq:transf_Lick_LIS}).\label{conversion.}} 
\end{table}

\begin{table}
\centering
\begin{tabular}{lrrrr}
\hline\hline
          &  a$_0$            & a$_1$           & a$_2$           &  a$_3$ \\
\hline
CN$_1$     &$ 0.011$ & $0.913$ & $-0.119$ & $ 0.400$\\
CN$_2$     &$ 0.016$ & $0.911$ & $ 0.025$ & $ 0.076$\\
Ca4227     &$ 0.217$ & $0.668$ & $ 0.140$ & $-0.014$\\
G4300      &$ 0.541$ & $0.540$ & $ 0.015$ & $-0.004$\\
Fe4383     &$ 0.106$ & $0.796$ & $ 0.015$ &         \\
Ca4455     &$ 0.021$ & $0.466$ & $ 0.187$ & $-0.016$\\
Fe4531     &$ 0.267$ & $0.972$ & $-0.032$ & $ 0.003$\\
C4668      &$ 0.616$ & $0.819$ & $ 0.017$ &         \\
H$\beta$   &$ 0.159$ & $0.953$ & $-0.002$ & $0.001$\\
Fe5015     &$-0.037$ & $0.994$ & $-0.018$ &         \\
Mg$_1$     &$-0.001$ & $0.884$ & $ 0.056$ & $0.358$\\
Mg$_2$     &$-0.004$ & $0.906$ & $ 0.054$ & $-0.051$\\
Mgb        &$ 0.116$ & $0.912$ & $ 0.009$ & $-0.001$\\
Fe5270     &$ 0.253$ & $0.649$ & $ 0.104$ & $-0.010$\\
Fe5335     &$ 0.214$ & $0.708$ & $ 0.120$ & $-0.015$\\
Fe5406     &$ 0.257$ & $0.587$ & $ 0.225$ & $-0.040$\\
Fe5709     &$-0.005$ & $1.024$ & $-0.108$ & $ 0.030$\\
Fe5782     &$ 0.055$ & $0.707$ & $ 0.182$ &       \\
Na5895     &$-0.057$ & $0.887$ & $ 0.025$ & $-0.001$\\
TiO$_1$    &$-0.004$ & $0.766$ & $ 0.736$ & $-0.841$\\
TiO$_2$    &$ 0.001$ & $0.919$ & $ 0.128$ & $-0.142$\\
H$\delta_A$&$-0.246$ & $0.977$ & $ 0.006$ &        \\
H$\gamma_A$&$-0.175$ & $0.986$ &          &        \\
H$\delta_F$&$ 0.027$ & $0.973$ & $ 0.016$ & $-0.001$\\
H$\gamma_F$&$-0.149$ & $1.010$ & $ 0.015$ & $-0.002$\\
\hline
\end{tabular}
\caption{Polynomials to transform from the Lick/IDS system to the new system
defined here for dwarf and intermediate-mass galaxies (LIS-8.4\AA).}
\label{conversion.gal}
\end{table}

\begin{table}
\centering
\begin{tabular}{lrrrr}
\hline\hline
          &  a$_0$            & a$_1$           & a$_2$           &  a$_3$ \\
\hline
CN$_1$     &$ 0.005$  & $0.875$ & $-0.091$ & $0.402$\\
CN$_2$     &$-0.002$  & $0.863$ & $-0.110$ & $0.477$\\
Ca4227     &$ 0.134$  & $0.380$ & $ 0.113$ & $-0.008$\\
G4300      &$ 0.251$  & $0.910$ & $ 0.019$ & $-0.004$\\
Fe4383     &$-0.203$  & $0.749$ & $-0.004$ & $ 0.001$\\ 
Ca4455     &$ 0.031$  & $0.331$ & $ 0.073$ & $ 0.005$\\
Fe4531     &$ 0.211$  & $0.814$ & $-0.025$ & $ 0.004$\\
C4668      &$ 0.500$  & $0.779$ & $ 0.006$ &         \\
H$\beta$   &$ 0.257$  & $0.861$ & $-0.003$ & $ 0.001$ \\
Fe5015     &$ 0.123$  & $0.738$ & $-0.004$ &       \\
Mg$_1$     &$-0.003$  & $0.885$ & $-0.164$ & $ 0.856$\\
Mg$_2$     &$-0.003$  & $0.874$ & $ 0.154$ & $-0.134$\\
Mgb        &$ 0.112$  & $0.738$ & $ 0.022$ & $-0.001$\\
Fe5270     &$ 0.224$  & $0.565$ & $ 0.076$ & $-0.007$\\
Fe5335     &$ 0.178$  & $0.496$ & $ 0.090$ & $-0.010$\\
Fe5406     &$ 0.160$  & $0.547$ & $ 0.084$ & $-0.015$\\
Fe5709     &$-0.018$  & $0.929$ & $-0.064$ & $-0.013$\\
Fe5782     &$ 0.038$  & $0.490$ & $ 0.254$ & $-0.056$\\
Na5895     &$-0.022$  & $0.801$ & $ 0.034$ & $-0.002$\\
TiO$_1$    &$-0.004$  & $0.749$ & $ 0.798$ & $-0.902$\\
TiO$_2$    &$ 0.000$  & $0.909$ & $ 0.144$ & $-0.155$\\
H$\delta_A$ &$-0.017$ & $0.914$ & $ 0.008$ &         \\
H$\gamma_A$ &$-0.126$ & $0.971$ &          &         \\
H$\delta_F$ &$-0.074$ & $0.909$ & $0.013$  & $-0.001$\\
H$\gamma_F$& $-0.199$ & $0.925$ & $0.014$  & $-0.002$\\
\hline
\end{tabular}
\caption{Polynomials to transform from the Lick/IDS system to the new system
defined here for massive galaxies (LIS-14.0\AA).}
\label{conversion.gal3}
\end{table}

\end{document}